\documentclass[a4paper,11pt]{article}
\pdfoutput=1 

\usepackage{jheppub} 

\usepackage[T1]{fontenc} 

\usepackage[dvipsnames]{xcolor}

\title{\boldmath Towards constraining Dark Matter at the LHC: Higher order QCD
  predictions for $t\bar{t}+Z(Z\to \nu_\ell \bar{\nu}_\ell)$}

\author[a]{G. Bevilacqua,}
\author[b]{H. B. Hartanto,}
\author[c]{M. Kraus,}
\author[d]{T. Weber }
\author[d]{and M. Worek }

\affiliation[a]{MTA-DE Particle Physics Research Group, University of
Debrecen, H-4010 Debrecen, PBox 105, Hungary} 
\affiliation[b]{Institute for
Particle Physics Phenomenology, Department of Physics, 
Durham University, Durham, DH1 3LE, UK} 
\affiliation[c]{Humboldt-Universit\"at zu
Berlin, Institut f\"ur Physik, Newtonstra\ss{}e 15, D-12489 Berlin,
Germany}
\affiliation[d]{ Institute for Theoretical Particle Physics
and Cosmology, RWTH Aachen University, D-52056 Aachen, Germany}
 
\emailAdd{\\
giuseppe.bevilacqua@science.unideb.hu,
  heribertus.b.hartanto@durham.ac.uk,
  manfred.kraus@physik.hu-berlin.de, \\tweber@physik.rwth-aachen.de,\\
  worek@physik.rwth-aachen.de}

\abstract{
Triggered by ongoing dark matter searches in the top quark sector at
the Large Hadron Collider we report on the calculation of the
next-to-leading order QCD corrections to the Standard Model process
$pp\to t\bar{t}+ Z(\to \nu_\ell \bar{\nu}_\ell)$. This calculation is
based on matrix elements for $e^+\nu_e \, \mu^- \bar{\nu}_\mu \,
b\bar{b} \, \nu_\tau \bar{\nu}_\tau$ production and includes all
non-resonant diagrams, interferences, and off-shell effects of the top
quarks.  Non-resonant and off-shell effects due to the finite
$W$-boson width are also consistently taken into account.  As it is
common for such studies, we present results for both integrated and
differential cross sections for a few renormalisation and
factorisation scale choices and three different parton distribution
functions. Already with the fairly inclusive cut selection and
independently of the scale choice and the parton distribution function
non-flat differential ${\cal K}$-factors are obtained for $p_T^{miss},
\Delta \phi_{\ell\ell}, \Delta y_{\ell\ell}, \cos\theta_{\ell\ell}
,H_T, H^\prime_T$ observables that are relevant for new physics
searches. Good theoretical control over the Standard Model background is
a fundamental prerequisite for a correct interpretation of possible
signals of new physics that may arise in this channel. Thus, these
observables need to be carefully reexamined in the presence of more
exclusive cuts before any realistic strategies for the detection of
new physics signal can be further developed. Since from the
experimental point of view both $t\bar{t}$ and $t\bar{t}+Z(Z\to
\nu_\ell\bar{\nu}_\ell)$ comprise the same final states, we
additionally study the impact of the enlarged missing transverse
momentum on various differential cross section distributions. To this
end normalised differential distributions for $pp\to e^+ \nu_e \,
\mu^- \bar{\nu}_\mu \, b \bar{b} \, \nu_\tau \bar{\nu}_\tau$ and
$pp\to e^+ \nu_e \, \mu^- \bar{\nu}_\mu \, b \bar{b}$ are compared.}

\dedicated{\rm TTK-19-18, HU-EP-19/06, P3H-19-014, IPPP/19/54}


\begin{document} 
\maketitle
\flushbottom

%
\section{Introduction}
%

\begin{figure}[t!]
\begin{center}
  \includegraphics[width=0.99\textwidth]{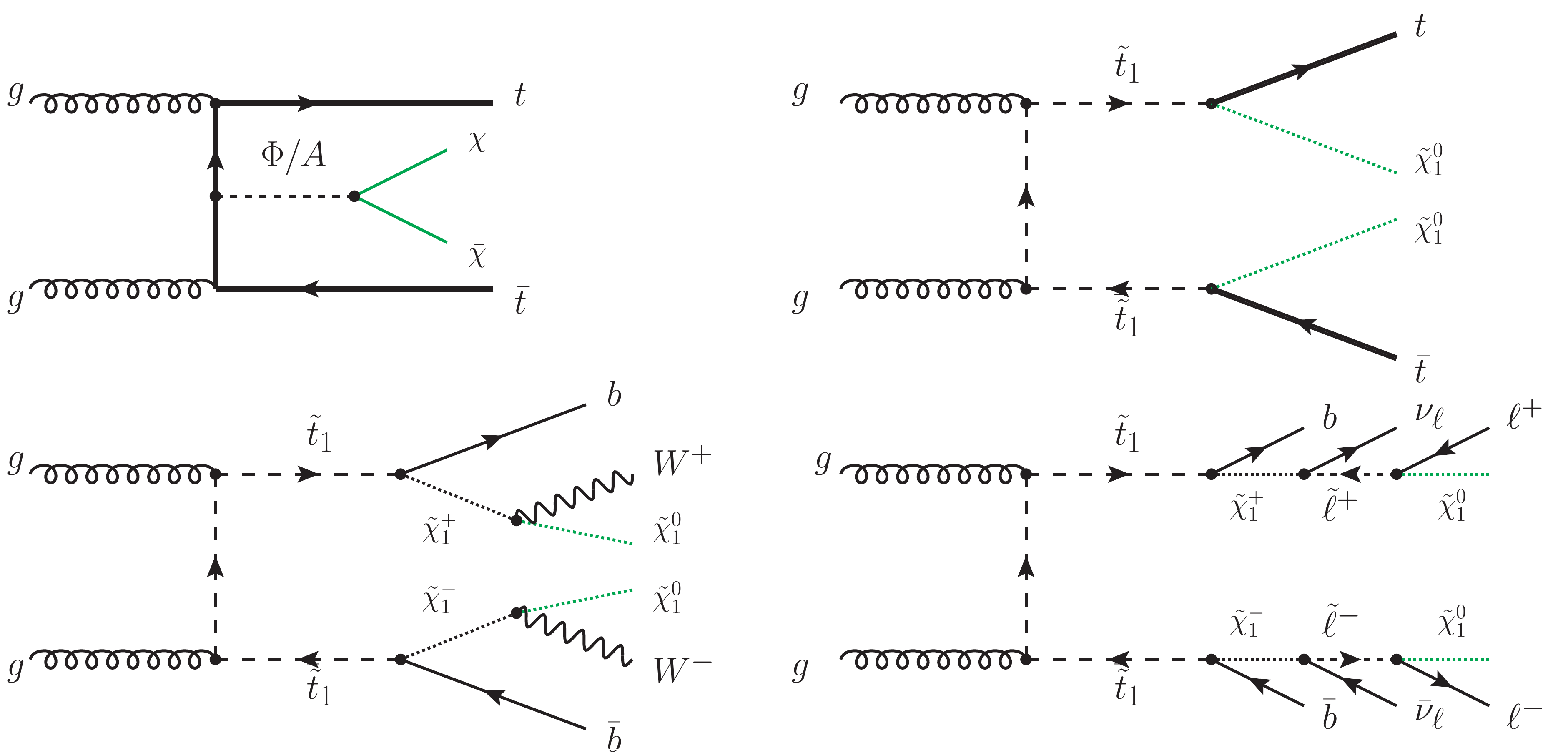}
\end{center}
\vspace{-0.6cm}
\caption{\it Representative Feynman diagrams for DM production (the
first diagram) and for supersymmetric models with supersymmetric
partners of the top quarks that might contribute to beyond the SM
$pp\to t\bar{t}+p_T^{miss}$ production at the LHC.}
\label{fig:bsm}
\end{figure}

Even though the Standard Model (SM) is currently the best description
of all known elementary particles including interactions among them,
it falls short of being a complete theory of fundamental
interactions. On the one hand this self-consistent theory has
demonstrated huge successes in explaining (almost) all experimental
results and precisely predicted a wide variety of phenomena, on the
other hand it leaves many important questions unanswered. Among
others, the theory incorporates only three out of the four fundamental
forces, omitting gravity and does not contain any viable dark matter
(DM) particle that possesses all of the required properties deduced
from observational cosmology. Thus, it is not surprising that searches
for new physics beyond the SM are continuously carried out.  Moreover, the
hunt for the complete picture or at least answers to some of our
questions, like for example, what dark matter is, is ongoing. Many
experiments aimed at direct detection and the study of dark matter
particles are actively undertaken, but none of them has been
successful up until now. Therefore, if dark matter exists, unlike
normal matter, it must barely interact with the known constituents of
the SM. An alternative approach to the direct detection of dark matter
particles in nature is to produce them in a laboratory. One of the
candidates for a dark matter particle, as predicted by many
theoretical models, is a weakly-interacting massive particle (WIMP).
It is believed that this hypothetical particle is light enough to be
produced at the Large Hadron Collider (LHC). At the LHC both ATLAS and
CMS search for WIMP DM pair production in $pp$ collisions. Since the DM
particle does not interact with the SM particles it would not be
detected directly. Simplified benchmark models for DM
\cite{Arina:2016cqj} assume, however, the existence of a mediator
particle, which should couple both to the SM particles and to the dark
sector.  A possible example for such a mediator is a spin zero
particle that can be either a colour neutral scalar or pseudo-scalar
particle. In the former case additionally mixing between the scalar
mediator and the SM Higgs boson is assumed to be zero. Even though
the nature of dark matter remains largely unknown, the couplings of the
mediator to the SM fermions are strongly constrained by precision
flavour measurements. Thus, the flavour structure of the new physics
sector can not be generic, otherwise the non-standard contributions in
flavour changing neutral current transitions would not be suppressed
to a level consistent with experimental data. At this point the
Minimal Flavour Violation (MFV) hypothesis \cite{DAmbrosio:2002vsn} is
often quoted, according to which the interaction between any new
neutral spin zero state and SM fermions must be proportional to the
fermion masses via Yukawa couplings. In other words, the SM Yukawa
couplings are the only flavour symmetry breaking terms that are
allowed in models beyond the SM if quark flavour mixing is to be
protected. Because only the top quark has the Yukawa coupling of the
order of one ($Y_t= \sqrt{2} m_t/v \approx 1$, where $v$ is the Higgs
vacuum expectation value) DM couples preferentially to top quarks in
models with MFV. Thus, colour neutral mediators should be abundantly 
produced via loop induced gluon fusion or in association with
$t\bar{t}$ pairs. The signature for the former would exhibit missing
transverse momentum $(p_T^{miss})$ from non interacting DM particles
that would be difficult to  extract from  the overwhelming QCD
background. The signature for the latter would reveal event topologies
consistent with the presence of top quarks, i.e. two oppositely
charged leptons (electron and/or muons), jets identified as
originating from bottom quarks and large missing transverse momentum,
see the first Feynman diagram in Figure \ref{fig:bsm}. Processes with
similar final states might also occur in supersymmetric models
including 
supersymmetric partners of the top quarks. In such models the direct decay
of top squarks into the top quark and a neutralino might occur or top
squarks can undergo a cascade decay through charginos and sleptons.
In $R$-parity conserving models, the lightest neutralino is stable and
all supersymmetric cascade-decays end up decaying into this particle
which is undetected by ATLAS and CMS and whose existence can only be
inferred by looking for unbalanced momentum. As a heavy, stable
particle, the lightest neutralino is an excellent candidate to form
the universe's cold dark matter. Representative Feynman diagrams are
shown in Figure \ref{fig:bsm}.
%
%

DM production in association with a top-quark pair ($pp\to t\bar{t}
+\chi \tilde{\chi}\to t\bar{t}+p_T^{miss}$, where $\chi$ stands for
the WIMP) and top squark pair production ($pp\to \tilde{t}
\bar{\tilde{t}}\to t\bar{t}+\tilde{\chi}^0_1 \tilde{\chi}^0_1 \to
t\bar{t}+p_T^{miss}$, where $\tilde{\chi}^0_1$ is considered to be the
stable neutralino, i.e. the lightest supersymmetric particle (LSP))
have both been explored by the ATLAS and CMS collaborations within the
8 TeV \cite{Aad:2014vea,Khachatryan:2015nua} and 13 TeV
\cite{Sirunyan:2017xgm,Aaboud:2017nfd,
Aaboud:2017rzf,Sirunyan:2017leh, Sirunyan:2018ell,Sirunyan:2019gfm}
data sets.  The exclusion limits at 8 TeV have been based on an
effective field theory approach, whereas the 13 TeV ones have been
interpreted in the context of the simplified supersymmetric models
with pair produced top squarks and in the context of simplified DM
models with DM particle coupled to top quarks.  Up until now no
significant deviations with respect to the SM predictions have been
observed. In all cases direct mass exclusion limits for new particles
have been placed.  In the case of DM scalar and pseudo-scalar mediator
masses below $290$ GeV and $300$ GeV respectively have been excluded
at $95\%$ confidence level. These exclusion limits, as provided by the
CMS experiment \cite{Sirunyan:2019gfm}, are currently the most
stringent limits derived at the LHC.  One should mention at this
point, however, that many assumptions enter such exclusion limits,
among other a dark matter particle of $1$ TeV and mediator couplings
to fermions and dark matter particles equal to unity have been usually
assumed. Moreover, ATLAS and CMS's experimental results have been used
to derive limits on a parameter space in particular in the effective
field theory approach, see
e.g. \cite{Cheung:2010zf,Lin:2013sca,Haisch:2015ioa, Haisch:2016gry}.

Independently of the underlying theoretical model the $pp\to
t\bar{t}+p_T^{miss}$ final state in the di-lepton top quark decay
channel, where both $W$ gauge bosons from $t\to W b$ decay further
into $W \to \ell \nu_\ell$, is the most promising channel to look for
new physics.  The advantage of this channel in comparison to the
semi-leptonic one lies in the fact that measurements of charged
leptons (electrons and/or muons) are particularly precise at the LHC
due to the excellent lepton energy resolution of the ATLAS and CMS
detectors. Additionally, angular distributions of charged leptons are
of huge importance since the $CP$ nature of the coupling between the
mediator and top quarks is encoded in the spin correlations of the top
quark pair that can be probed via top quark decay products.
Therefore, it is not surprising that the di-lepton channel is
currently scrutinised by experimentalists using data recorded by the
ATLAS and CMS collaborations in 2016. The new physics signal, however,
needs to be extracted from the SM background processes. There are
three distinct classes of major SM backgrounds that can resemble the
features of the $t\bar{t}+p_T^{miss}$ signal. The biggest (reducible)
background (in absolute cross section value) comes from the $t\bar{t}$
production process. Other processes that can be classified as the top
quark background comprise $t\bar{t}j$, $tW$ and $t\bar{t}W$. Here
neutrinos from $W\to \ell \nu_\ell$ decays contribute to
$p_T^{miss}$. Reducible non top quark backgrounds, on the other hand,
comprise di-boson productions, $W^+W^-, W^\pm Z$ and $ZZ$ as well as
production of $W$ and $Z$ gauge boson in association with QCD light
jets. For these background processes less jet activity is expected
than for the signal process which can be further combined with the
lack of bottom flavour jets. The common feature of both type of
backgrounds, however, lies in the fact that background events populate
low regions of the most relevant observables for the
$t\bar{t}+p_T^{miss}$ signature in the di-lepton channel, that consist
of $p_T^{miss}$ and $M_{T2}$ \cite{Bai:2012gs}. Selecting events with
a large amount of $p_T^{miss}$, asking for events with at least one
$b$-jet and non-vanishing $m_{\ell\ell}$ as well as requiring that the
missing transverse momentum and the transverse momentum of the two
charged lepton system are well separated in the azimuthal angle,
$\Delta \phi(p_{T,\,\ell\ell} \,, p_T^{miss})$
\cite{Sirunyan:2017xgm}, is sufficient to suppress overwhelming top
backgrounds and other reducible background processes while keeping an
adequate number of signal events.

The last and most important SM background comprises the irreducible
$t\bar{t}+Z$ background process. Here the $p_T^{miss}$ signature
arises from $W\to \ell \nu_\ell$ and $Z\to \nu_\ell
\bar{\nu}_\ell$. The $t\bar{t}+Z$ production is the only process that
provides extra genuine $p_T^{miss}$, thus, substantially adds to the
tails of $p_T^{miss}$ and $M_{T2}$ distributions which are also
populated by signal events. Indeed, various studies have shown that
this residual background can survive all the selection cuts and the
experimental sensitivity depends strongly on the proper modelling of
$t\bar{t}Z$ production, see e.g. \cite{Haisch:2016gry}.  Let us
mention that in current analyses this background process is either
simulated at leading order (LO) only or next-to-leading
order (NLO) in QCD predictions for stable top quarks are combined
with parton shower programs following the \textsc{Powheg}
or the \textsc{MC@Nlo} matching procedure. Top quark decays are
treated in the parton shower approximation omitting $t\bar{t}$ spin
correlations among other effects.

The goal of this paper is, therefore, to provide the state-of-the-art
NLO QCD predictions for the SM $t\bar{t}Z$ background process in the
di-lepton top quark decay channel.  More precisely, NLO QCD
theoretical predictions to the $e^+ \nu_e \, \mu^- \bar{\nu}_\mu \, b
\bar{b} \, \nu_\tau \bar{\nu}_\tau$ final state are calculated for the
first time.  All double-, single- and non-resonant Feynman diagrams,
interferences, and off-shell effects of the top quarks are properly
incorporated at the NLO level in QCD. Also non-resonant and off-shell
effects due to the finite $W$-boson width are included. This
calculation constitutes the first fully realistic NLO QCD computation
for top quark pair production with additional missing $p_T$ in
hadronic collisions.

As a final comment, we note that NLO QCD corrections to
the inclusive $t\bar{t}Z$ production process (with on-shell top quarks
and the $Z$ gauge boson) have been calculated for the first time in
Ref. \cite{Lazopoulos:2008de} and afterwards recomputed in
Refs. \cite{Kardos:2011na,Garzelli:2011is,Garzelli:2012bn,
Bylund:2016phk}. NLO QCD theoretical predictions from
\cite{Garzelli:2011is,Garzelli:2012bn} have additionally been matched
with shower Monte Carlo (MC) programs using the \textsc{PowHel}
framework. The latter relies on \textsc{Powheg-Box} and allows for the
matching between the fixed order computation at NLO in QCD (as
provided by the \textsc{Helac-Nlo} MC program) and the parton shower
evolution, followed by hadronization and hadron decays (as described
by \textsc{Pythia} and \textsc{Herwig}). In
\cite{Garzelli:2011is,Garzelli:2012bn} top quark and $Z$ decays have
been treated in the parton shower approximation omitting $t\bar{t}$
spin correlations. Finally, in Ref.  \cite{Rontsch:2014cca} improved
calculations for $pp\to t\bar{t}Z$ have been presented. This time NLO
QCD corrections have been included to the production and
(semi-leptonic) decays of top quarks in the narrow-width approximation
(NWA), thus, also taking into account $t\bar{t}$ spin
correlations. Moreover, LO $Z\to \ell ^+\ell^-$ decays have been
considered. Besides NLO QCD corrections, further step towards a more
precise modeling of $t\bar{t}Z$ have been achieved by including
electroweak corrections \cite{Frixione:2015zaa} and soft gluon
resummation effects  \cite{Kulesza:2018tqz,Broggio:2017kzi}. 

The paper is organised as follows. In Section \ref{sec:1}, we briefly
summarise the framework of our calculation and discuss technical
aspects of the computation. Section \ref{sec:2} outlines the
theoretical setup for LO and NLO QCD results. Results for the total
cross sections and various differential cross sections are presented
in Section \ref{sec:3}. They are provided for the LHC centre-of-mass
system energy of $13$ TeV and for a few renormalisation and
factorisation scale choices. The theoretical uncertainties, that are
associated with neglected higher order terms in the perturbative
expansion and with different parameterisations of the parton
distribution functions, are also given. Additionally, we show
differential cross section distributions, which are of particular
interest for new physics searches. The latter comprise $p_T^{miss},
\Delta \phi_{\ell\ell}, \Delta y_{\ell\ell}, \cos\theta_{\ell\ell}
,H_T$ and $H_T^\prime$. From the experimental point of view both
$t\bar{t}Z$ and $t\bar{t}$ processes have the same signature,
two charged leptons $(\ell^\pm)$, two bottom flavoured jets $(j_b)$
and missing transverse momentum from escaping neutrinos
$(p_T^{miss})$. Thus, in Section \ref{sec:4} we study the impact of
the enlarged missing transverse momentum on various differential cross
section distributions. To this end normalised differential
distributions constructed from (anti-)top quark decay products for
both $pp\to e^+ \nu_e \, \mu^- \bar{\nu}_\mu \, b \bar{b} \, \nu_\tau
\bar{\nu}_\tau$ and $pp\to e^+ \nu_e \, \mu^- \bar{\nu}_\mu \, b
\bar{b}$ are compared and discussed in that Section. Finally, in
Section \ref{sec:5} our results are
summarised and our conclusions are outlined.

%
\section{Details of the calculation}
\label{sec:1}
%
%
%
\begin{figure}[t!]
\begin{center}
  \includegraphics[width=0.95\textwidth]{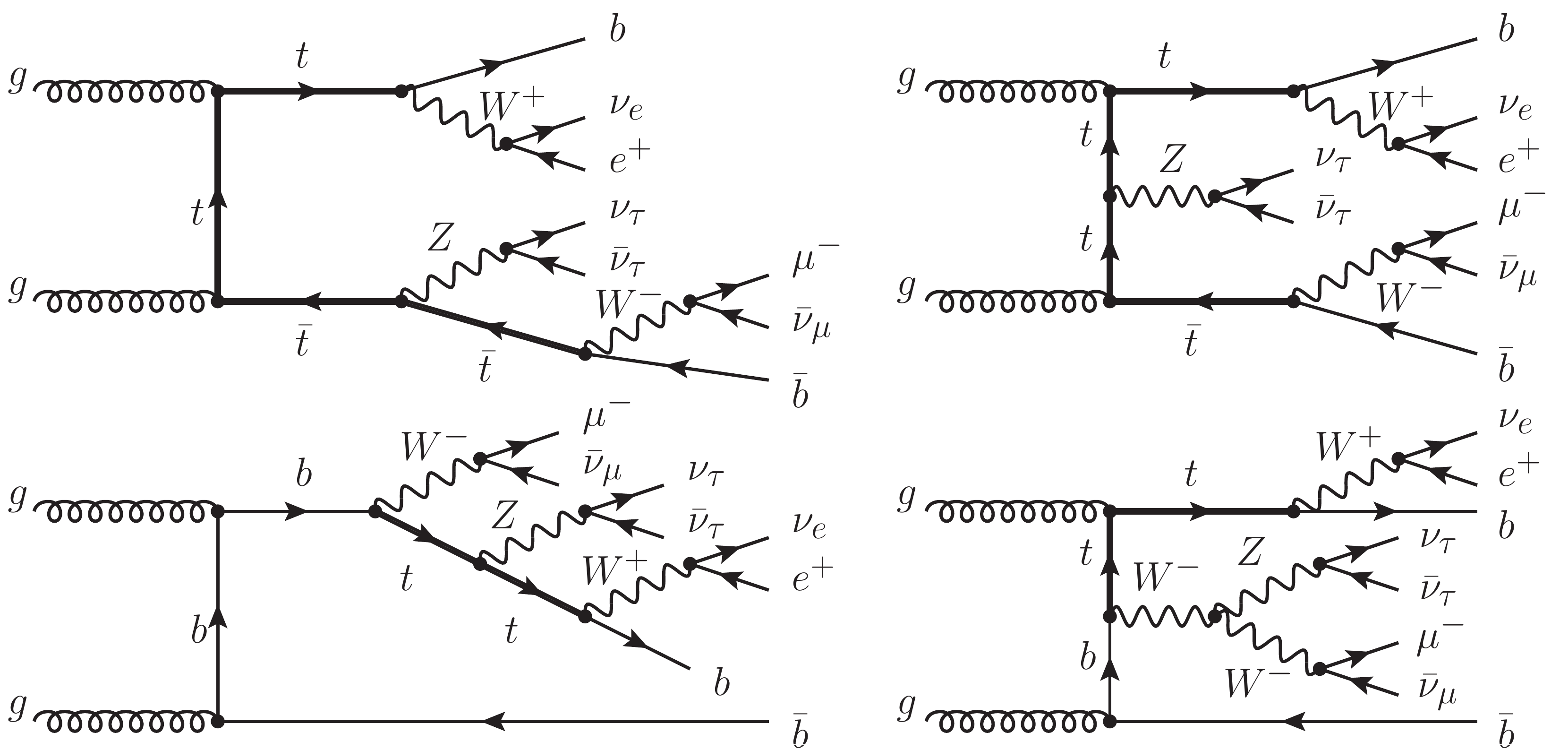}
   \includegraphics[width=0.95\textwidth]{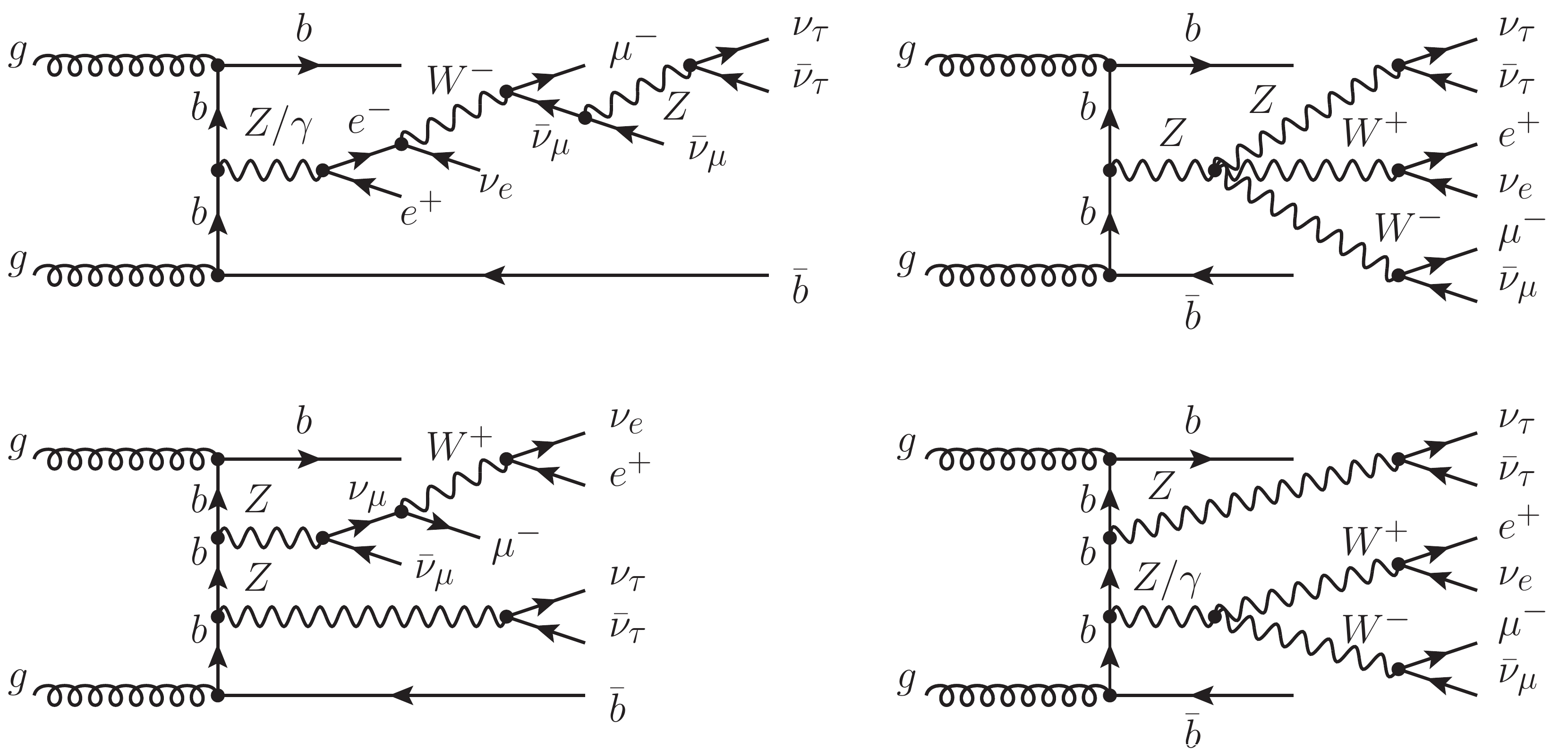}
\end{center}
\vspace{-0.6cm}
\caption{\it Representative Feynman diagrams with double- (first row),
single- (second row) and no top quark resonances (third row)
contributing to $pp \to e^+\nu_e \mu^- \bar{\nu}_\mu
b\bar{b}\nu_\tau\bar{\nu}_\tau$ production at the leading order in
perturbative expansion. Diagrams with a single $W$ boson resonance
that contribute to the off-shell effects of the $W$ gauge boson are
also presented (last row).}
\label{fig:fd}
\end{figure}
%
At the LO level in perturbative expansion the $e^+\nu_e \mu^-
\bar{\nu}_\mu b\bar{b}\nu_\tau\bar{\nu}_\tau$ final states are
produced via the scattering of either two gluons or one quark and the
corresponding anti-quark. The ${\cal O}(\alpha_s^2 \alpha^6)$
contributions can be subdivided into three classes, diagrams
containing two top quark propagators that can become resonant,
diagrams containing only one top quark resonance and finally diagrams
without any top quark resonance. Regarding the $W^\pm$ resonances one
can distinguish only two subclasses, double- and single-resonant gauge
boson contributions. Examples of Feynman diagrams for each class are
depicted in Figure~\ref{fig:fd}. In total, there are $1024$ LO
diagrams for the $gg\to e^+\nu_e \mu^- \bar{\nu}_\mu
b\bar{b}\nu_\tau\bar{\nu}_\tau$ partonic reaction and $540$ for each
$q\bar{q} \to e^+\nu_e \mu^- \bar{\nu}_\mu
b\bar{b}\nu_\tau\bar{\nu}_\tau$ subprocess where $q$ stands for up- or
down-type quarks. Even-though we do not employ Feynman diagrams in our
calculations we present their numbers as a measure of the complexity
of the calculation. Instead, the calculation of scattering amplitudes
is based on well-known off-shell iterative algorithms performed
automatically within the \textsc{Helac-Dipoles} package
\cite{Czakon:2009ss}, which avoids  multiple evaluation of recurring
building blocks. The results are cross checked with the
\textsc{Helac-Phegas} Monte Carlo (MC) program
\cite{Cafarella:2007pc}. Phase space integration is performed and
optimised with the help of \textsc{Parni} \cite{vanHameren:2007pt} and
\textsc{Kaleu} \cite{vanHameren:2010gg}.  Since the produced top
quarks are unstable particles, the inclusion of the decays is
performed in the complex mass scheme \cite{Denner:1999gp,
Denner:2005fg,Bevilacqua:2010qb,Denner:2012yc}. It fully respects
gauge invariance and is straightforward to apply. Since we are
interested in NLO QCD corrections, gauge bosons are treated within the
fixed width scheme.

The virtual corrections comprise the 1-loop corrections to the LO
reactions. These corrections can be classified into self-energy,
vertex, box-type, pentagon-type, hexagon-type and heptagon-type
corrections. In Table \ref{tab:one-loop} the number of one-loop
Feynman diagrams, that corresponds to each type of correction for the
dominant $gg \to e^+\nu_e \mu^- \bar{\nu}_\mu b\bar{b} \nu_\tau
\bar{\nu}_\tau$ partonic subprocess as obtained with \textsc{Qgraf}
\cite{Nogueira:1991ex}, is given. We have cross-checked our results
with the publicly available general purpose MC program
\textsc{MadGraph5-aMC@NLO} \cite{Alwall:2014hca}. Explicitly, we have
compared results for the virtual NLO contribution to the squared
amplitude, $2\Re \left({\cal M}^*_{tree}{\cal M}_{one-loop}\right)$,
for a few phase-space points for $gg$ and $u\bar{u}$ partonic
subprocesses and we have found perfect agreement in each case. In
evaluating virtual corrections, the \textsc{Helac-1Loop}
\cite{vanHameren:2009dr} MC library is used, that
incorporates \textsc{CutTools} \cite{Ossola:2007ax} and
\textsc{OneLOop} \cite{vanHameren:2010cp} as its cornerstones. The
\textsc{CutTools} program contains an implementation of the OPP method
for the reduction of one-loop amplitudes at the integrand level
\cite{Ossola:2006us}.  The \textsc{OneLOop} library, on the other
hand, is dedicated to the evaluation of the one-loop scalar
functions. Renormalisation is done in the usual way by evaluating
tree-level diagrams with counter-terms. For our process, we chose to
renormalise the strong coupling in the $\overline{\rm MS}$ scheme with five
active flavours and the top quark decoupled, while the mass renormalisation
is performed in the on-shell scheme.
%
\begin{table}[t!]
\begin{center}
\begin{tabular}{||c|c||}
\hline \hline 
\textsc{One-Loop}& \textsc{Number Of} \\[0.2cm]
\textsc{Correction} &\textsc{Feynman Diagrams} \\[0.2cm]
  \hline \hline
 \textsc{Self-Energy} & 17424  \\[0.2cm]
 \textsc{Vertex} &  21544 \\[0.2cm]
 \textsc{Box-Type}                & 11726   \\[0.2cm]
 \textsc{Pentagon-Type} & 4650  \\[0.2cm]
   \textsc{Hexagon-Type}              & 1038\\[0.2cm]
\textsc{Heptagon-Type}  & 90    \\ [0.2cm]
  \hline \hline
  \textsc{Total Number}  & 56472 \\[0.2cm]
   \hline \hline
 \end{tabular}
\end{center}
\caption{\label{tab:one-loop} \it
  The number of one-loop Feynman diagrams for the dominant $gg\to
e^+\nu_e \mu^- \bar{\nu}_\mu b\bar{b}\nu_\tau\bar{\nu}_\tau$ partonic
subprocess at ${\cal O}(\alpha_s^3 \alpha^6)$ for the $pp\to e^+ \nu_e
\mu^- \bar{\nu}_\mu b\bar{b} \nu_\tau \bar{\nu}_\tau +X$ process. The
Higgs boson exchange contributions are not considered and the
Cabibbo-Kobayashi-Maskawa mixing matrix is kept diagonal.}
\end{table}

The real emission corrections to the LO process arise from tree-level
amplitudes with one additional parton, i.e. an additional gluon, or a
quark anti-quark pair replacing a gluon.  For the calculation of the
real emission contributions, the package \textsc{Helac-Dipoles} is
employed. It implements the dipole formalism of Catani and Seymour
\cite{Catani:1996vz,Catani:2002hc} for arbitrary helicity eigenstates
and colour configurations of the external partons \cite{Czakon:2009ss}
and the Nagy-Soper subtraction scheme \cite{Bevilacqua:2013iha}, which
makes use of random polarisation and colour sampling of the external
partons. Having two independent subtraction schemes in
\textsc{Helac-Dipoles} allow us to cross check the correctness of the
real corrections by comparing the two results. All partonic
subprocesses that are taken into account for the real emission
contributions are listed in Table \ref{tab:real-emission}, together
with the number of the corresponding Feynman diagrams, the number of
Catani-Seymour dipoles and Nagy-Soper subtraction terms. In each case,
three times less terms are needed in the Nagy-Soper subtraction scheme
compared to the Catani-Seymour scheme. The difference corresponds to
the total number of possible spectators in the process under scrutiny,
which are relevant in the Catani-Seymour case, but not in the
Nagy-Soper case.
%
\begin{table}[t!]
\begin{center}
\begin{tabular}{||c|c|c|c||}
\hline \hline
\textsc{Partonic}& \textsc{Number Of} &\textsc{Number Of}&
                                                           \textsc{Number
                                                           Of}\\[0.2cm]
\textsc{Subprocess} &\textsc{Feynman Diagrams} & \textsc{CS Dipoles}  
& \textsc{NS Subtractions} \\[0.2cm]
\hline \hline
$gg\to e^+ \nu_e \mu^- \bar{\nu}_\mu b\bar{b} \nu_\tau \bar{\nu}_\tau g$
& 6880  & 27  &  9 \\[0.2cm]
$gq\to e^+ \nu_e \mu^- \bar{\nu}_\mu b\bar{b}  \nu_\tau \bar{\nu}_\tau q$ 
& 3520 & 15 & 5\\[0.2cm]
$g\bar{q}\to e^+ \nu_e \mu^- \bar{\nu}_\mu b\bar{b} \nu_\tau \bar{\nu}_\tau\bar{q}$ 
& 3520 & 15 & 5 \\[0.2cm]
$q\bar{q}\to e^+ \nu_e \mu^- \bar{\nu}_\mu b\bar{b} \nu_\tau \bar{\nu}_\tau  g$
&  3520& 15 &  5\\[0.2cm]
\hline \hline
 \end{tabular}
\end{center}
\caption{\label{tab:real-emission} \it The list of partonic
subprocesses contributing to the subtracted real emission at ${\cal
O}(\alpha_s^3 \alpha^6)$ for the $pp\to e^+ \nu_e \mu^- \bar{\nu}_\mu
b\bar{b} \nu_\tau \bar{\nu}_\tau +X$ process where $q = u, d, c,
s$. Also shown are the number of Feynman diagrams, as well as the
number of Catani-Seymour and Nagy-Soper subtraction terms that
correspond to these partonic subprocesses.}
 \end{table}
%
 
Let us note that \textsc{Helac-1Loop} and \textsc{Helac-Dipoles} are
part of the \textsc{Helac-NLO} framework
\cite{Bevilacqua:2011xh} and that further technical details are described in
Ref.~\cite{Bevilacqua:2010qb,Bevilacqua:2015qha,Bevilacqua:2016jfk,
Bevilacqua:2018woc}. Let us also add that our theoretical predictions
are stored in the form of (modified) Les Houches Event Files (LHEFs)
\cite{Alwall:2006yp} or ROOT \cite{Antcheva:2009zz} Ntuples. Building
on \cite{Bern:2013zja}, each event is stored with accessory matrix
element and PDF information to allow re-weighting for different scale
or PDF choices.  Storing events shows clear  advantages  when
different observables are to be studied, or different kinematical cuts
are to be applied, since no additional time-consuming running of the
code is required.

%
\section{Setup for numerical predictions}
\label{sec:2}
%
%

We consider the process $pp \to e^+ \nu_e \, \mu^- \bar{\nu}_\mu \, b
\bar{b} \, \nu_\tau \bar{\nu}_\tau$ for the LHC Run II centre-of-mass
system energy of ${\sqrt{s} = 13}$ TeV. We only simulate decays of the
weak bosons to different lepton generations to avoid virtual photon
singularities stemming from $\gamma \to \ell^+ \ell^-$. These
interference effects are at the $0.2\%$ level for inclusive cuts, as
checked by an explicit leading order calculation 
performed with the help of the \textsc{Helac-Phegas} MC framework. The
$\ell^\pm\ell^\mp$ cross section (with $\ell_{1,2}=e,\mu$ since $\tau$
leptons are always studied separately) can be obtained by multiplying
the result with a lepton-flavor factor of $4$. However, we
additionally must count $3$ different decays of the $Z$ gauge boson
($Z\to \nu_\ell \bar{\nu}_\ell$ with
$\nu_\ell=\nu_e,\nu_\mu,\nu_\tau$).  Thus, the complete cross section
can be realised by multiplying the results presented in the paper by
$12$. We keep the Cabibbo-Kobayashi-Maskawa mixing matrix
diagonal. The following SM parameters are given within the $G_\mu$
scheme that takes into account electroweak corrections related to the
running of $\alpha$
\begin{equation*}
\begin{array}{lll}
 m_{W}=80.385 ~{\rm GeV} \,, 
&\quad \quad \quad \quad&\Gamma_{W} = 2.0988 ~{\rm GeV}\,, 
\vspace{0.2cm}\\
  m_{Z}=91.1876  ~{\rm GeV} \,, 
  &&\Gamma_{Z} = 2.50782 ~{\rm GeV}\,, \vspace{0.2cm}\\
  G_{ \mu}=1.166378 \times 10^{-5} ~{\rm GeV}^{-2}\,,&
  &\sin^2\theta_W =1- m^2_W/m^2_Z\,.
\end{array}
\end{equation*}
Leptonic $W$ gauge boson decays do not receive NLO QCD corrections. To
take some effects of higher order corrections into account for the
gauge boson widths the NLO QCD values are used for LO and NLO matrix
elements.  The electroweak coupling is derived from the Fermi constant
$G_\mu$ according to
\begin{equation}
\alpha= \frac{\sqrt{2} \, G_\mu m^2_W \sin^2\theta_W}{\pi} \,.
\end{equation}  
The top quark mass is set to $m_t = 173.2$ GeV. All other QCD partons
including $b$ quarks as well as leptons are treated as massless. Since
we treat $b$ quarks as massless partons there are no Higgs-exchange
diagrams at tree level. Moreover, due to the negligibly small
dependence on the Higgs boson mass, closed fermion loops which involve
top quarks coupled to Higgs bosons, are neglected. The top quark
width, as calculated from \cite{Jezabek:1988iv,Chetyrkin:1999ju}, is
taken to be $\Gamma_{t}^{\rm LO} = 1.47848$ GeV at LO and
$\Gamma_{t}^{\rm NLO} = 1.35159$ GeV at NLO. The value of $\alpha_s$
used for the top quark width $\Gamma^{\rm NLO}_{t}$ calculation is
taken at $m_t$.  This $\alpha_s$ is independent of $\alpha_s(\mu_0)$
that goes into the matrix element and PDF calculations. The latter is
used to describe the dynamics of the whole process, the former only
the top quark decays. Our calculation, like any fixed-order one,
contains a residual dependence on the renormalisation scale $(\mu_R)$
and the factorisation scale $(\mu_F)$ arising from the truncation of
the perturbative expansion in $\alpha_s$. As a consequence,
observables depend on the values of $\mu_R$ and $\mu_F$ that are
provided as input parameters. We assume that the default scale $\mu_R
= \mu_F = \mu_0$ is the same for both the renormalisation and
factorisation scales. The scale systematics, however, is evaluated by
varying $\mu_R$ and $\mu_F$ independently in the range
\begin{equation}
  \begin{split}
    \frac{1}{2} \, \mu_0  &  \, \le  \, \mu_R\,,\mu_F \, \le \,
    2 \,  \mu_0\,, \\[0.2cm]
    \frac{1}{2}  & \, \le \,
    \frac{\mu_R}{\mu_F}  \, \le \,  2 \,,
    \end{split}
\end{equation}
which in practice amounts to consider  the following pairs 
\begin{equation}
\label{scan}
\left(\frac{\mu_R}{\mu_0}\,,\frac{\mu_F}{\mu_0}\right) = \Big\{
\left(2,1\right),\left(0.5,1  
\right),\left(1,2\right), (1,1), (1,0.5), (2,2),(0.5,0.5)
\Big\} \,.
\end{equation}
We search for the minimum and maximum of the resulting cross
sections. For $\mu_0$ we consider two cases, the kinematic independent
scale choice (fixed scale) and the kinematic dependent scale choice
(dynamical scale). In the case of the integrated $pp \to e^+\nu_e \,
\mu^- \bar{\nu}_\mu \, b\bar{b} \, \nu_\tau \bar{\nu}_\tau$ cross
section both choices would be suitable to properly describe the
production process. For the differential cross section distributions,
however, the fixed scale would adequately describe the phase-space
regions close to the $t\bar{t}$ threshold but will fail at the tails
of various dimensionful distributions. A proper dynamical scale
choice, on the other hand, should characterise accurately all phase
space regions. Specifically, we employ the following fixed scale
\begin{equation}
    \mu_0 = m_{t}+\frac{m_Z}{2}\,, 
\end{equation}
commonly used in the studies of $t\bar{t}Z$ production, see e.g.
\cite{Lazopoulos:2008de,Kardos:2011na,
Garzelli:2012bn,Rontsch:2014cca}, whereas for the dynamical scale a
few choices will be examined. Firstly, we concentrate on the total
transverse energy of the system, $H_T$, that is blind to the fact that
in $ e^+\nu_e \, \mu^- \bar{\nu}_\mu \, b\bar{b} \, \nu_\tau
\bar{\nu}_\tau$ production Feynman diagrams with one or two top quark
resonances might appear.  Thus, the first dynamical scale choice is
constructed according to
\begin{equation}
  \mu_0 = \frac{ H_T}{3}\,.
 \end{equation}
 Here $H_T$  is calculated on an event-by-event basis in line with
\begin{equation}
H_T=p_{T, \,e^+}+p_{T, \,\mu^-} +p_{T}^{miss}
+ p_{T,\,b_1} + p_{T,\,b_2} \,,
 \end{equation} 
 where $b_1$ and $b_2$ are bottom flavoured jets and the
$p_{T}^{miss}$ is the total missing transverse momentum from escaping
neutrinos defined according to
\begin{equation}
p_T^{miss}=|\,\vec{p}_{T,\,\nu_e} +\vec{
p}_{T,\, \bar{\nu}_\mu}+ \vec{p}_{T,\,\nu_\tau}+ \vec{p}_{T,\,
\bar{\nu}_\tau}| \,.
\end{equation}
In the next step the information about the underlying resonant nature
of the process is used. To this end the following resonance aware
dynamical scale choices, that we denote $E_T,\, E_T^\prime$ and
$E_T^{\prime\prime}$, are going to be inspected
\begin{equation}
  \begin{split}
  \mu_0 &= \frac{E_T}{3}  =\frac{1}{3} \left(
    m_{T,\,t} + m_{T,\, \bar{t}} + p_{T,\,Z}
  \right)\,,\\[0.2cm]
  \mu_0 &= \frac{E_T^\prime}{3}=  \frac{1}{3}\left(m_{T,\,t} +
    m_{T,\, \bar{t}} + m_{T,\,Z}
\right)\,,\\[0.2cm]
\mu_0 &= \frac{E_T^{\prime\prime}}{3}= \frac{1}{3} \left(
  m_{T,\,t} + m_{T,\,\bar{t}} \right) \,.
\end{split}
\end{equation}
Here $m_{T,\,i}$ is defined in accordance with $m_{T,\,i}
=\sqrt{p_{T,\,i}^2+m_i^2}$, where $i$ stands for $i=t,\bar{t},Z$.  The
top and anti-top quark as well as the $Z$ gauge boson are
reconstructed from their decay products assuming exact $W$ and $Z$
gauge bosons reconstruction and $b$-jet tagging efficiency of $100\%$,
i.e.  $p(t) = p(j_{b_1})+p(e^+)+ p(\nu_e)$, $p(\bar{t}\,)
=p(j_{b_2}\,)+p(\mu^-)+p(\bar{\nu}_\mu)$ and $ p(Z) = p(\nu_\tau) +
p(\bar{\nu}_\tau)$, where $j_{b_1}$ originates from the $b$-quark and
$j_{b_2}$ from anti-b quark. To construct final state jets the IR-safe
{\it anti}$-k_T$ jet algorithm \cite{Cacciari:2008gp} is employed with
the resolution parameter $R=0.4$. The {\it anti}$-k_T$ jet algorithm
iterates recombinations of the final state partons with
pseudo-rapidity $|\eta| <5$ until no partons are left and jets are
created.  We require at least two jets for our process, of which
exactly two must be bottom flavoured jets. Moreover, we asked for two
charged leptons and a large missing transverse momentum. We impose the
following cuts on the transverse momenta and the rapidity of two
recombined $b$-jets, which we assume to be always tagged
\begin{equation}
  p_{T,\,b} > 40 ~{\rm GeV}\,,    \quad \quad \quad |y_b| < 2.5\,,
  \quad \quad \quad 
  \Delta R_{b\bar{b}} > 0.4\,.
\end{equation}
The last cut, i.e. the separation between the $b$-jets, is implied by
the jet algorithm. Basic selection cuts are applied to charged leptons to
ensure that they are observed inside the detector and well separated
from each other and from $b$-jets
\begin{equation}
 p_{T,\,\ell}>30 ~{\rm GeV}\,,    \quad \quad \quad  |y_\ell|<2.5\,,
 \quad \quad \quad 
  \Delta R_{\ell \ell} > 0.4 \quad \quad \quad  
  \Delta R_{\ell b} > 0.4 \,,
\end{equation}
where $\ell$ stands for the charged lepton: $\mu^-,e^+$. We
additionally put a requirement on the missing transverse momentum
$p^{miss}_{T} >50$ GeV. Finally, we place no restriction on the
kinematics of the extra (light) jet.

The running of the strong coupling constant $\alpha_s$ with two-loop
(one-loop) accuracy at NLO (LO) is provided by the LHAPDF interface
\cite{Buckley:2014ana}. The number of active flavours is $N_F = 5$.
Contributions induced by the bottom-quark parton density are
neglected.  At LO the $b\bar{b}$ partonic subprocess contributes at
the level of $1.1\%$ to the $q\bar{q}$ initial state. However, the
full $pp$ cross section is dominated by the the $gg$ channel $(67\%)$,
thus, the $b\bar{b}$ contribution to the $pp\to e^+\nu_e \,\mu^-
\bar{\nu}_\mu \, b\bar{b} \, \nu_\tau \bar{\nu}_\tau$ production
process amounts to $0.4\%$ only and can be safely
disregarded. Following recommendations of PDF4LHC for the usage of
parton distribution functions (PDFs) suitable for applications at the
LHC Run II \cite{Butterworth:2015oua} we employ CT14
\cite{Dulat:2015mca}, which is the default PDF set in our studies,
NNPDF3.0 \cite{Ball:2014uwa} and MMHT14 \cite{Harland-Lang:2014zoa}.

We would like to stress that the above parameters and cuts on final
states correspond to one particular setup. It is clear that there are
many interesting phenomenological analyses that might be performed for
the $ e^+\nu_e \,\mu^- \bar{\nu}_\mu \, b\bar{b} \, \nu_\tau
\bar{\nu}_\tau$ process using our system with different setup. The
latter could be chosen either in the context of the SM or having in
mind searches for various new physics scenarios. Obviously, in each
case a slightly different event selection would be required to
optimise the search. Hence, in this publication we are not able to
provide theoretical predictions for the irreducible background for
each proposed BSM model. Instead, our main goal here is to demonstrate
the size of higher order corrections to the $ e^+\nu_e \,\mu^-
\bar{\nu}_\mu \, b\bar{b} \, \nu_\tau \bar{\nu}_\tau$ final state at
the LHC in the presence of the inclusive cut selection that resembles
as closely as possible the ATLAS and/or CMS detector
responses. However, we shall also discuss the impact of NLO QCD
corrections on a few observables that are relevant for new physics
searches. If non-flat differential ${\cal K}$-factors are acquired for
these observables already with a fairly inclusive cut selection and
independently of the scale choice, these observables need to be
carefully reexamined in the presence of more exclusive cuts before any
realistic strategies for the detection of new physics signal can be
further developed.

%
\section{NLO QCD predictions for the LHC Run II
  energy of 13 TeV}
\label{sec:3}
%
%
%
\subsection{Integrated cross section and its scale dependence
  for the fixed scale}
%
%

\begin{figure}[t!]
\begin{center}
   \includegraphics[width=0.49\textwidth]{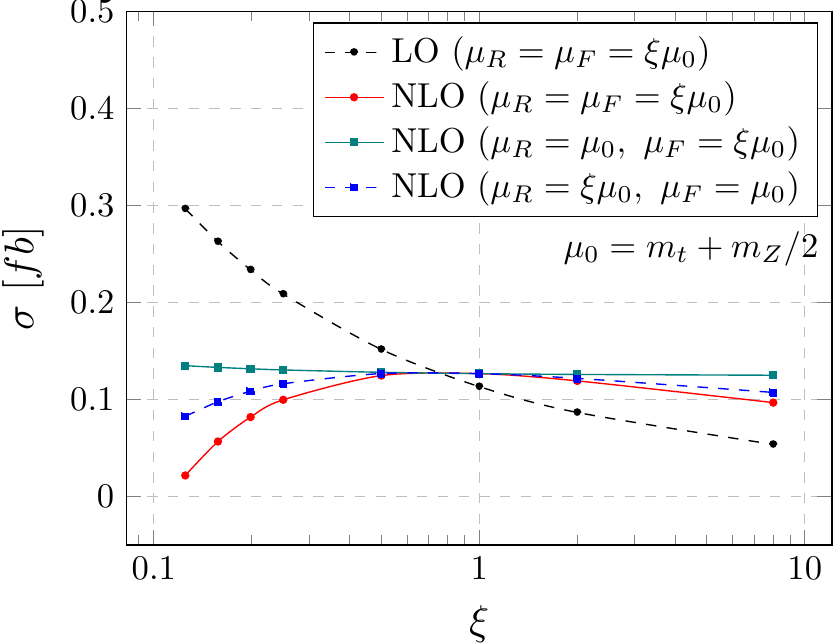}
\end{center}
\vspace{-0.6cm}
\caption{\it
  Scale dependence of the LO and NLO integrated cross section for the
$pp\to e^+ \nu_e \mu^- \bar{\nu}_\mu b\bar{b} \, \nu_\tau
\bar{\nu}_\tau +X$ process at the LHC run II with $\sqrt{s}=13$
TeV. Renormalisation and factorisation scales are set to
$\mu_R=\mu_F=\xi \mu_0$ where $\mu_0=m_t+m_Z/2$ and $\xi \in
\left\{0.125,\dots, 8\right\}$.  The LO and NLO CT14 PDF sets are
employed. Also shown is the variation of $\mu_R$ with fixed $\mu_F$
and the variation of $\mu_F$ with fixed $\mu_R$.}
\label{fig:scale-fixed}
\end{figure}

With the input parameters and cuts specified in Section~\ref{sec:2},
we arrive at the following predictions for
$\mu_R=\mu_F=\mu_0=m_t+m_Z/2$
\begin{equation}
  \begin{split}
\sigma^{\rm LO}_{pp \to e^+\nu_e\mu^-\bar{\nu}_\mu b\bar{b} \nu_\tau
  \bar{\nu}_\tau} ({\rm CT14}, \mu_0=m_t+m_Z/2)
&= 0.1133^{+ 0.0384\, (33\%)}_{-0.0266\, (23\%)} \, {\rm fb}\,, \\[0.2cm]
\sigma^{\rm NLO}_{pp \to e^+\nu_e\mu^-\bar{\nu}_\mu b\bar{b}  \nu_\tau
  \bar{\nu}_\tau} ({\rm CT14}, \mu_0=m_t+m_Z/2)
&=  0.1266^{+0.0014 \, (1.1\%)}_{- 0.0075\, (5.9\%)} \, {\rm fb}\,.
\end{split}
\end{equation}  
The full $pp$ cross section receives positive and moderate NLO
corrections of $12\%$.  The theoretical uncertainties resulting from
the scale variation and taken in a very conservative way as a maximum
of the lower and upper bounds are $33\%$ at LO and $5.9\%$ at
NLO. Thus, a reduction of the theoretical error by a factor of almost
$6$ is observed when higher order corrections are incorporated. In the
case of truly asymmetric uncertainties, however, it is always more
appropriate to symmetrise the errors.  After symmetrisation the scale
uncertainty at LO does not change substantially, i.e. it is reduced
down to $29\%$. However, at the NLO in QCD the reduction is
considerable as far as $3.5\%$. Therefore, by going from LO to NLO we
have reduced the theoretical error by a factor of $8$. Should we instead
vary $\mu_R$ and $\mu_F$ simultaneously, up and down by a factor of 2
around $\mu_0$, the uncertainties would remain unchanged. This is due
to the fact that the scale variation is driven solely by the changes
in $\mu_R$ as can be observed in Figure \ref{fig:scale-fixed}, where
the graphical presentation of the behaviour of LO and NLO cross
sections upon varying the scale by a factor $\xi \in \left\{
  0.125,\dots, 8\right\}$ is shown for  CT14 PDF sets.
%
%

Let us mention at this point that despite its relatively small cross
section, a good theoretical control over the $pp\to e^+ \nu_e \mu^-
\bar{\nu}_\mu b\bar{b} \, \nu_\tau \bar{\nu}_\tau$ production process
is phenomenologically relevant. This irreducible SM background at NLO
in QCD is of the order of $1.5$ fb, where a factor $12$ has been used
to obtain the complete cross section for the process. For comparison,
typical predictions of new physics scenarios such as models with the
vector, axial-vector, pseudoscalar, and scalar interaction between the
top quark and the dark matter particle are also at the same level,
see e.g. Refs.~\cite{Boughezal:2012zb,
  Boughezal:2013pja,Haisch:2015ioa, Bauer:2017ota}. Thus, our NLO
analysis of $pp\to e^+ \nu_e \mu^- \bar{\nu}_\mu b\bar{b} \, \nu_\tau
\bar{\nu}_\tau$ at the LHC is a necessary step towards a correct
interpretation of the possible signals of new physics that may arise
in this channel.

%
\subsection{Integrated cross section and its scale dependence
for the dynamical scale}
%
%
%
\begin{table}[t!]
\begin{center}
\begin{tabular}{||c|c|c|c||}
  \hline \hline
 \textsc{Scale Choice}&
  $\sigma^{\rm LO}_{pp\to e^+\nu_e\mu^-\bar{\nu}_\mu b\bar{b}  \nu_\tau
  \bar{\nu}_\tau}$ [fb]
  & $\sigma^{\rm NLO}_{pp \to e^+\nu_e\mu^-\bar{\nu}_\mu b\bar{b}  \nu_\tau
  \bar{\nu}_\tau}$  [fb] &  ${\cal K}=\sigma^{\rm NLO}/\sigma^{\rm LO}$ \\[0.2cm] 
  \hline \hline
  $\mu_0=H_T/3$              & $0.1260^{+0.0438\, (35\%)}_{-0.0302\, (24\%)}$
 & $0.1270^{+0.0009\, (0.7\%)}_{-0.0086\, (6.8\%)}$  & 1.01\\[0.2cm]
  \hline 
 $\mu_0=E_T/3$              & $0.1110^{+0.0368 \, (33\%)}_{-0.0258\, (23\%)}$
  & $0.1272^{+0.0020\, (1.6\%)}_{-0.0086\, (6.8\%)}$  & 1.14\\[0.2cm]
  \hline 
  $\mu_0=E^\prime_T/3$ & $0.1087^{+0.0359\, (33\%)}_{-0.0251\, (23\%)}$ &
 $0.1268^{+0.0019\, (1.5\%)}_{-0.0081\, (6.4\%)}$                  & 1.17\\[0.2cm]
  \hline 
  $\mu_0=E^{\prime \prime}_T/3$ &
$0.1227^{+0.0423\, (34\%)}_{-0.0293\, (24\%)}$
 &
$0.1286^{+0.0013\, (1.0\%)}_{-0.0060\, (4.7\%)}$
                         & 1.05\\[0.2cm] 
\hline \hline 
 \end{tabular}
\end{center}
\caption{\label{tab:scales-dynamic} \it
  LO and NLO cross sections for the $pp\to e^+ \nu_e \mu^-
\bar{\nu}_\mu b\bar{b} \, \nu_\tau \bar{\nu}_\tau +X$ process at the
LHC run II with $\sqrt{s}=13$ TeV. Results for various scale choices
are presented. Also included are theoretical errors as obtained from
the scale variation. In the last column the ${\cal K}$ factor, the
ratio of the NLO to LO cross section, is given. The LO and  NLO
CT14 PDF sets are employed. }
\end{table}
%

\begin{figure}[t!]
\begin{center}
  \includegraphics[width=0.49\textwidth]{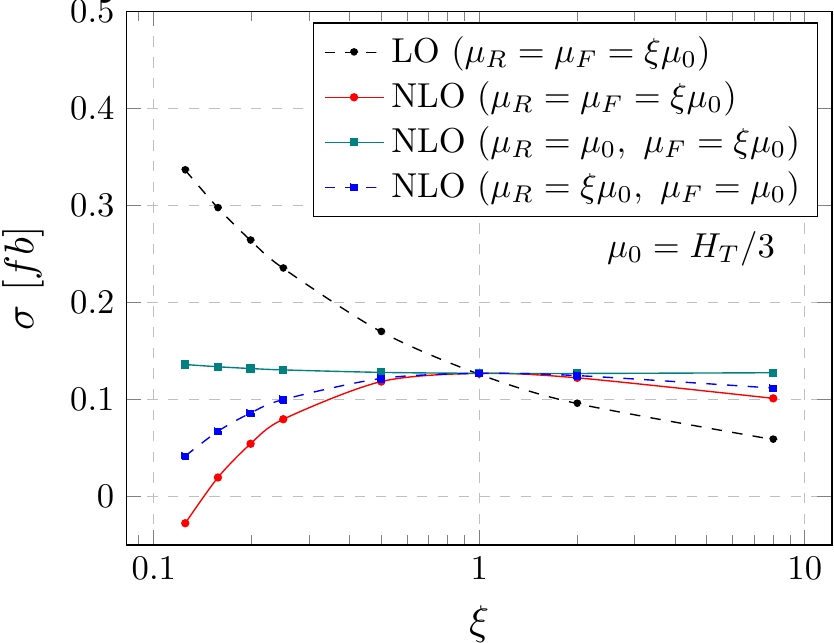}
  \includegraphics[width=0.49\textwidth]{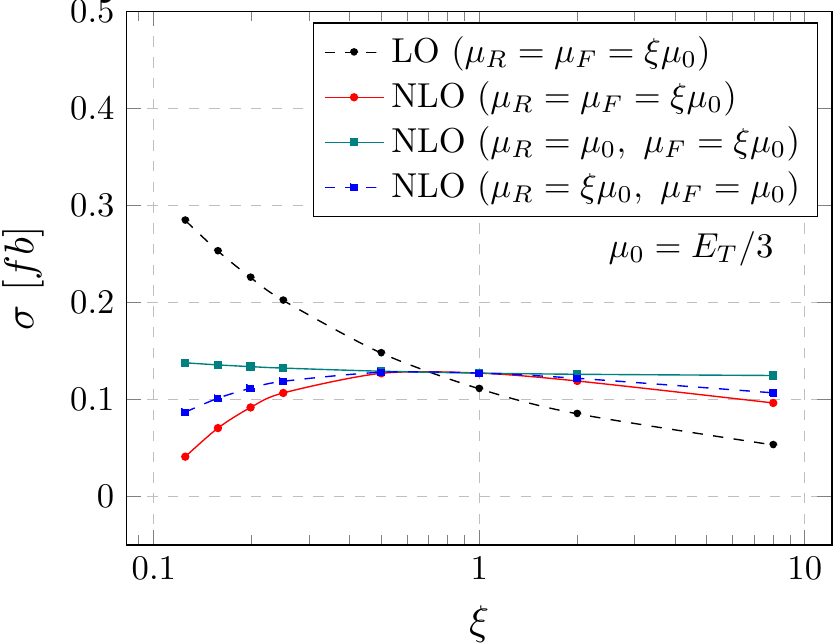}
  \includegraphics[width=0.49\textwidth]{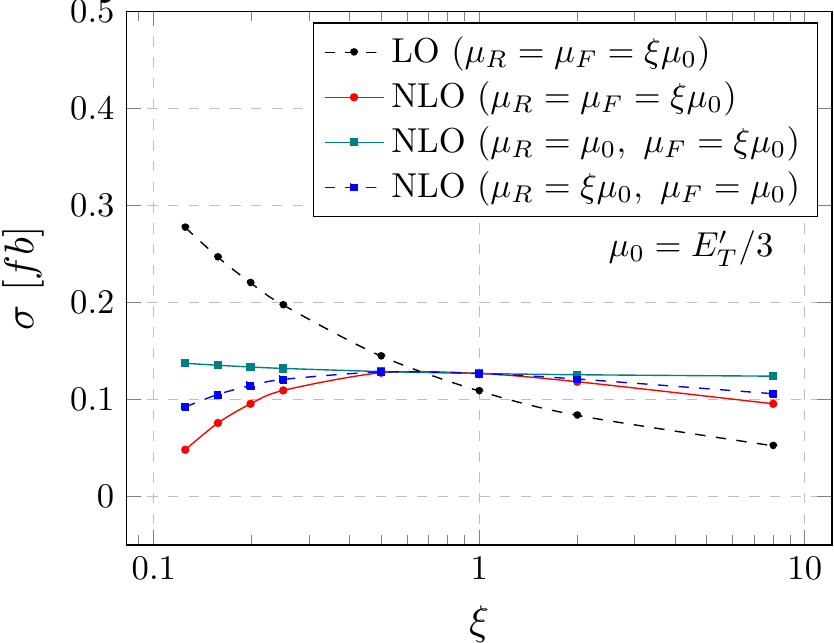}
  \includegraphics[width=0.49\textwidth]{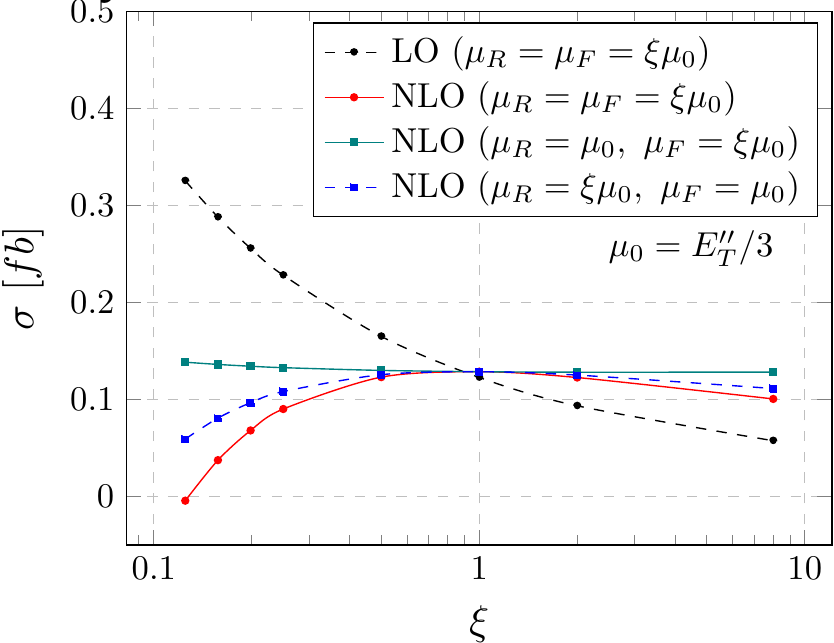}
\end{center}
\vspace{-0.6cm}
\caption{\it
 Scale dependence of the LO and NLO integrated cross
section for the $pp\to e^+ \nu_e \mu^- \bar{\nu}_\mu b\bar{b} \,
\nu_\tau \bar{\nu}_\tau +X$ process at the LHC run II with
$\sqrt{s}=13$ TeV. Renormalisation and factorisation scales are set
to $\mu_R=\mu_F=\xi \mu_0$ where  $\mu_0=H_T/3\,,E_T/3\,,
E_T^\prime/3,E_T^{\prime\prime}/3$ and $\xi \in
\left\{0.125,\dots, 8\right\}$.  The LO and NLO CT14 PDF
sets are employed. For each case of $\mu_0$ also shown is the
variation of $\mu_R$ with fixed $\mu_F$ and the variation of $\mu_F$
with fixed $\mu_R$.}
\label{fig:scale-dynamic}
\end{figure}

In the following we inspect our dynamical scale choices. Results for
four cases, $\mu_0=H_T/3,E_T/3,E^\prime_T/3$ and
$\mu_0=E^{\prime\prime}_T/3$ are summarised in Table
\ref{tab:scales-dynamic}. They have been evaluated with the LO and NLO
CT14 PDF sets. Also shown are theoretical errors as obtained from the
scale variation and the corresponding ${\cal K}$ factors. The latter are the ratios
of the NLO to LO cross sections. All LO and NLO results agree very
well with the corresponding predictions for the fixed scale within the
quoted theoretical errors. Specifically, the agreement of
$0.05\sigma-0.2\sigma$ $(0.03\sigma-0.4\sigma)$ has been obtained at
LO (NLO). Overall, no substantial reduction of theoretical uncertainties can
be observed for integrated cross sections once a kinematic dependent
scale is chosen. Specifically, for the LO cross section,
after the symmetrisation of the theoretical error is applied, the
theoretical error of the order of $28\%-29\%$ can be reported. For the
NLO case, on the other hand, the $2.8\%-4.1\%$ range has been
obtained. For all $\mu_0$ the $pp\to e^+\nu_e\mu^-\bar{\nu}_\mu
b\bar{b} \nu_\tau \bar{\nu}_\tau$ process receives positive and small
$(1\%-5\%)$ to moderate $(14\%-17\%)$ NLO QCD corrections. Thus,
judging just by the integrated cross section the case could be made
that both $\mu_0=H_T/3$ and $\mu_0=E^{\prime\prime}_T/3$ combine two
advantages, the smallest theoretical error and a small size of the 
higher order corrections as compared to the fixed scale choice. Of
course, the importance of the dynamic scale does not lie in the
calculation of the integrated cross section, which, after all, is a
quite inclusive observable, hence less sensitive to the details of the
scale choice. A place in which the dynamic scale must prove its
usefulness is the correct description of various differential cross
sections over a wide range of phase space,  which are relevant from the
point of view of top quark physics phenomenology. For completeness, in
Figure \ref{fig:scale-dynamic} we present again the scale dependence of
the LO and NLO integrated cross section for each case of $\mu_0$. The
LO and the NLO CT14 PDF sets are employed. Also here a similar
pattern as for $\mu_0=m_t+m_Z/2$ can be noticed, i.e. the independence
of the NLO cross section from the variation of $\mu_F$ while keeping
fixed the value of $\mu_R$. Generally, from the point of view of the
integrated $\sigma_{pp\to e^+\nu_e\mu^-\bar{\nu}_\mu b\bar{b} \nu_\tau
\bar{\nu}_\tau}$ cross section each scale is a valid choice that might
be used in real phenomenological applications.
%
%

%
\subsection{Choosing the scale for differential cross sections}
%
%
%
\begin{figure}[t!]
\begin{center}
  \includegraphics[width=0.49\textwidth]{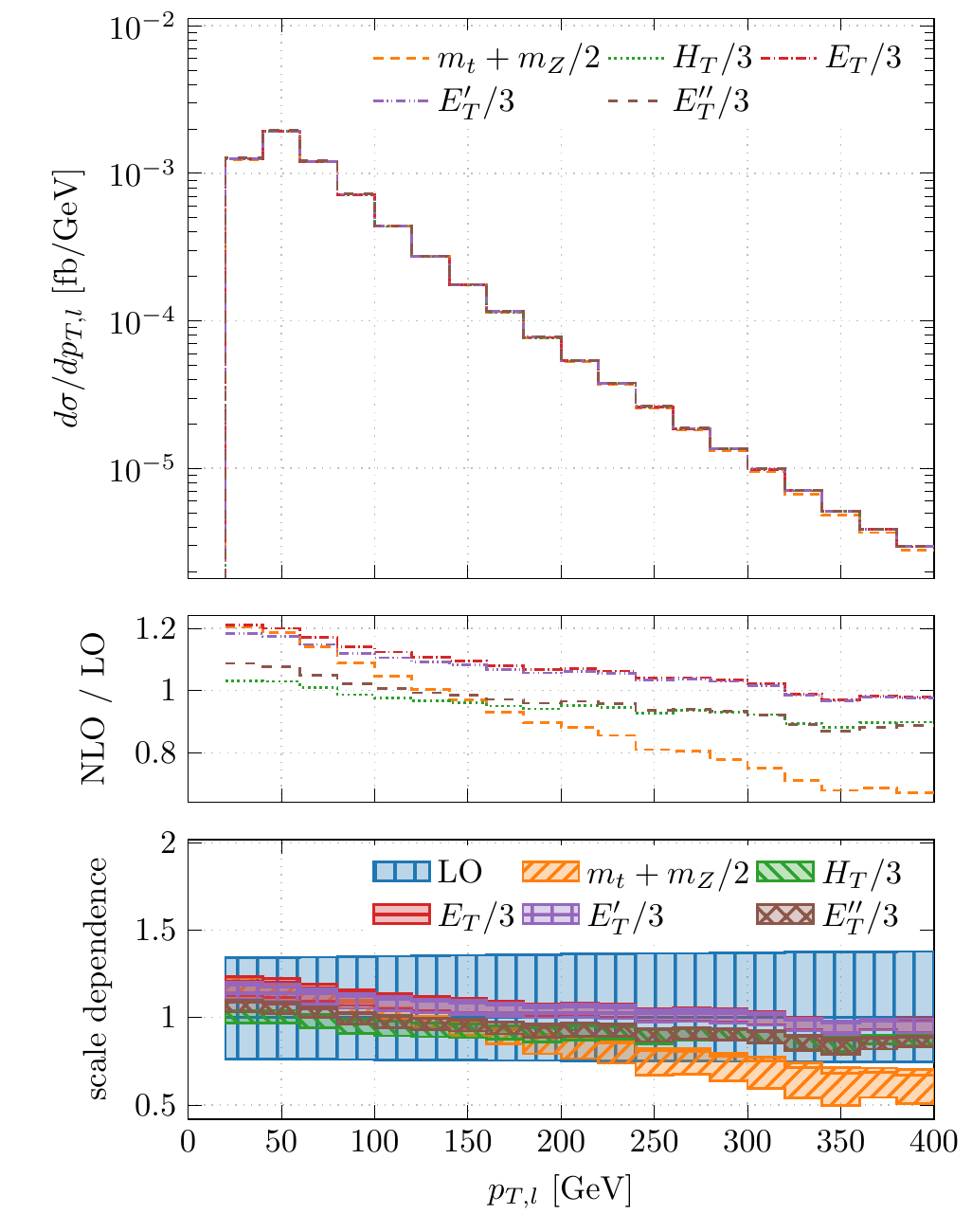}
  \includegraphics[width=0.49\textwidth]{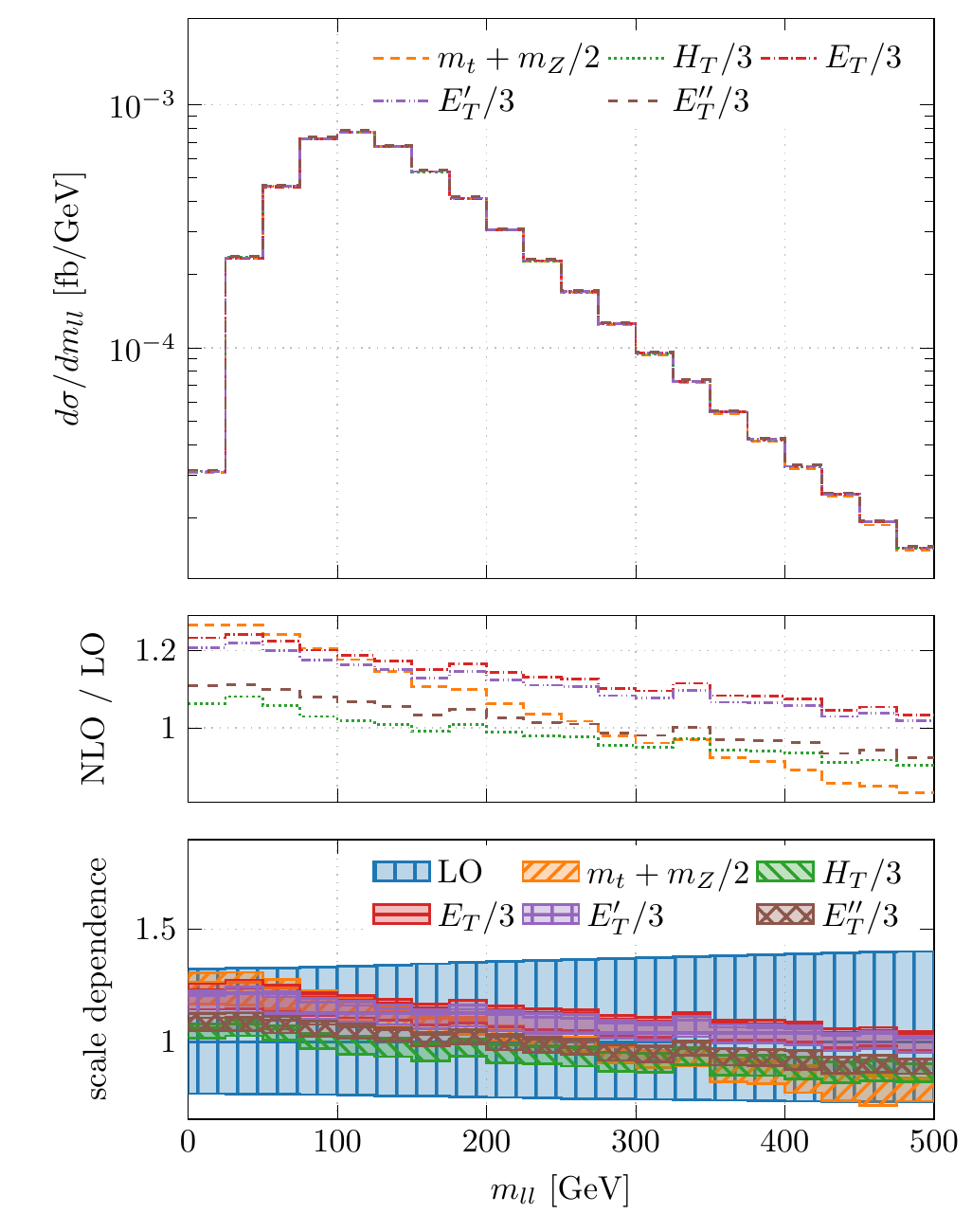}
  \includegraphics[width=0.49\textwidth]{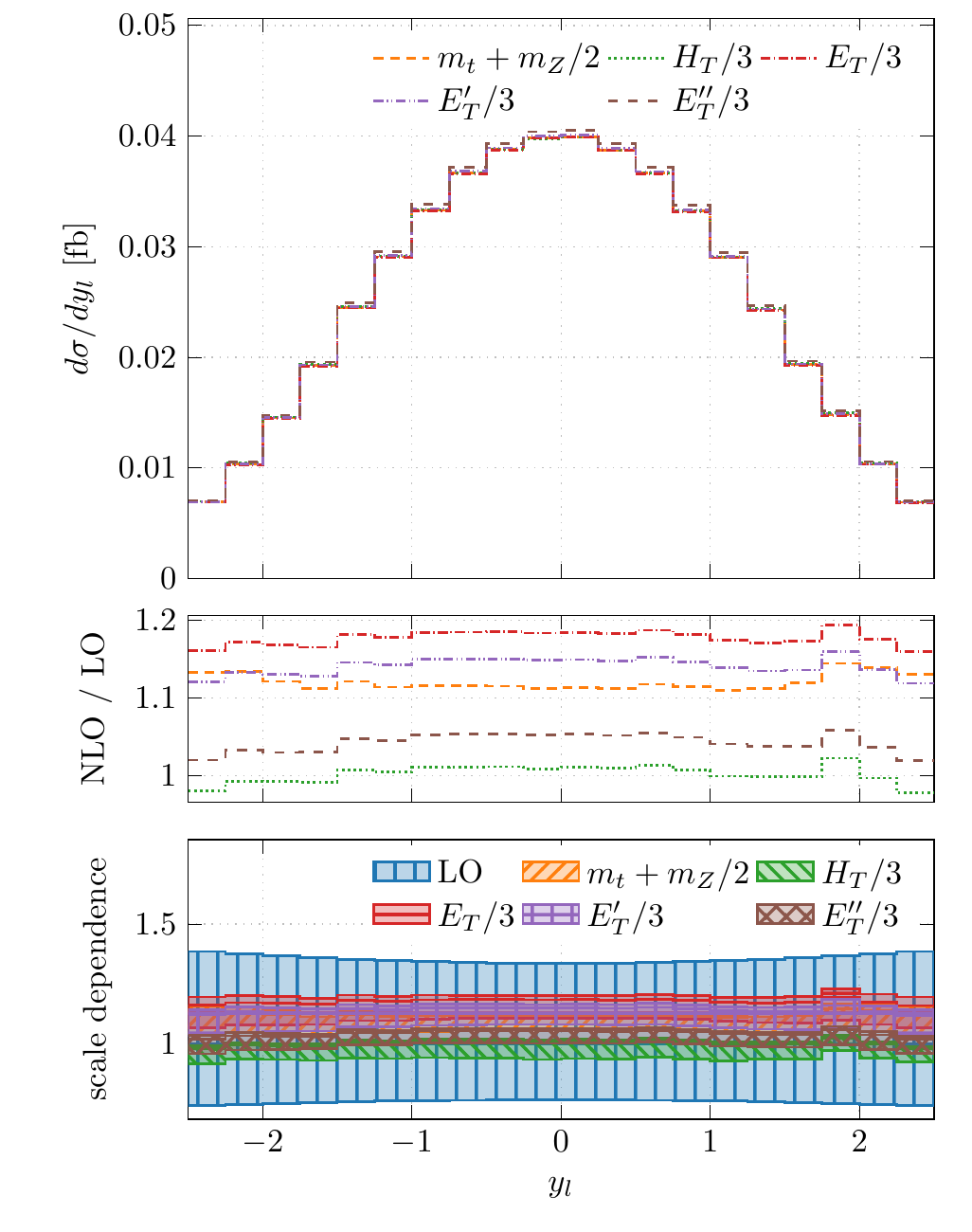}
   \includegraphics[width=0.49\textwidth]{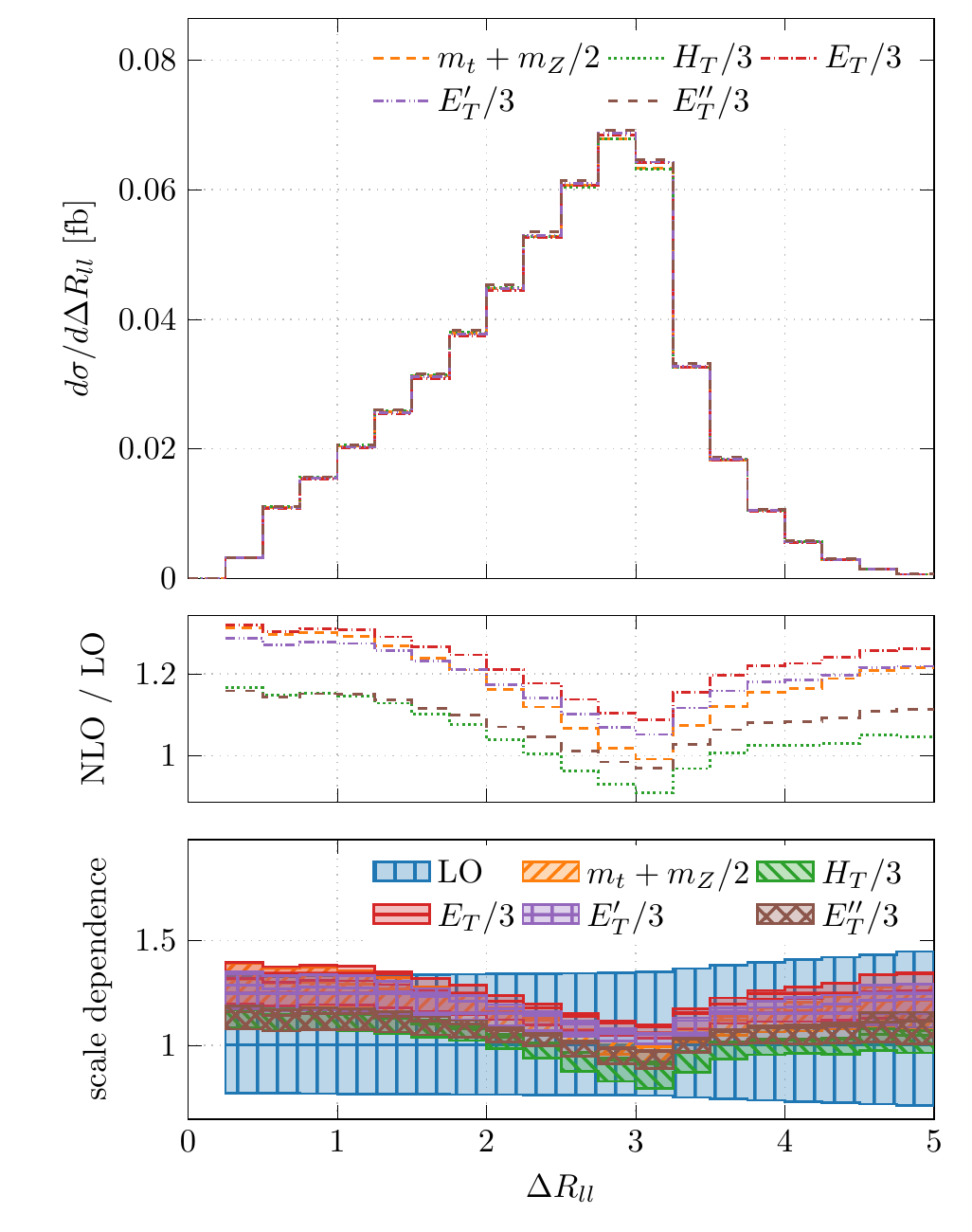}
\end{center}
\vspace{-0.6cm}
\caption{\it
  The $pp\to e^+ \nu_e \mu^- \bar{\nu}_\mu b\bar{b} \, \nu_\tau
\bar{\nu}_\tau +X$ differential cross section distribution as a
function of (averaged) $p_{T,\, \ell}$, $m_{\ell\ell}$, (averaged)
$y_\ell$ and $\Delta R_{\ell\ell}$ at the LHC run II with
$\sqrt{s}=13$ TeV. The upper plots show absolute NLO QCD predictions
for various values of $\mu_0$ where $\mu_R=\mu_F=\mu_0$. The middle
panels display differential ${\cal K}$ factors.  The lower panels
present differential ${\cal K}$ factors together with the uncertainty
band from the scale variation for various values of $\mu_0$. Also
given is the relative scale uncertainties of the LO cross section for
$\mu_0=m_t+m_Z/2$.  The LO and the NLO CT14 PDF sets are employed. }
\label{fig:leptons}
\end{figure}
\begin{figure}[t!]
\begin{center}
  \includegraphics[width=0.49\textwidth]{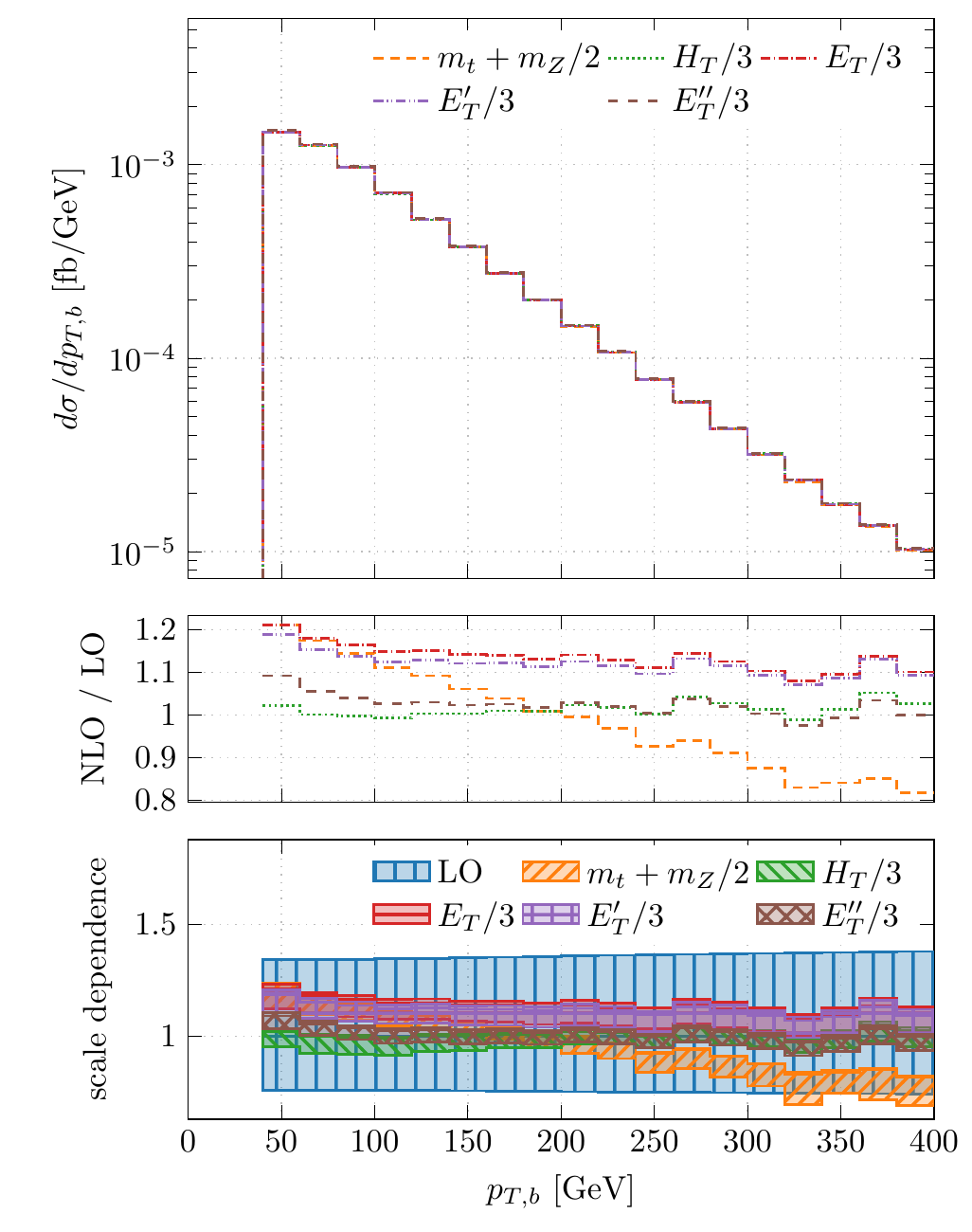}
  \includegraphics[width=0.49\textwidth]{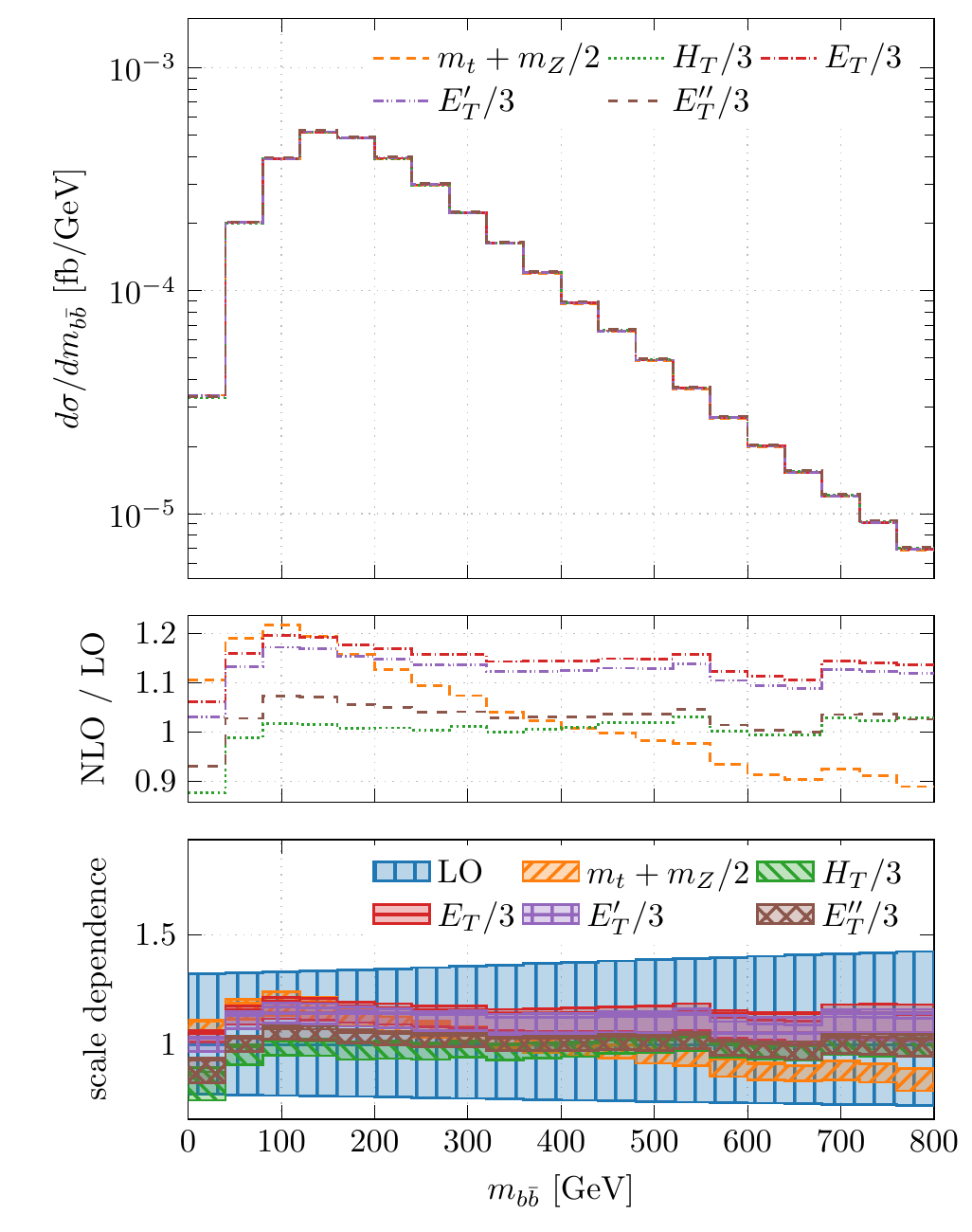}
  \includegraphics[width=0.49\textwidth]{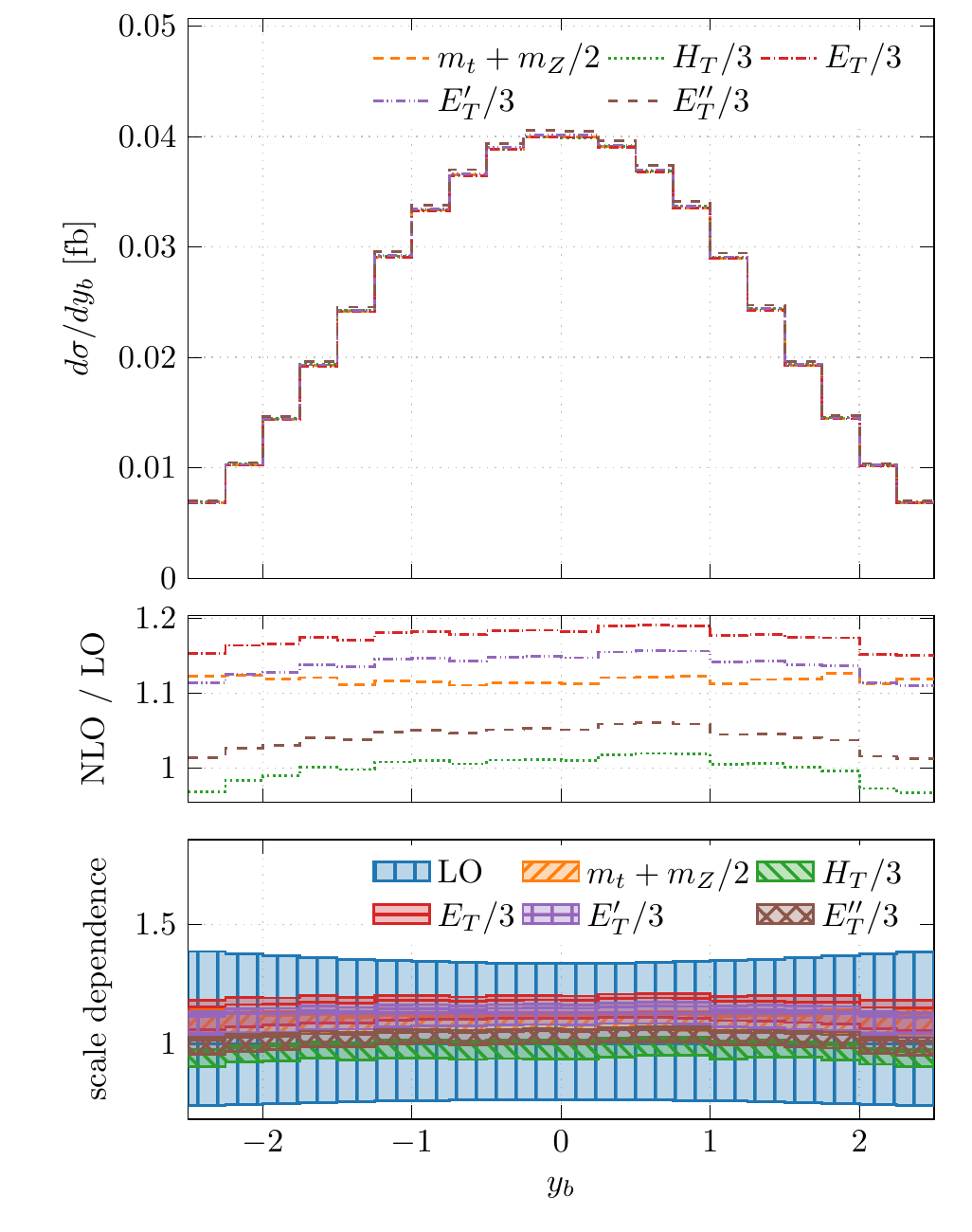}
   \includegraphics[width=0.49\textwidth]{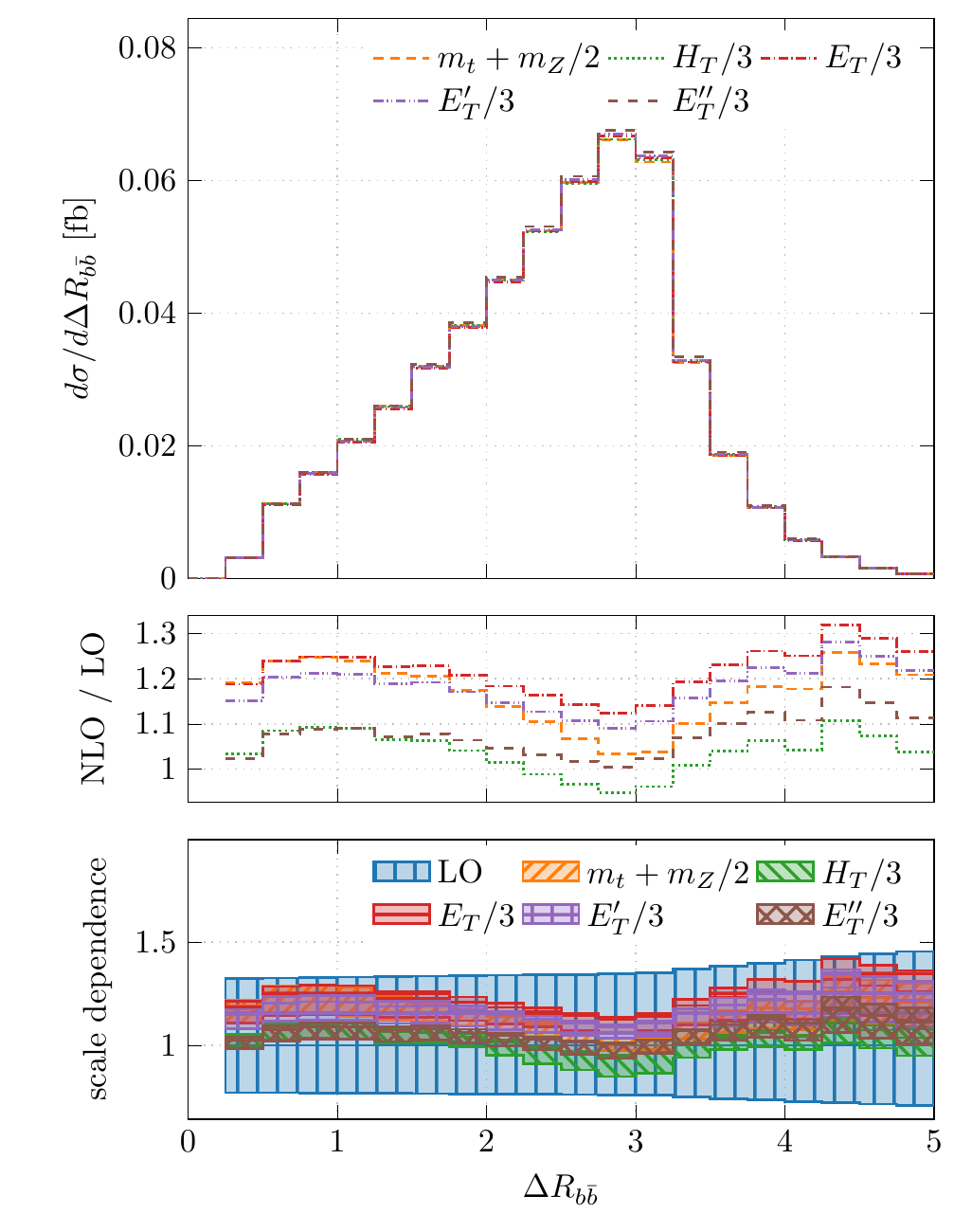}
\end{center}
\vspace{-0.6cm}
\caption{\it
 The $pp\to e^+ \nu_e \mu^- \bar{\nu}_\mu b\bar{b} \, \nu_\tau
\bar{\nu}_\tau +X$ differential cross section distribution as a
function of (averaged) $p_{T,\, b}$, $m_{b\bar{b}}$, (averaged) $y_b$
and $\Delta R_{b\bar{b}}$ at the LHC run II with $\sqrt{s}=13$
TeV. The upper plots show absolute NLO QCD predictions for various
values of $\mu_0$ where $\mu_R=\mu_F=\mu_0$. The middle panels display
differential ${\cal K}$ factors.  The lower panels present
differential ${\cal K}$ factors together with the uncertainty band
from the scale variation for various values of $\mu_0$. Also given is
the relative scale uncertainties of the LO cross section for
$\mu_0=m_t+m_Z/2$.  The LO and the NLO CT14 PDF sets are employed. }
\label{fig:b-jets}
\end{figure}

In the following we examine the size of NLO QCD corrections to various
differential cross section distributions with the different scale choices
that we have proposed in Section~\ref{sec:2}. The observables that we
are going to present are obtained with our default CT14 PDF sets as
well as for the cuts and parameters specified in the previous section. We
start with the standard observables like for example the averaged
transverse momentum and rapidity of the charged lepton
$(p_{T,\,\ell}\,, y_\ell)$, and the averaged transverse momentum and
rapidity of the $b$-jet $(p_{T,\, b}\,, y_b)$, the invariant mass of
two $b$-jets $(m_{bb})$, the invariant mass of two charged leptons
$(m_{\ell\ell})$ as well as the separation in the rapidity-azimuthal
angle plane between the two $b$-jets and the two charged leptons
$(\Delta R_{bb}\,, \Delta R_{\ell\ell})$. By examining these observables
we would like to establish which of the proposed dynamical scales is
the most suitable for the description of the $pp\to e^+\nu_e \mu^-
\bar{\nu}_{\mu} b\bar{b}\nu_\tau \bar{\nu}_\tau$ production process at
the differential level in the presence of rather inclusive cuts on the
final states. Ideally, we would be interested in the scale choice,
which guarantees us small NLO QCD corrections in the whole plotted
range for all observables and at the same time reduces theoretical
uncertainties as compared to results obtained with the fixed scale
choice. Thus, for comparison purposes we also present differential
cross section results with $\mu_R=\mu_F=\mu_0=m_t+m_Z/2$.

We start with the leptonic observables that are depicted in Figure
\ref{fig:leptons}. For each observable we present three plots. The
upper plots show absolute NLO QCD predictions for various values of
$\mu_0$ where $\mu_R=\mu_F=\mu_0$. The middle panels provide
differential ${\cal K}$ factors defined as ${\cal K}=d\sigma^{\rm
NLO}(\mu_0)/d\sigma^{\rm LO}(\mu_0)$. The lower panels display the
same differential ${\cal K}$ factors together with the uncertainty
bands from the scale variation. The latter are defined according to
${\cal K}(\mu)=d\sigma^{\rm NLO}(\mu)/d\sigma^{\rm
LO}(\mu_0)$. Additionally, the LO blue bands are given to illustrate
the relative scale uncertainty of the LO cross section. The latter are
defined according to ${\cal K}(\mu)=d\sigma^{\rm LO}(\mu)/d\sigma^{\rm
LO}(\mu_0)$ for $\mu_0=m_t+m_Z/2$. For the dimensionful observables
$p_{T,\,\ell}$ and $m_{\ell\ell}$ we can observe perturbative instabilities in high
energy tails of distributions in the case of $\mu_0=m_t+m_Z/2$ as can be
seen from the lower panels. In these regions negative NLO QCD
corrections of the order of $33\% \, (17\%)$ are visible for
$p_{T,\,\ell}$ $(m_{\ell\ell})$. These results can be compared with
the results for the dynamical scale choices where also negative but
rather moderate higher order corrections of the order of $10\%-11\%$
$(8\%-10\%)$ have been found for the tails of $p_{T,\,\ell}$
$(m_{\ell\ell})$ differential cross section distributions respectively
with $\mu_0=H_T/3$ and $\mu_0=E_T^{\prime\prime}/3$. Even though for
$\mu_0=E_T/3$ and $\mu_0=E^\prime_T/3$ we have obtained even smaller
NLO QCD corrections in these regions, i.e. of the order of $\pm
(2\%-3\%)$ only, the size of distortions is much larger for these
scale choices.  Consequently for $\mu_0=H_T/3$ and
$\mu_0=E_T^{\prime\prime}/3$ flatter differential ${\cal K}$-factors
are registered for these two observables. For a dimensionless
observable $y_{\ell}$ on the other hand almost constant corrections
are obtained in the whole plotted range independently of the scale
choice.  What makes the result different for the various  scale
choices is the size of NLO QCD corrections. For the fixed scale as well
as for $\mu_0=E_T/3$ and $\mu_0=E^\prime_T/3$ they are positive and in
the range of $12\%-15\%$ while for $\mu_0=H_T/3$
$(\mu_0=E^{\prime\prime}/3)$ the maximum corrections received are of
the order of $\pm 2\%$. Finally, for $\Delta R_{\ell\ell}$ substantial
distortions are noticed that are up to $32\%$, $26\%$, $23\%$, $24\%$
and $19\%$ respectively for $\mu_0=m_t+m_Z/2, \,H_T/3,\,E_T/3,\,
E_T^\prime/3$ and $E_T^{\prime\prime}/3$.

In the next step we concentrate on theoretical uncertainties for these
leptonic observables as estimated from scale variations at the NLO
level in QCD. For the averaged transverse momentum of the charged
lepton substantial scale variations are noticed at the end of the
plotted spectrum, i.e. around 400 GeV for $\mu_0=m_t+m_z/2$. In these
regions theoretical uncertainties taken conservatively as a maximum of
the lower and upper bounds are $\pm 25\%$ ($\pm 14\%$ after
symmetrisation).  On the other hand for all presented dynamical scale
choices they are reduced down to $\pm 7\%$ ($\pm 4 \%$ after
symmetrisation). In the latter case they are almost constant in the
whole plotted range. These numbers can be compared to the LO scale
uncertainties that for the fixed scale choice are up to $\pm 41\% \,
(\pm 34\%)$. In the case of the invariant mass of the positron and the
muon the difference between the fixed scale choice and the dynamical
ones is milder. For $\mu_0=m_t+m_Z/2$ theoretical uncertainties up to
$\pm 11\% \, (\pm 6\%)$ have been reached, whereas for $\mu_0=H_T/3$
up to $\pm 9\% \, (\pm 4.5\%)$. The latter is reduced down to $\pm 7\%
\, (\pm 3.5\%)$ for other scales. This is a substantial reduction
taking into account that at LO  theoretical uncertainties up to
$\pm 41\% \, (\pm 34\%)$ have been evaluated.  As expected
dimensionless observables have rather constant scale dependence
independent of the scale choice. For $\Delta R_{\ell\ell}$ we have
obtained theoretical uncertainties around $\pm 10\%\, (\pm 8\%)$,
whereas for $y_{\ell}$ we have instead up to $\pm 8\%\, (\pm 6\%)$ for
$\mu_0=m_t+m_Z/2,E_T/3$ and $\mu_0=E^\prime_T/3$ as well as up to $\pm
6\%\, (\pm 3\%)$ for $\mu_0=H_T/3$ and $\mu_0= E^{\prime
\prime}/3$. These outcomes can be compared to $\pm 43\%\, (\pm 36 \%)$
and $\pm 37\%\, (\pm 31\%)$ uncertainties at the LO level respectively
for $\Delta R_{\ell\ell}$ and $y_{\ell}$ with $\mu_0=m_t+m_Z/2$.

Similar conclusions can be drawn for the $b$-jet kinematics that is
shown in Figure ~\ref{fig:b-jets}. For the averaged $p_T$ distribution
of the bottom jet at the end of the plotted range negative NLO QCD
corrections of the order of $18\%$ are acquired for
$\mu_0=m_t+m_Z/2$. This can be compared with positive $20\%$
corrections at the beginning of the spectrum which resulted in
distortions of the order of $40\%$. The situation is substantially
improved for the case of $\mu_0=H_T/3$ where positive higher order QCD
corrections below $5\%$ are attained for $p_{T,\,b}\in
\left[40,400\right]$ GeV. For the remaining three scale choices the
similar size of distortions of the order of $10\%$ have been
observed. For the invariant mass of the two $b$-jet system the best
scale choice seems to be again $\mu_0=H_T/3$ for which rather constant
corrections, with the exception of the beginning of the spectrum, are
visible. The former are of the order of $+3\%$, the latter are up to
$-12\%$. This can be contrasted with results for $\mu_0=m_t+m_Z/2$
where we have obtained NLO corrections ranging from $+22\%$ down to
$-11\%$. Looking at the dimensionless observable like for example the
averaged rapidity of the $b$-jet we have noticed almost constant NLO
QCD corrections in the considered range $y_b\in \left[-2.5,
2.5\right]$. The smallest corrections of the order of $-3\%$ and
$+2\%$ have been obtained respectively for $\mu_0=H_T/3$ and
$\mu_0=E^{\prime \prime}/3$. On the other hand the largest
corrections, up to even $+18\%$, have been received for
$\mu_0=E_T/3$. For the fixed scale choice they are only up to
$+13\%$. Finally, for the last standard observable that we have
studied, which is $\Delta R_{b\bar{b}}$, we can recommend
$\mu_0=H_T/3$ for which NLO QCD corrections maximally up to $+10\%$
and distortions up to $15\%$ have been gained. By comparison the
distortions are the most severe for the fixed scale choice. They
amount even up to $22\%$.

In the following we move to the NLO theoretical uncertainties for
observables that describe the  kinematics of the bottom jets. Once more we
notice that for the fixed scale choice represented by
$\mu_0=m_t+m_Z/2$ and for dimensionful observables like for example
the averaged transverse momentum of the $b$-jet, $p_{T,\, b}$,
theoretical uncertainties are outside the LO bands at the end of the
plotted range. Nevertheless taken conservatively they are rather
moderate of the order of $\pm 15\%$ ($\pm 8\%$ after symmetrisation)
in that region. The size and perturbative behaviour is modified when
the dynamical scale choice is applied. For $\mu_0=H_T/3$ and
$\mu_0=E^{\prime\prime}/3$ theoretical errors are reduced by more than
a factor of $2$, down to $\pm 7\% \, (\pm 4\%)$ and $\pm 6\%\, (\pm
4\%)$ respectively.  We also note that employing the dynamical scale
choices caused NLO bands to lie within the LO ones as one would expect
from a well-behaved perturbative expansion in $\alpha_s$. For the
invariant mass of the two $b$-jet system we have a similar
behaviour. For results with $\mu_0=m_t+m_Z/2$ the theoretical
uncertainties up to $\pm 11\% \, (\pm 6\%)$ have been obtained,
whereas in the case of $\mu_0=H_T/3,\, E^{\prime \prime}_T/3$
decreased theoretical uncertainties up to only $\pm 8\% \, (\pm 5\%)$
have been estimated. In both cases the improvement with respect to the
LO theoretical uncertainties is dramatic since we have $\pm 41\%\,
(\pm 34 \%)$ for $p_{T,\, b}$ and $\pm 44\% \, (\pm 36\%)$ for
$m_{b\bar{b}}$. Considering angular distributions like $y_b$ and
$\Delta R_{b\bar{b}}$ we have rather constants theoretical
uncertainties below $8 \%$ independent of the scale choice for the
former and below $10 \%$ for the later. After symmetrisation is
applied they go below $5\%$ and $8 \%$ respectively.  Whereas at the
LO level they are up to $\pm 37 \% \, (\pm 31 \%)$ and $\pm 44 \%\,
(\pm 36\%)$.

Combining information about the size of NLO QCD corrections and NLO
QCD theoretical uncertainties we conclude that either scale
$\mu_0=H_T/3$ or $\mu_0=E^{\prime \prime}/3$ should be employed at the
differential level for the adequate description of the standard
observables in the $e^+\nu_e \mu^- \bar{\nu}_\mu b\bar{b} \nu_\tau
\bar{\nu}_\tau$ production process at the LHC with a centre of mass
system energy of $\sqrt{s}=13$ TeV in the presence of rather inclusive
cuts on the measured final states.

%
\subsection{Impact of higher order corrections on new
  physics observables}
%
%

%
\begin{figure}[t!]
\begin{center}
  \includegraphics[width=0.49\textwidth]{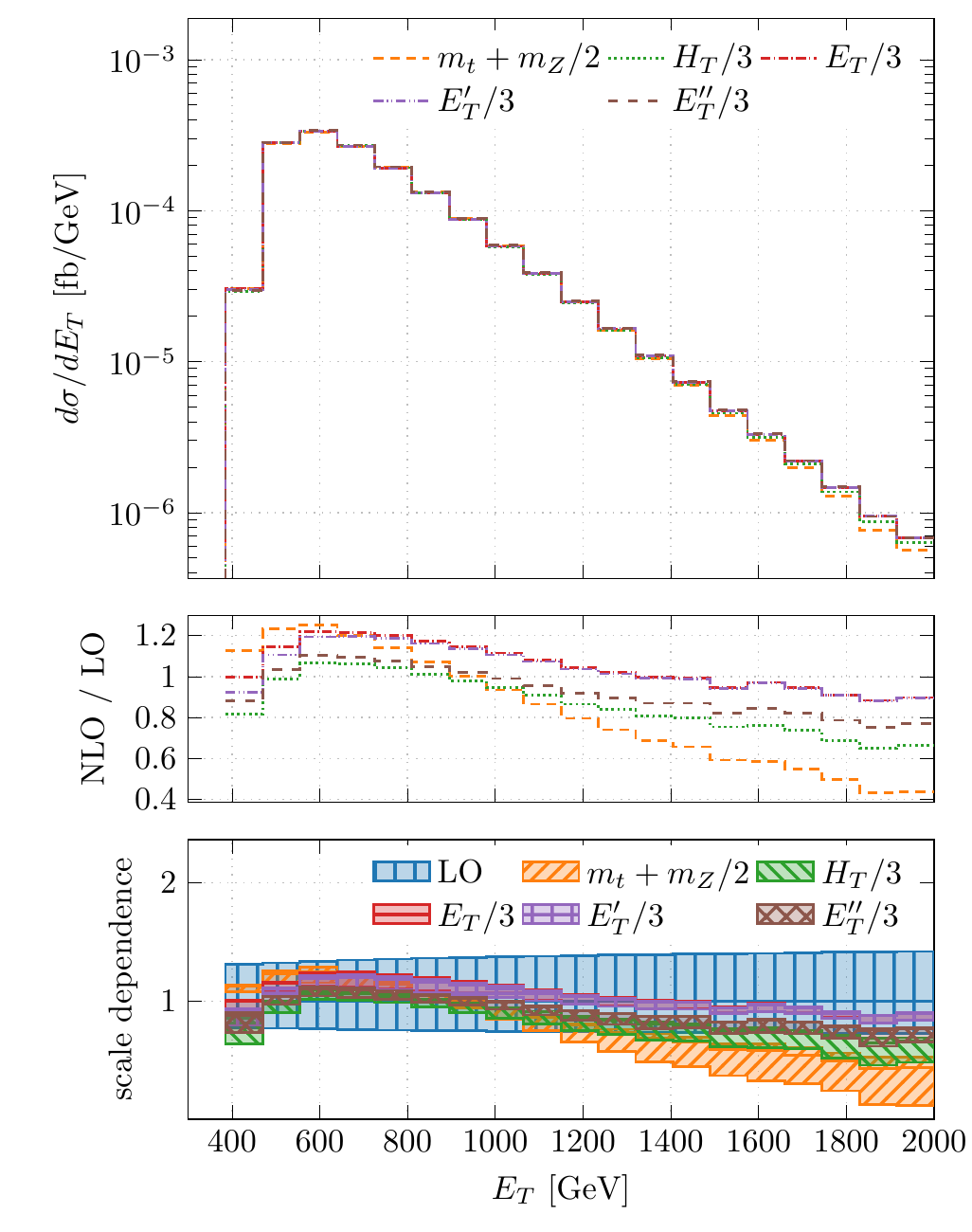}
  \includegraphics[width=0.49\textwidth]{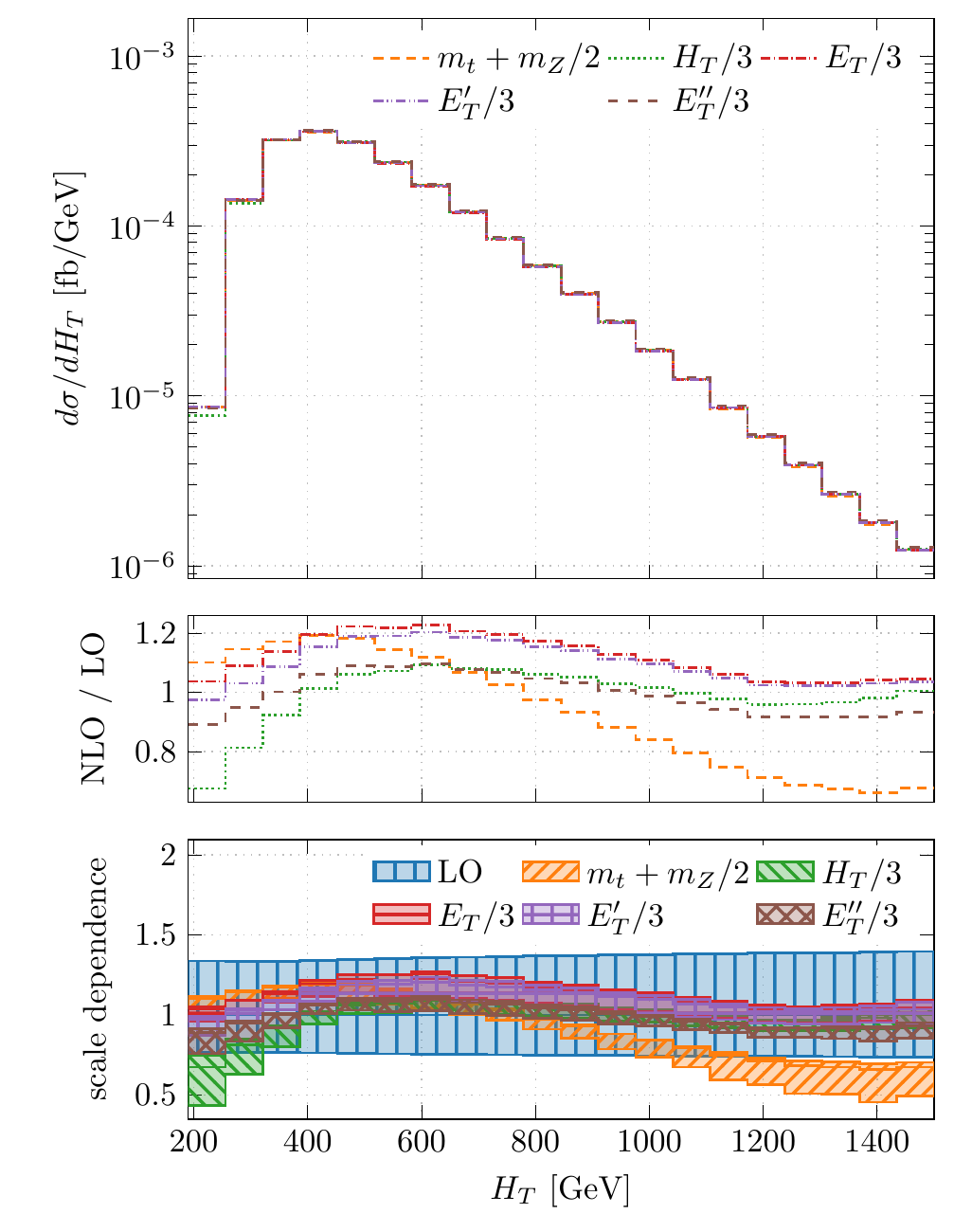}
  \includegraphics[width=0.49\textwidth]{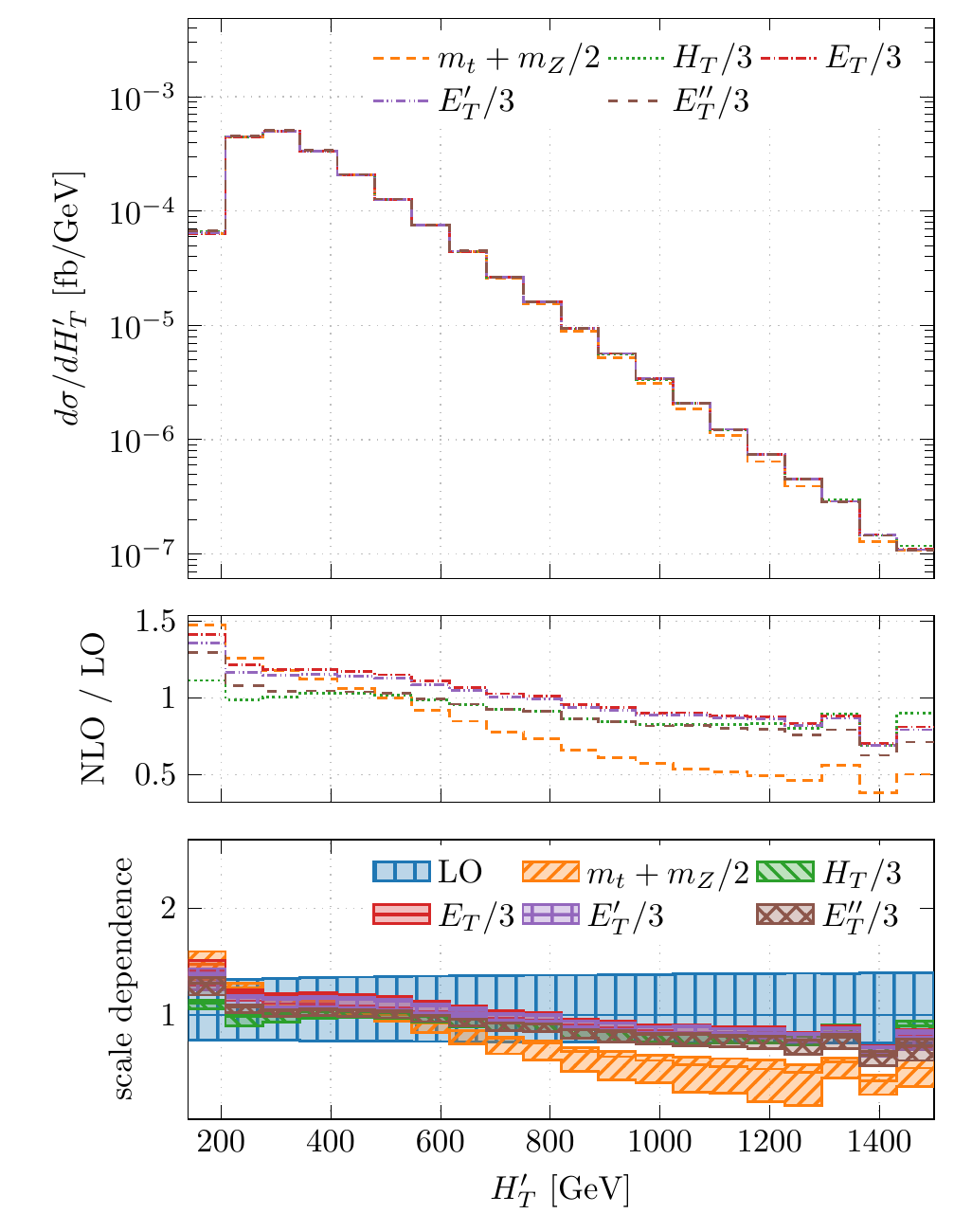}
\end{center}
\vspace{-0.6cm}
\caption{\it
 The $pp\to e^+ \nu_e \mu^- \bar{\nu}_\mu b\bar{b} \, \nu_\tau
\bar{\nu}_\tau +X$ differential cross section distribution as a
function of $E_T$, $H_T$ and $H_T^\prime$ at the LHC run II with
$\sqrt{s}=13$ TeV. The upper plots show absolute NLO QCD predictions
for various values of $\mu_0$ where $\mu_R=\mu_F=\mu_0$. The middle
panels display differential ${\cal K}$ factors.  The lower panels
present differential ${\cal K}$ factors together with the uncertainty
band from the scale variation for various values of $\mu_0$. Also
given is the relative scale uncertainties of the LO cross section for
$\mu_0=m_t+m_Z/2$.  The LO and the NLO CT14 PDF sets are employed. }
\label{fig:new_physics1}
\end{figure}
\begin{figure}[t!]
\begin{center}
  \includegraphics[width=0.49\textwidth]{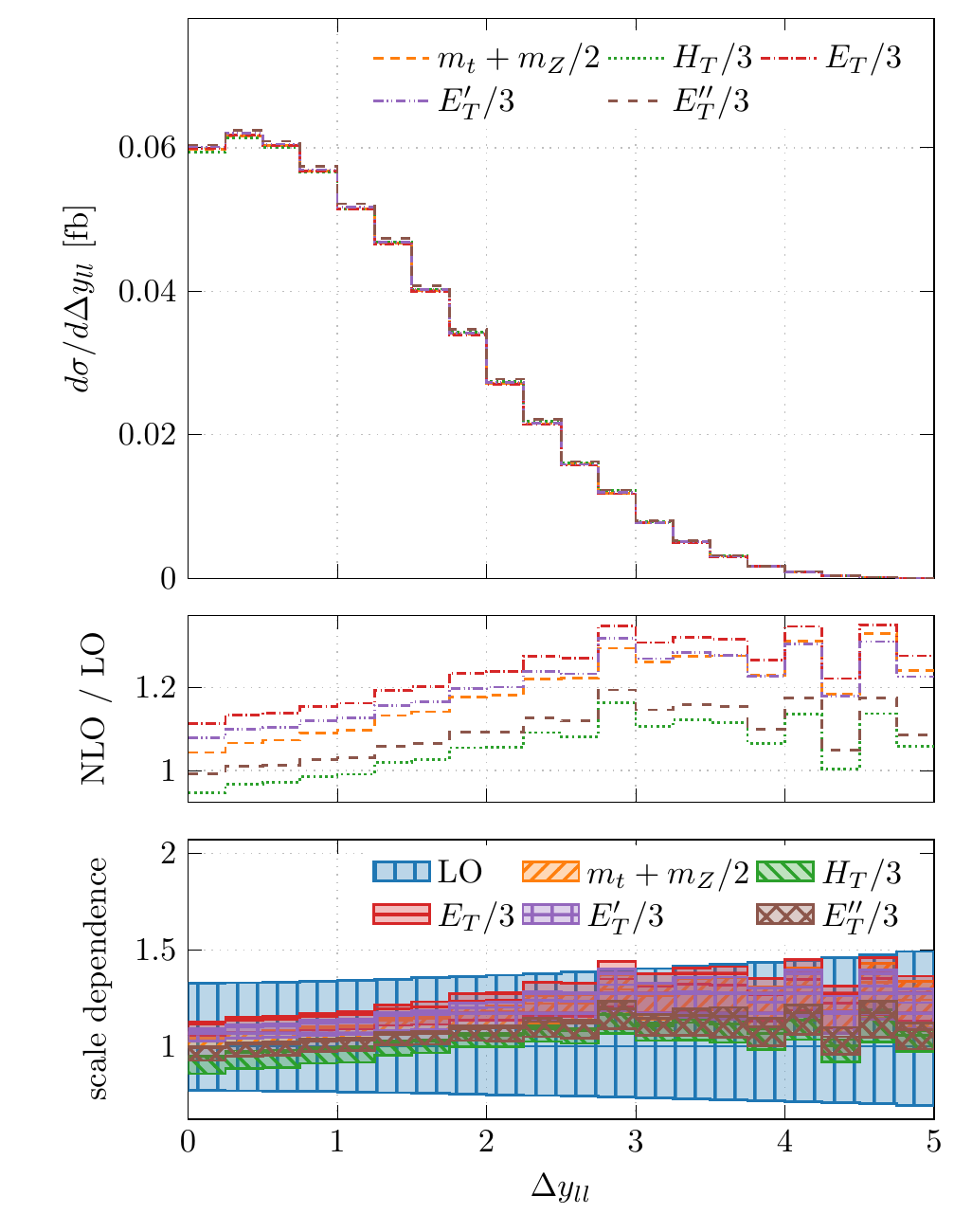} 
  \includegraphics[width=0.49\textwidth]{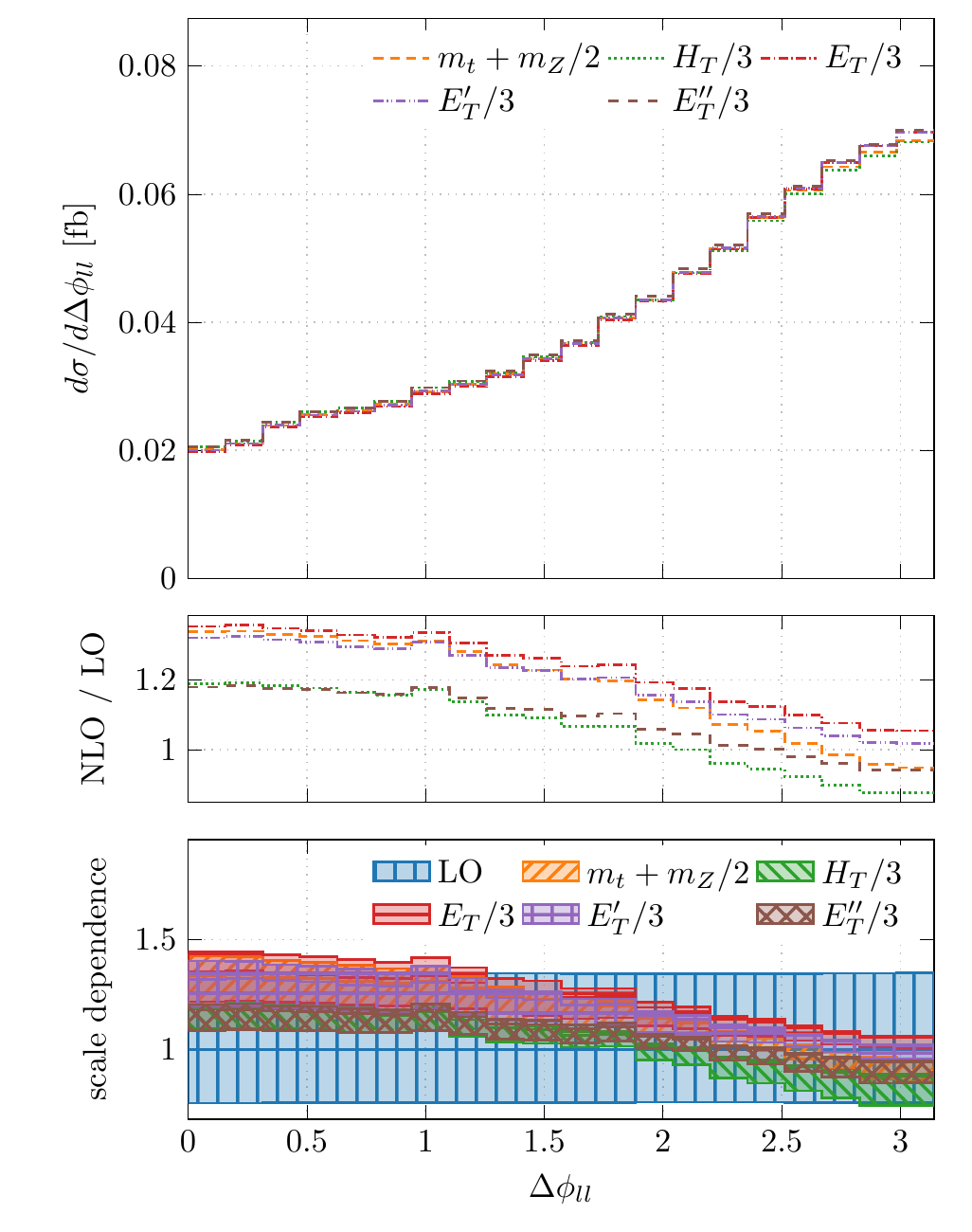}
   \includegraphics[width=0.49\textwidth]{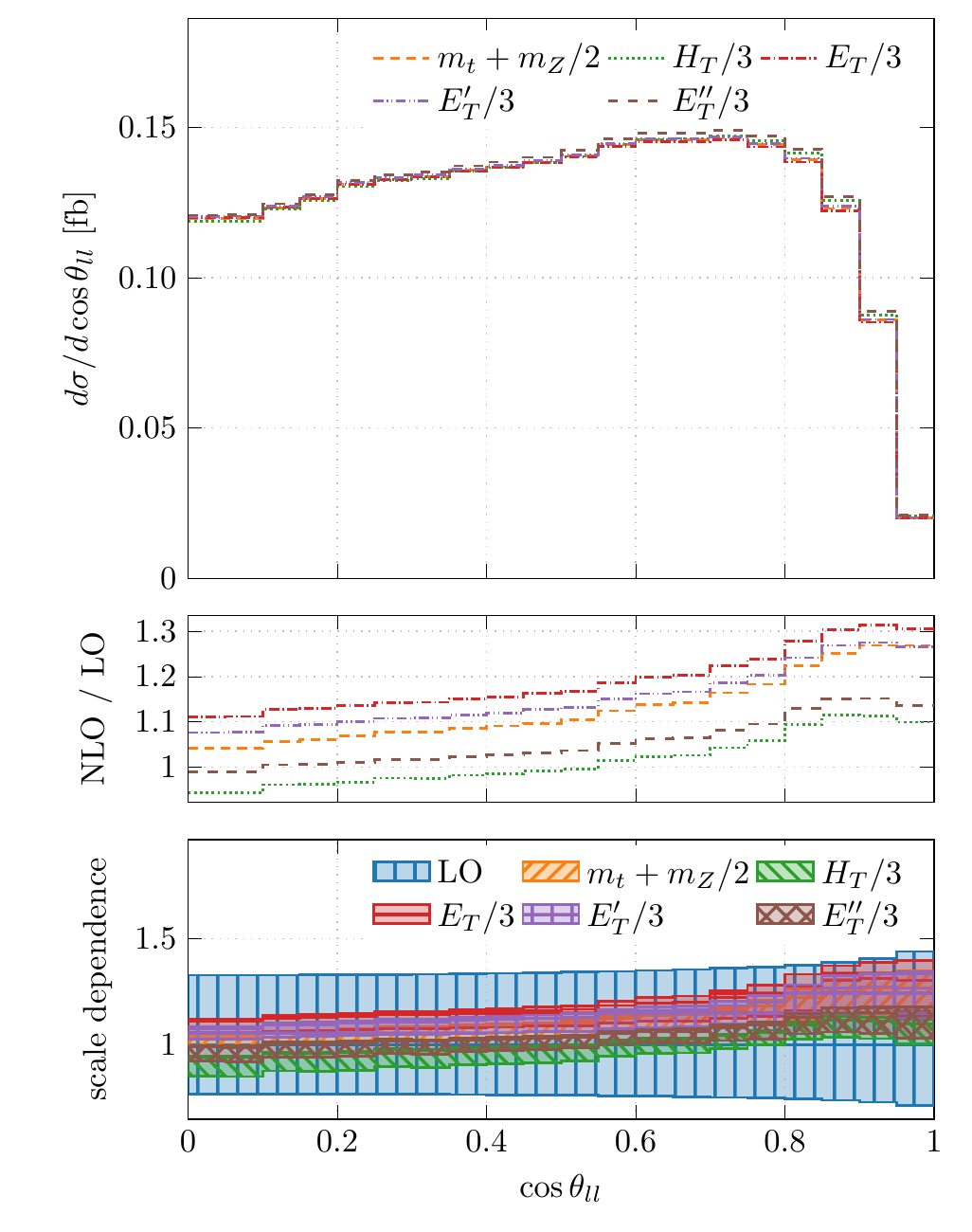}
\end{center}
\vspace{-0.6cm}
\caption{\it
 The $pp\to e^+ \nu_e \mu^- \bar{\nu}_\mu b\bar{b} \, \nu_\tau
\bar{\nu}_\tau +X$ differential cross section distribution as a
function of $\Delta y_{\ell\ell}= |y_{\ell_1}-y_{\ell_2}|$, $\Delta
\phi_{\ell\ell} = |\phi_{\ell_1} -\phi_{\ell_2}|$ and $\cos
\theta_{\ell\ell}= \tanh (\Delta y_{\ell\ell}/2)$ at the LHC run II
with $\sqrt{s}=13$ TeV. The upper plots show absolute NLO QCD
predictions for various values of $\mu_0$ where
$\mu_R=\mu_F=\mu_0$. The middle panels display differential ${\cal K}$
factors.  The lower panels present differential ${\cal K}$ factors
together with the uncertainty band from the scale variation for
various values of $\mu_0$. Also given is the relative scale
uncertainties of the LO cross section for $\mu_0=m_t+m_Z/2$.  The LO
and the NLO CT14 PDF sets are employed.}
\label{fig:new_physics2}
\end{figure}
%
Moving forward we employ our recommended scale choices
$\mu_0=H_T/3$ and $\mu_0=E^{\prime \prime}/3$ to discuss the size of
NLO QCD corrections to a few observables that are relevant in the
context of dark matter searches.  Among others we have identified six
observables, three dimensionful and three dimensionless. Specifically,
we have studied the total transverse energy, $E_T$, as given by
\begin{equation}
E_T = \sqrt{p_T^2(t) + m_t^2} + \sqrt{p^2_T(\bar{t}\,)
      +m_t^2} + \sqrt{p_T^2(Z) + m_Z^2}\,,
\end{equation}  
and the total transverse momentum of the $t\bar{t}Z$ system,
$H_T$. The total transverse momentum  build only from the
visible final states and denoted as $H_T^\prime$ is also
investigated. The latter two are defined according to 
\begin{equation}
  \begin{split}
H_T&= p_{T,\,{b_1}}+p_{T,\,{b_2}}+p_{T,\,e^+}+p_{T,\,\mu^-}
+p_T^{miss}\,,\\[0.2cm]
H^\prime_T&= p_{T,\,{b_1}}+p_{T,\,{b_2}}+p_{T,\,e^+}+p_{T,\,\mu^-} \,.
\end{split}
\end{equation}
We investigate additionally the rapidity separation of the two
charged leptons, $\Delta y_{\ell\ell}=|y_{\ell_1}-y_{\ell_2}|$, the
azimuthal angle difference between the two leptons, $\Delta
\phi_{\ell \ell} = |\phi_{\ell_1} - \phi_{\ell_2}|$ and $\cos
\theta_{\ell\ell}$ constructed  according to the following formula
\begin{equation}
  \cos\theta_{\ell\ell}= \tanh (\Delta y_{\ell\ell}/2)\,.
\end{equation}
The angular distributions of the charged leptons resulting from top
decays carry information about the spin correlations between the
final-state top quarks. Thus, they can be used for example to study
the CP nature of the coupling between the mediator particle and top
quarks in various dark matter scenarios, see
e.g. \cite{Haisch:2016gry}. Proper modelling for these observables
within the SM is a fundamental requirement for a correct
interpretation of the possible signals of new physics that may arise
in the $pp\to t\bar{t} + p_T^{miss}$ channel.

In Figure \ref{fig:new_physics1} we present the differential cross
section distribution as the function of $E_T$, $H_T$ and
$H_T^\prime$. For comparison reasons also for these observables
predictions for all scale choices for $\mu_0=\mu_F=\mu_R$ are
depicted. In the case of $E_T$, negative and substantial NLO QCD
corrections up to $34\%$, $23\%$ and $56\%$ are obtained around $2$
TeV respectively for $\mu_0=H_T/3$, $\mu_0=E^{\prime \prime}_T/3$ and
$\mu_0=m_t+m_Z/2$. Overall shape distortions are of the order of
$15\%$, $11\%$ for the dynamical scale choices and around $69\%$ for
the fixed scale choice. NLO theoretical uncertainties from the scale
dependence are up to $\pm 27\%$ ($\pm 16\%$ after symmetrisation),
$\pm 17\%$ ($\pm 10\%$) and $\pm 72\%$ ($\pm 46\%$) respectively for
$\mu_0=H_T/3$, $\mu_0=E^{\prime \prime}_T/3$ and $\mu_0=m_t+m_Z/2$. A
similar pattern could be seen for $H_T$ and $H_T^\prime$. In the
former case for $\mu_0=E^{\prime \prime}_T/3$ NLO QCD corrections are
negative and moderate up to $7\%$ at around $1.5$ TeV. When comparing
the threshold region above $190$ GeV with the end of the plotted
range, shape distortions of the order of $4\%$ are only detected for
this scale. On the other hand, for other two choices $\mu_0=H_T/3$ and
$\mu_0=m_t+m_Z/2$ large and negative NLO QCD corrections at the level
of $32\%$ have been perceived, respectively either at the beginning or
at the end of the $H_T$ spectrum. Consequently, shape distortions are
of the order of $33\%$ and $43\%$ for the dynamical and fixed values
of $\mu_0$. For the scale dependence we can reach $\pm 15\%$ $(\pm
9\%)$ close to the threshold and $\pm 8 \%$ $(\pm 5\%)$ around $1.5$
TeV for $\mu_0=E^{\prime \prime}_T/3$. Whereas predictions with
$\mu_0=H_T/3$ and $\mu_0=m_t+m_Z/2$ have substantially larger
theoretical errors up to $\pm 35\%$ $(\pm 23\%)$ for $\mu_0=H_T/3$ (at
the beginning of the spectrum of $H_T$) and up to $\pm 27\%$ $(\pm
16\%)$ for $\mu_0=m_t+m_Z/2$ (at the end of the plotted spectrum of
$H_T$). Finally, we examine the simplified version of $H_T$, namely
$H_T^\prime$. As already advertised the latter comprises only visible
final states, i.e. charged leptons as well as the two bottom jets and
it is frequently used by experimental groups to look for new physics
in top quark pair production. The observable received rather large
higher order corrections at the end of the plotted spectrum as
compared to $H_T$. Specifically, we have negative NLO QCD corrections
of the order of $50\%$ for the fixed scale choice and up to $29\%$ for
$\mu_0=E^{\prime \prime}_T/3$. On the other hand, for $\mu_0=H_T/3$
negative but moderate corrections up to only $10\%$ are observed. NLO
shape distortions are of the order of $97\%$, $58\%$ and $21\%$
respectively for $\mu_0=m_t+m_Z/2$, $\mu_0= E_T^{\prime\prime}/3$, and
$\mu_0= H_T/3$. Clearly, the differential ${\cal K}$-factors are far
from flat, showing major changes in the shape of the observables when
the QCD corrections of the order of $\alpha_s$ are
incorporated. When investigating scale uncertainties for the
$H_T^\prime$ observable we noticed their similarities to the case of
$H_T$. In detail, we have estimated theoretical errors up to $\pm
34\%\, (\pm 24\%)$, $\pm 19\% \, (\pm 14\%)$, and $\pm 10\%\, (\pm
7\%)$ individually for $\mu_0=m_t+m_Z/2$, $\mu_0=
E_T^{\prime\prime}/3$, and $\mu_0= H_T/3$.

Leptonic angular distributions, i.e. $\Delta y_{\ell\ell}$, $\Delta
\phi_{\ell\ell}$ and $\cos\theta_{\ell\ell}$, are depicted in Figure
\ref{fig:new_physics2}.  For the rapidity difference of the two
charged leptons we observe small corrections of the order of a few
percent for $\mu_0=H_T/3$ and $\mu_0=E_T^{\prime\prime}/3$. By
comparison for $\mu_0=m_t+m_Z/2$ they can reach even $16\%$. Overall
shape distortions at NLO for these scale choices are $2.5\%$, $5\%$
and $12\%$, respectively. For the opening angle between the two charged
leptons, on the other hand, already for both dynamical scale choices
positive corrections up to $19\%$ are visible, whereas for the fixed
scale choice they are up to $34\%$ in the same region. Also shape
distortions are substantially larger for this observable, i.e. they
are at the level of $31\%$, $23\%$ and $39\%$. Finally,
$\cos\theta_{\ell\ell}$ has received rather moderate NLO QCD
corrections up to $10\%$ for $\mu_0=H_T/3$ and up to $14\%$ for
$\mu_0=E_T^{\prime\prime}/3$. Also in this case higher order
corrections for the fixed scale choice are substantially larger,
reaching $27\%$. Shape distortions are at the level of $17\% \,
(\mu_0=H_T/3)$, $15\% \, (\mu_0=E_T^{\prime\prime}/3)$ and $23\% \,
(\mu_0=m_t+m_Z/2)$. When examining the scale dependence for these
observables we can see a similar size of theoretical errors regardless
of what scale we choose. The theoretical uncertainties are also
similar in size for all three angular observables. Specifically, for
$\Delta y_{\ell \ell}$ they are up to $\pm 9\%$ ($\pm 6\%$ after
symmetrisation) for the dynamical scale choice and up to $\pm 12\%$
$(\pm 10 \%)$ for the fixed scale choice. In the case of $\Delta
\phi_{\ell \ell}$ they are below $\pm 10\%$ $(\pm 5 \%)$ for
$\mu_0=E_T^{\prime\prime}/3$ and $(\mu_0=m_t+m_Z/2)$, while for
$(\mu_0=H_T/3)$ they are slightly higher up to $\pm 15\%$ $(\pm
8\%)$. Lastly, for $\cos\theta_{\ell \ell}$ scale dependence is of the
order of $\pm 10\%$ $(\pm 5\%)$. Well behaved as they are, these
leptonic observables can be now safely exploited to probe new physics
at the LHC.

%
\subsection{Total missing transverse momentum distribution}
%

%
\begin{figure}[t!]
\begin{center}
  \includegraphics[width=0.50\textwidth]{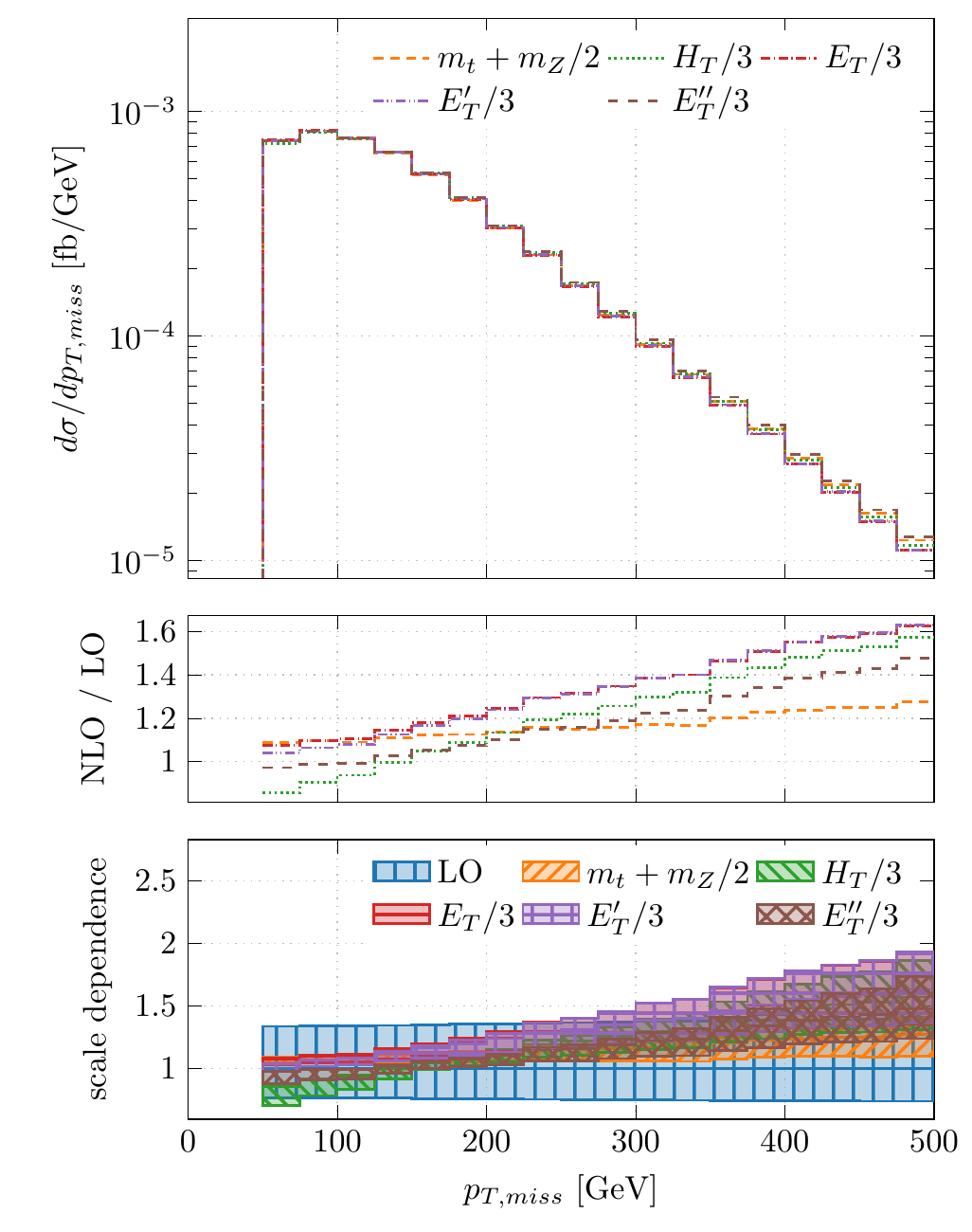}
 \end{center}
\vspace{-0.6cm}
\caption{\it
 The $pp\to e^+ \nu_e \mu^- \bar{\nu}_\mu b\bar{b} \, \nu_\tau
\bar{\nu}_\tau +X$ differential cross section distribution as a
function of $p_T^{miss}$ at the LHC run II with $\sqrt{s}=13$ TeV. The
upper plots show absolute NLO QCD predictions for various values of
$\mu_0$ where $\mu_R=\mu_F=\mu_0$. The middle panels display
differential ${\cal K}$ factors.  The lower panels present
differential ${\cal K}$ factors together with the uncertainty band
from the scale variation for various values of $\mu_0$. Also given is
the relative scale uncertainties of the LO cross section for
$\mu_0=m_t+m_Z/2$.  The LO and the NLO CT14 PDF sets are employed. }
\label{fig:ptmiss-1}
\end{figure}
\begin{figure}[t!]
  \begin{center}
  \includegraphics[width=0.50\textwidth]{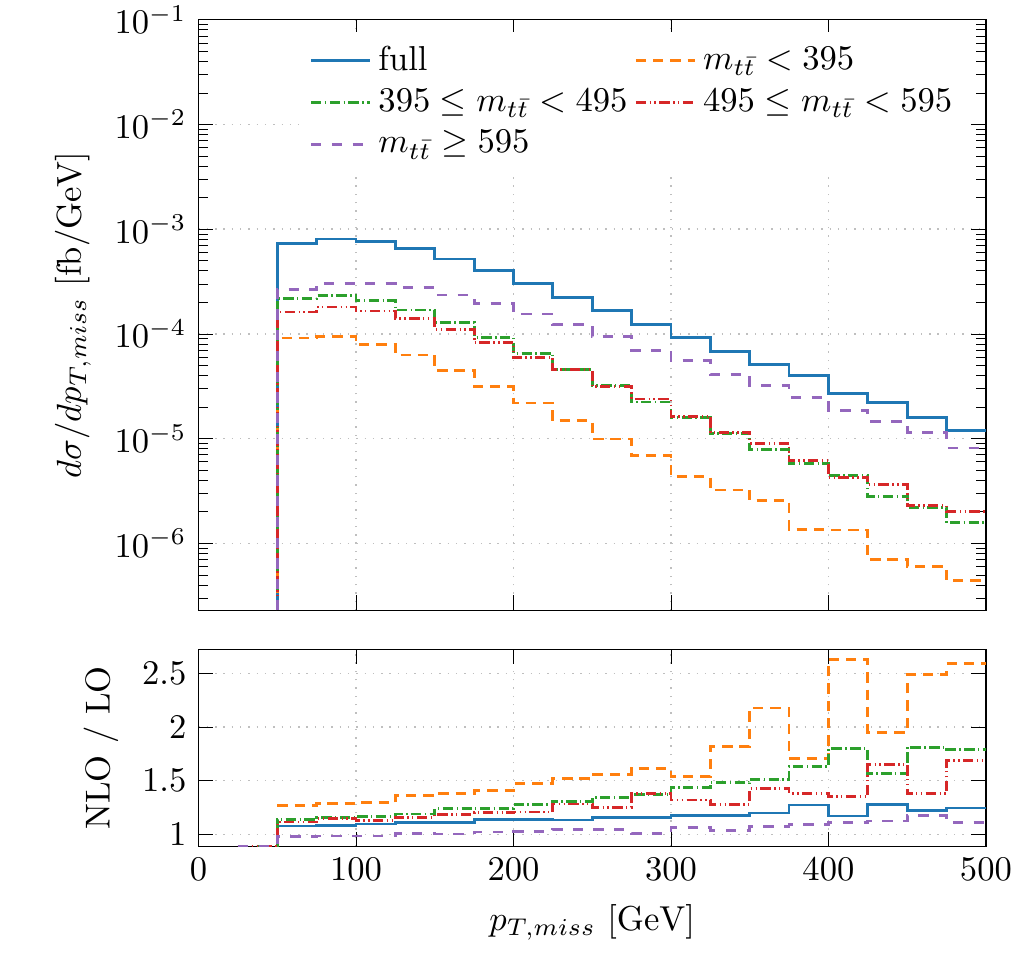}
 \end{center}
\vspace{-0.6cm}
\caption{\it The $pp\to e^+ \nu_e \mu^- \bar{\nu}_\mu b\bar{b} \,
\nu_\tau \bar{\nu}_\tau +X$ double differential cross section
distribution as a function of $p_T^{miss}$ and $m_{t\bar{t}}$ at the
LHC run II with $\sqrt{s}=13$ TeV. The upper plot shows absolute NLO
QCD predictions. The lower panel presents differential ${\cal K}$
factors. The CT14 PDF sets and $\mu_0=m_t+m_Z/2$ are employed.}
\label{fig:ptmiss-2}
\end{figure}
\begin{figure}[t!]
  \begin{center}
  \includegraphics[width=0.49\textwidth]{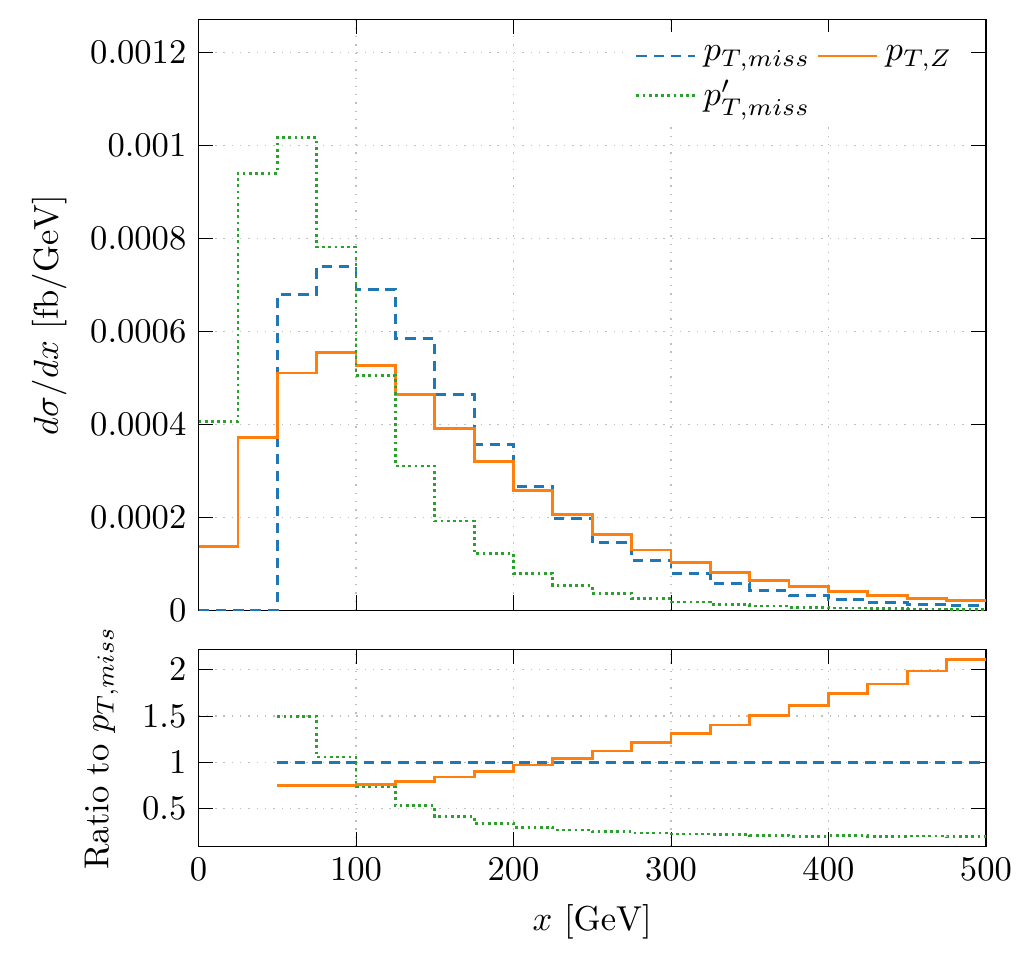}
  \includegraphics[width=0.49\textwidth]{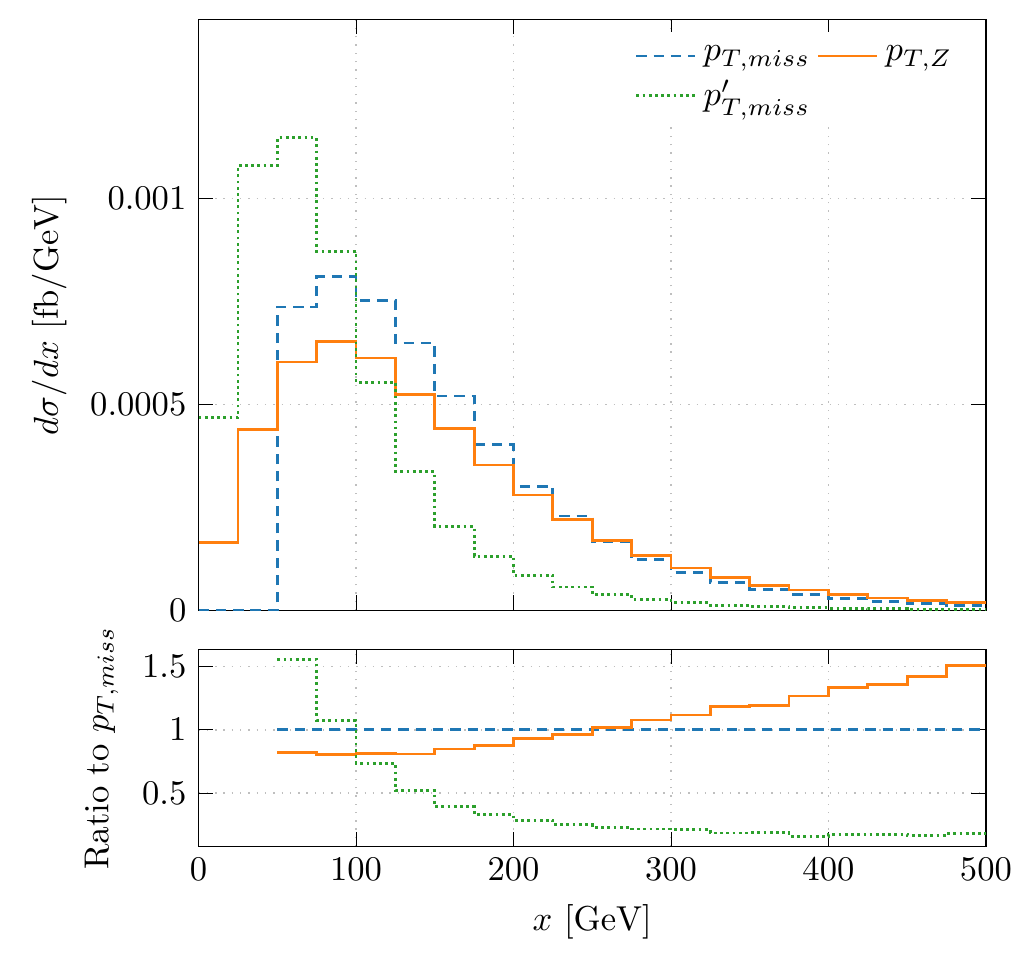}
 \end{center}
\vspace{-0.6cm}
\caption{\it The $pp\to e^+ \nu_e \mu^- \bar{\nu}_\mu b\bar{b} \,
\nu_\tau \bar{\nu}_\tau +X$ LO (left panel) and NLO (right panel)
differential cross section distribution as a function of $p_T^{miss}$,
$p_{T,\,Z}$ and $p_T^{\prime \, miss}$ at the LHC run II with
$\sqrt{s}=13$ TeV. The upper plots show absolute 
QCD predictions. The lower panels present ratios to $p_T^{miss}$. The
CT14 PDF sets and $\mu_0=m_t+m_Z/2$ are employed. }
\label{fig:ptmiss-3}
\end{figure}
\begin{figure}[t!]
  \begin{center}
  \includegraphics[width=0.49\textwidth]{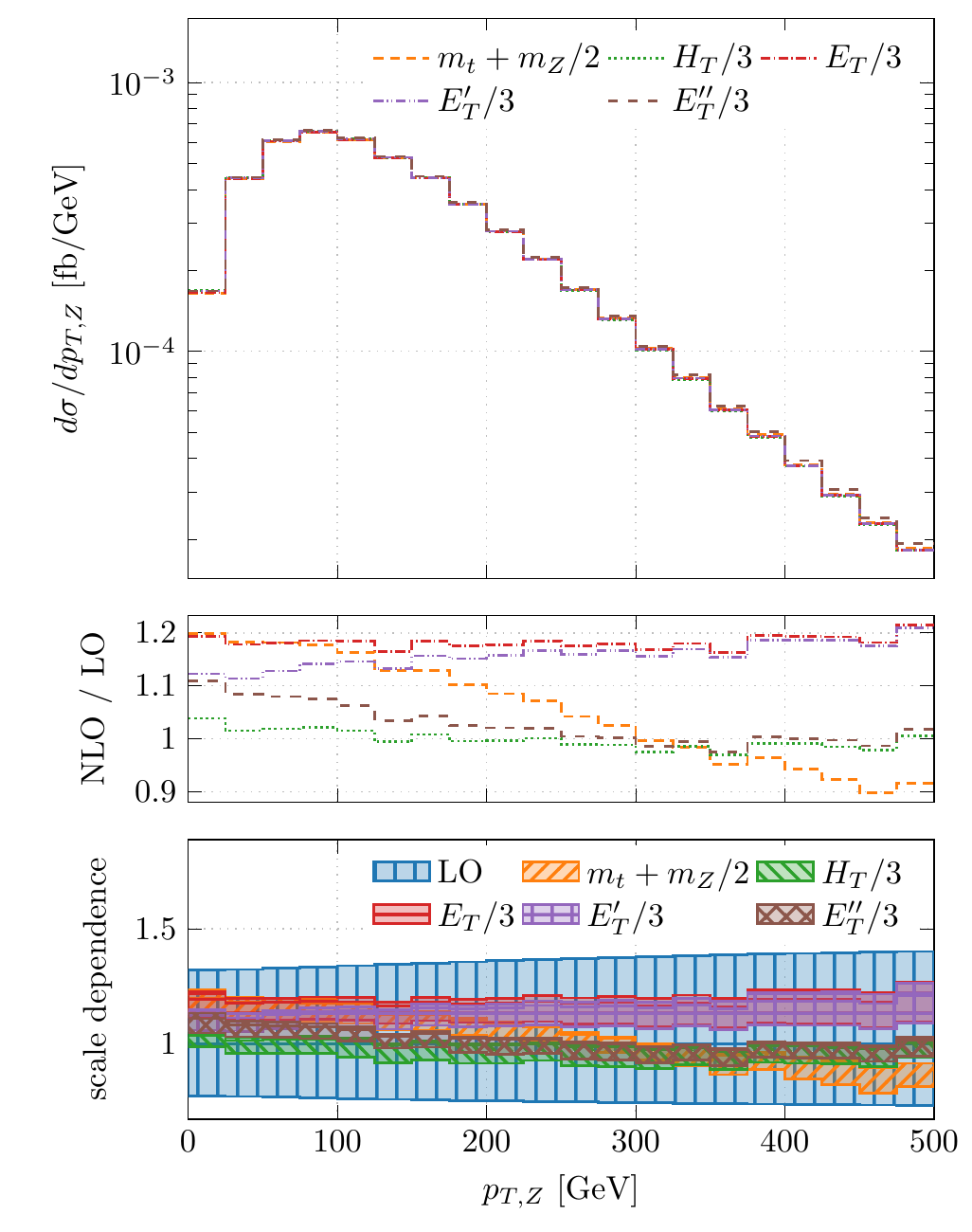}
    \includegraphics[width=0.49\textwidth]{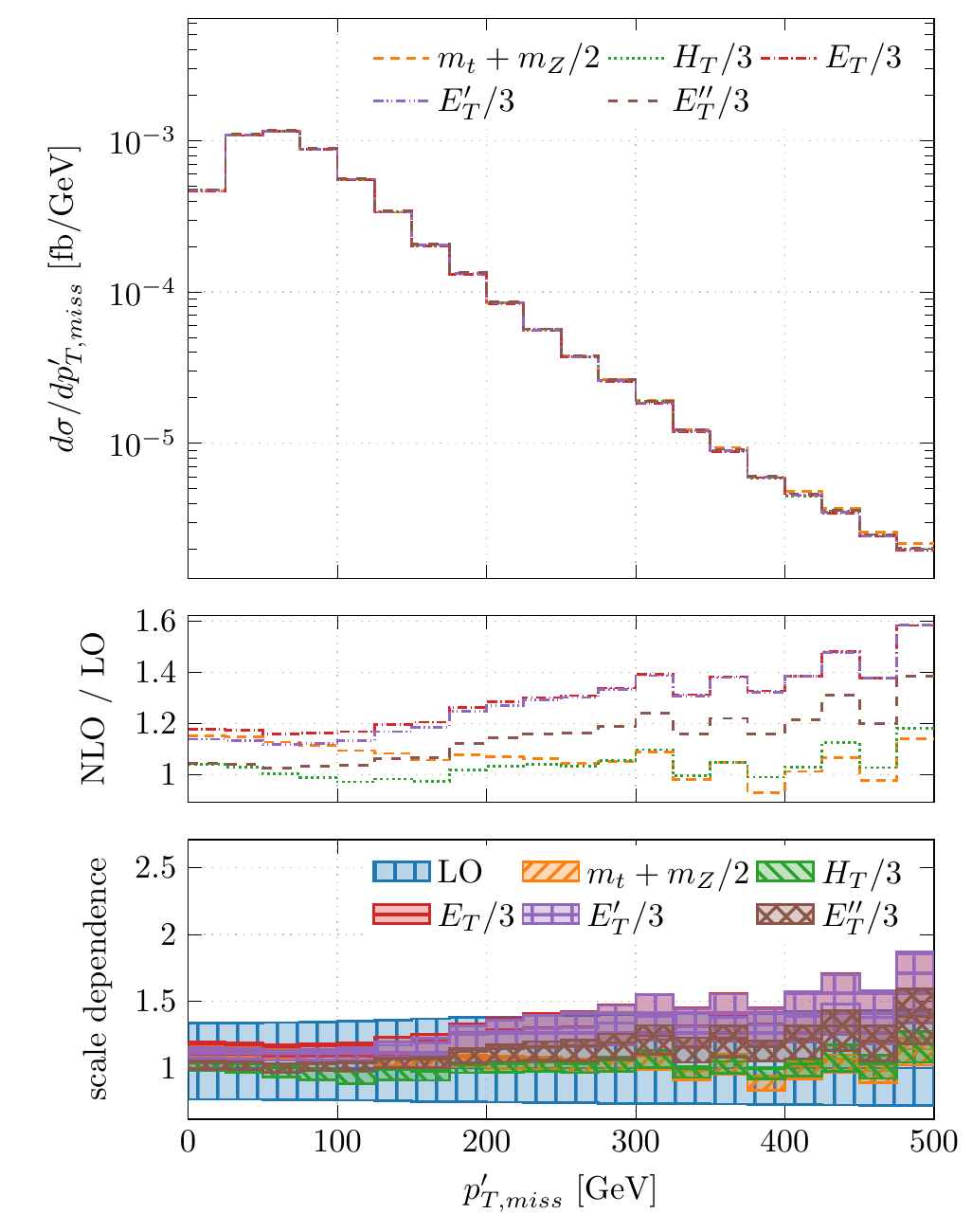}
 \end{center}
\vspace{-0.6cm}
\caption{\it
The $pp\to e^+ \nu_e \mu^- \bar{\nu}_\mu b\bar{b} \, \nu_\tau
\bar{\nu}_\tau +X$ differential cross section distribution as a
function of $p_{T,\,Z}$ and $p_T^{\prime \, miss}$ at the LHC run II
with $\sqrt{s}=13$ TeV. The upper plots show absolute NLO QCD
predictions for various values of $\mu_0$ where
$\mu_R=\mu_F=\mu_0$. The middle panels display differential ${\cal K}$
factors.  The lower panels present differential ${\cal K}$ factors
together with the uncertainty band from the scale variation for
various values of $\mu_0$. Also given is the relative scale
uncertainties of the LO cross section for $\mu_0=m_t+m_Z/2$.  The LO
and the NLO CT14 PDF sets are employed. }
\label{fig:ptmiss-4}
\end{figure}
%
Among all infrared safe observables in $e^+ \nu_e \mu^- \bar{\nu}_\mu
b\bar{b} \, \nu_\tau \bar{\nu}_\tau$ production the total missing
transverse momentum, denoted as $p_T^{miss}$, plays a special
role. Thus, we discuss it separately in this section.  The observation
of an excess in $p_T^{miss}$ represents an important signature in
various BSM and DM models. The signals from new physics need to be
extracted from the SM background, hence an accurate modelling of the
$p_T^{miss}$ observable in the $t\bar{t} + p_T^{miss}$ channel,
particularly in the high-$p_T$ region, is crucial.  Given that neutral
weakly interacting particles (such as neutrinos, dark matter
candidates or the lightest SUSY neutralino) escape from the LHC
detectors, the presence of such particles can only be inferred through
the observation of a momentum imbalance in the final state. In the
process under consideration there are four particles contributing to
the total missing transverse momentum: two neutrinos
$(\nu_e,\bar{\nu}_\mu)$ ascribed to the top quark decays and two
neutrinos $(\nu_\tau,\bar{\nu}_\tau)$ originated from the decay of the
$Z$ gauge boson. Although these particles have different origin and
different kinematical constraints, there is no physical way to
distinguish them at the LHC and one must consider all of them under
the total missing transverse momentum $p_T^{miss}$.

The NLO differential cross section as a function of $p_T^{miss}$ is
shown in Figure \ref{fig:ptmiss-1}. We observe substantial NLO QCD
corrections when our recommended scale choice, based either on
$\mu_0=H_T/3$ or $\mu_0=E_T^{\prime\prime}/3$, is employed. Such
corrections are positive and amount respectively to $57\%$ and $48\%$
at the end of the plotted spectrum, i.e. for $p_T^{miss} \approx 500$
GeV.  We also observe that such corrections induce severe shape
distortions in the $p_T^{miss}$ distribution, of the order of $71\%$
for $\mu_0=H_T/3$ and $51\%$ for
$\mu_0=E_T^{\prime\prime}/3$. Comparing with the fixed scale
predictions based on the choice $\mu_0=m_t+m_Z/2$, one can see that
the latter behaves much better. Specifically, in the region
$p_T^{miss} \approx 500$ GeV the NLO QCD corrections amount to $27\%$
and shape distortions up to $19\%$. This behaviour contrasts with the
behaviour for other observables, where the dynamical scale choice has
been shown to guarantee reduced shape distortions.

In the attempt to understand why the fixed scale choice performs
better for the $p_T^{miss}$ observable we analyse the double
differential NLO cross section distribution expressed as a function of
$p_T^{miss}$ and $m_{t\bar{t}}$, see Figure \ref{fig:ptmiss-2}. One
expects that the fixed scale choice performs well whenever the phase
space regions close to the $t\bar{t}$ threshold $(m_{t\bar{t}}\approx
2m_t)$ provide the dominant contribution to the observable under
consideration. However, Figure \ref{fig:ptmiss-2} shows that this is
not the case for $p_T^{miss}$. Not only the region with $m_{t\bar{t}}\approx
2m_t$ is not enhanced in any special way, but the contributions to
$p_T^{miss}$, which have the largest impact come from regions far away
from the threshold, especially for $p_T^{miss} \in
\left[400-500\right]$ GeV.

Having established that the threshold region is not responsible for
the better performance of the scale choice $\mu_0=m_t+m_Z/2$, we move
to another possible explanation. To this end we investigate two additional 
observables:
\begin{equation}
p_{T,\, Z}=|\, \vec{
  p}_{T,\,\nu_\tau}+\vec{p}_{T,\, \bar{\nu}_\tau}|\,,
\quad \quad \quad \quad \quad \quad \quad 
p_T^{\prime \,
miss}= | \,\vec{p}_{T,\,\nu_e}+\vec{p}_{T,\, \bar{\nu}_\mu}|\,.
\end{equation}
Although not directly measurable at the LHC, these could help us to
understand the different behaviour of $p_T^{miss}$ under fixed and
dynamical scale choices. The first one $p_{T,\, Z}$, corresponds to
the transverse momentum of the $Z$ boson reconstructed from its
invisible decay products $(Z\to \nu_\tau \bar{\nu}_\tau)$. The second
one $p_T^{\prime \,miss}$ represents the missing transverse momentum
restricted to the invisible particles coming from the decays of the
top quarks only ($t \to e^+ \nu_e b$ and $\bar{t} \to \mu^-
\bar{\nu}_\mu \bar{b} $). Given the different origin of the neutrinos
involved in their definition, it is not surprising that these two
observables exhibit different kinematics. Moreover,  one should not expect
that they are affected by higher order corrections in a similar
way. We also note that the total missing transverse momentum,
$p_T^{miss}$, is not given as a simple sum of $p_{T,\, Z}$ and
$p_T^{\prime \,miss}$ but rather as a convolution of some type. In
Figure \ref{fig:ptmiss-3} we present a comparison of LO and NLO
differential cross section as a function of $p_T^{miss}$, $p_{T, \,Z}$
and $p_T^{\prime \, miss}$ based on the fixed scale choice
$\mu_0=m_t+m_Z/2$, with the goal of outlining possible differences in
kinematics. We are interested in regions above the $p_T^{miss}$ cut of
$50$ GeV even though both $p_{T, \,Z}$ and $p_T^{\prime \, miss}$ can
have lower values. One can observe that the $p_T^{miss}$ spectrum is
harder than $p_T^{\prime \, miss}$, but softer than $p_{T,
  \,Z}$. Additionally, shape differences between $p_T^{miss}$ and $p_{T, \,Z}$
are quite substantial. Figure \ref{fig:ptmiss-4}  shows that $p_{T,
  \,Z}$ and  $p_T^{\prime \, miss}$ are affected by NLO QCD
corrections in a different way. In the case of $p_{T, \,Z}$ the
dynamical scale choice $\mu_0=H_T/3$ and $\mu_0=E^{\prime\prime}_T/3$
result in negative and rather small corrections of the order of
$1\%-2\%$ in the tail of the distribution. On the other hand, for the
fixed scale choice we observe negative corrections of $8\%$ in the
same region. Dynamical scales feature also reduced shape distortions
in comparison with the fixed scale ($3\%$ for $\mu_0=H_T/3$, $9\%$
for $\mu_0=E^{\prime\prime}/3$ and $28\%$ for $\mu_0=m_t+m_Z/2$). A
different pattern reveals itself when we turn to the case of
$p_T^{\prime\, miss}$. In this case the dynamical scales result  in
sizeable NLO QCD corrections in the tail of the distribution ($18\%$  for
$\mu_0=H_T/3$ and $38\%$ for $\mu_0=E^{\prime\prime}_T/3$) to be
compared with the more satisfactory performance of the fixed scale
($14\%$). Moreover, the fixed scale choice provides negligible shape
distortions in the tail of the order of $1\%$. Thus, like for
$p_T^{miss}$  also for $p_T^{\prime \, miss}$ the scale
$\mu_0=m_t+m_Z/2$ seems to perform better. 

To conclude, our differential analysis of  $p_T^{miss}$, $p_T^{\prime
  \, miss}$ and $p_{T, \,Z}$ reveals that the first two observables
have spectra which are much softer than the one of $p_{T, \,Z}$. For
the latter, as well as for other dimensionful observables  that we
have studied in this paper, the prescription of using a dynamical
scale seems the most adequate to describe the  high $p_T$
tails. Instead, for $p_T^{miss}$ as well as for $p_T^{\prime \, miss}$
such prescription results in too large scales. In this case a fixed
scale choice is simply more adequate. 

 To close this part, we report on the size of the theoretical
error as derived from the scale dependence study. In the case of
$p_{T}^{miss}$ for $\mu_0=m_t+m_Z/2$ the theoretical uncertainties are
up to $\pm 14\% \, (\pm 12\%)$. They are slightly increased for our
chosen dynamical scales up to $\pm 18\% \, (\pm 17\%)$. As usual the
values in parenthesis correspond to the theoretical errors after the
symmetrisation of errors is performed. This is of course a significant
reduction in the theoretical error considering that at the LO 
one can obtain errors up to $\pm 43\% \, (\pm 36\%)$. Even though
$p_{T}^{\prime \, miss}$ and $p_{T,\, Z}$ can not be separately
measured at the LHC we give theoretical errors for them as well for
completeness. Specifically,  for the former observable we have estimated
errors up to $\pm 11\%\, (\pm 9 \%)$ and for the latter we have
received $\pm 11\% \, (\pm 5\%)$ with $\mu_R=\mu_F=\mu_0$ set to
$\mu_0= m_t+m_Z/2$. The dynamical scale choices have left us
with a theoretical error of the order of $\pm 15\% \, (\pm 15 \%)$
and $\pm 7\% \, (\pm 4\%)$ respectively for $p_{T}^{\prime \, miss}$
and $p_{T,\, Z}$. Once more, there is a significant improvement when
comparing to the LO results where such errors have been estimated to
be up to $\pm 45\%\, (\pm 37\%)$ and $\pm 42 \%\, (\pm 34 \%)$
respectively for $p_{T}^{\prime \, miss}$ and $p_{T,\, Z}$.

%
\subsection{PDF uncertainties}
%
%
\begin{figure}[th!]
\begin{center}
  \includegraphics[width=0.49\textwidth]{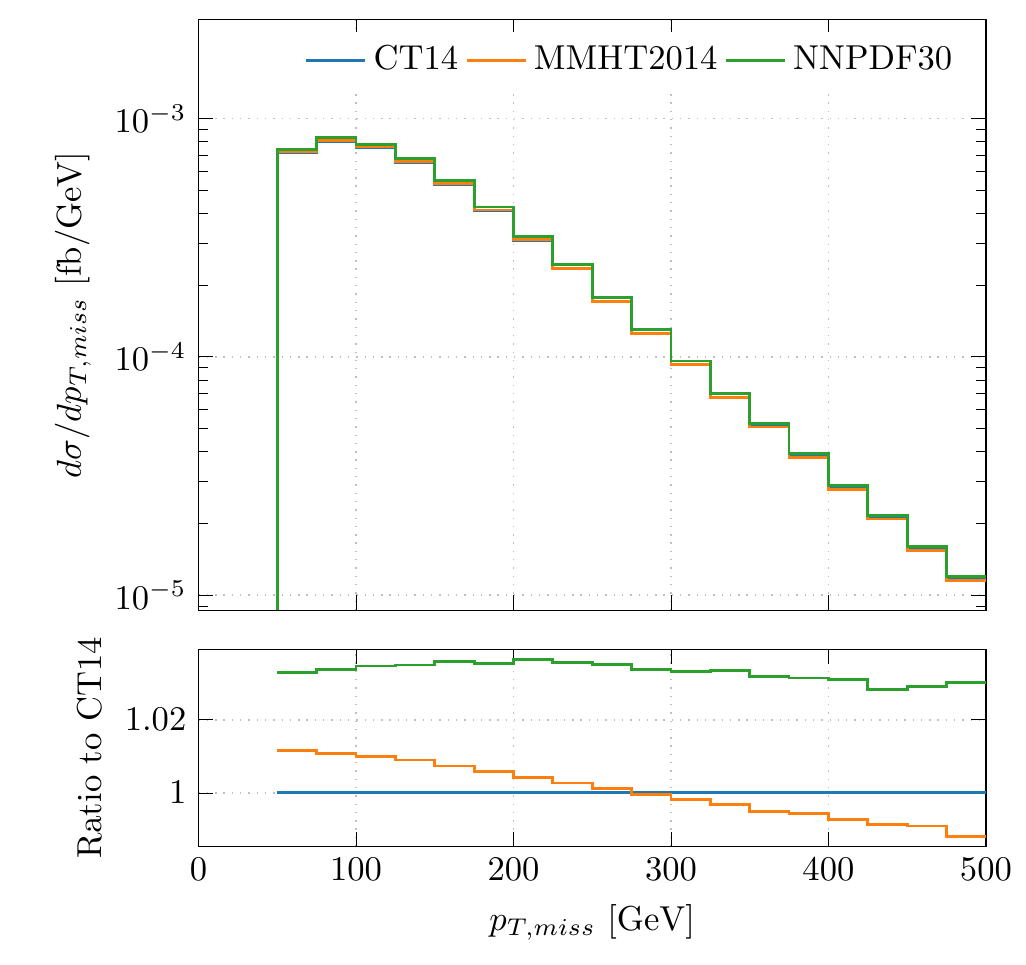}
  \includegraphics[width=0.49\textwidth]{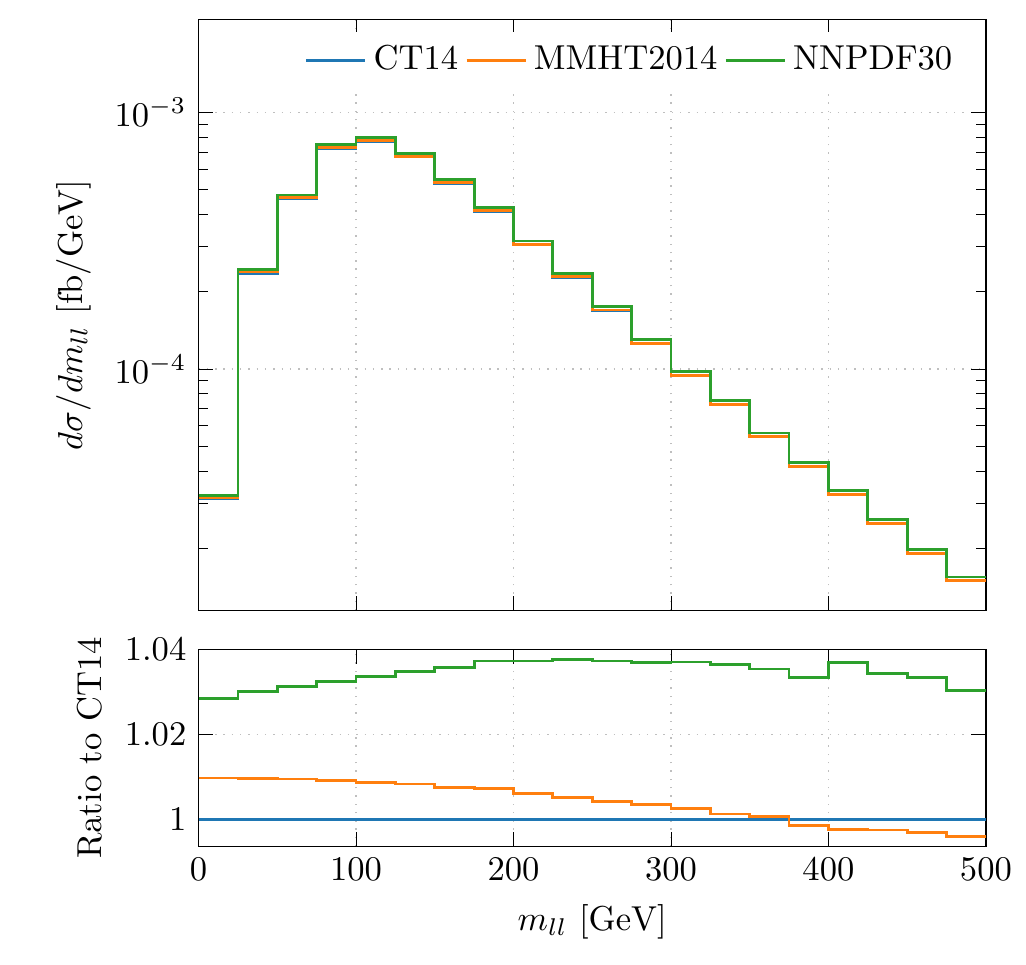}
  \includegraphics[width=0.49\textwidth]{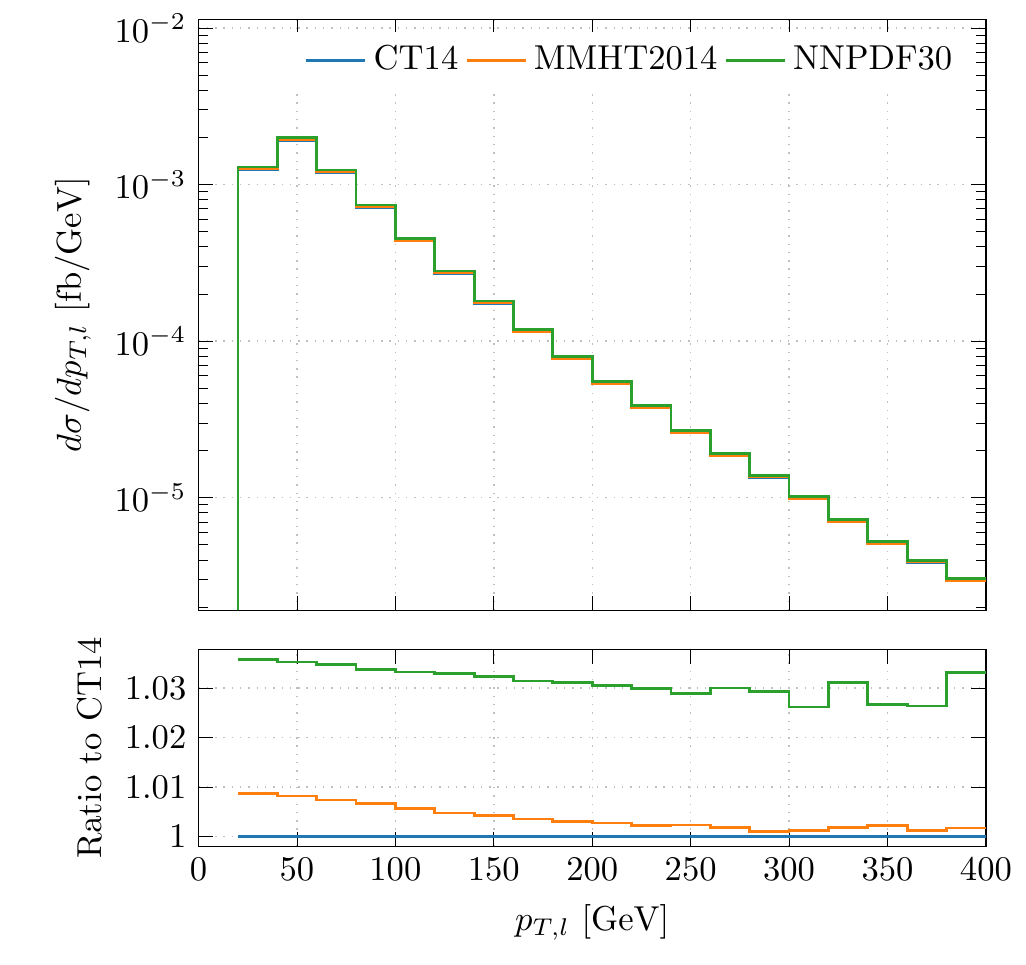}
     \includegraphics[width=0.49\textwidth]{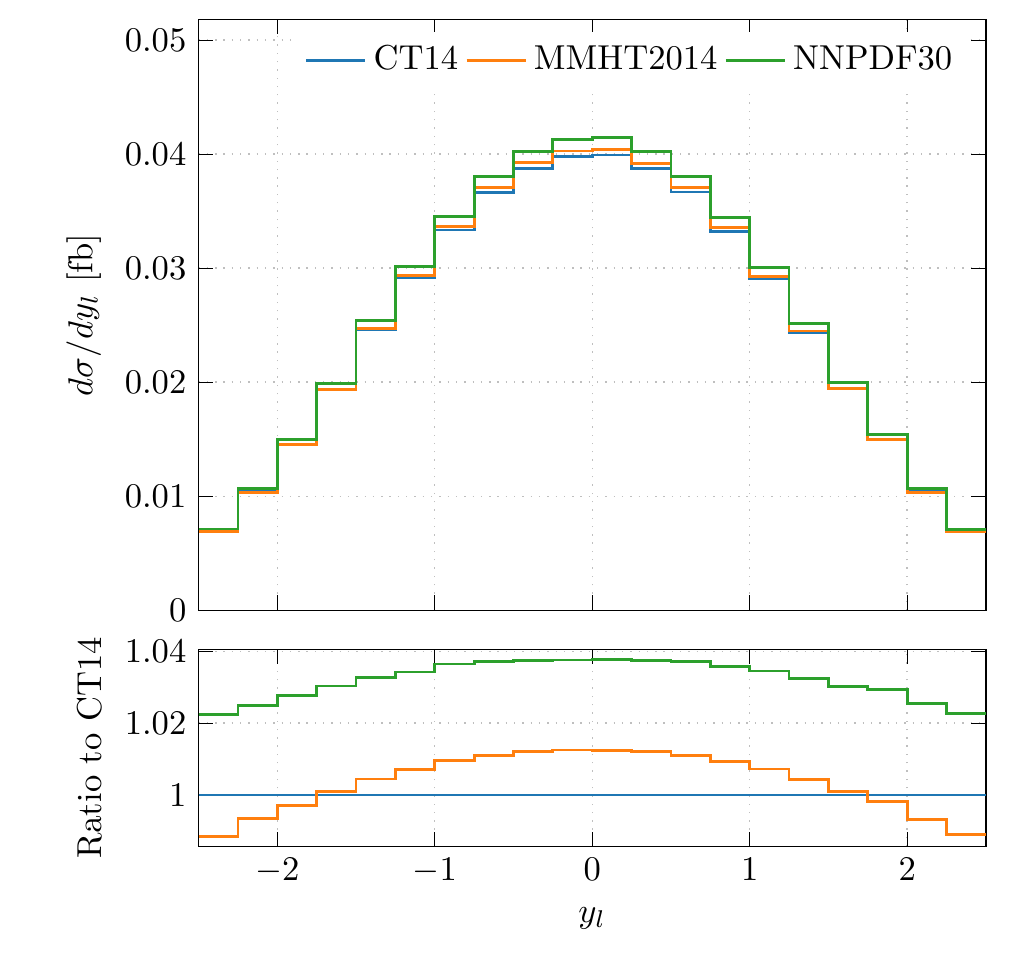}
  \includegraphics[width=0.49\textwidth]{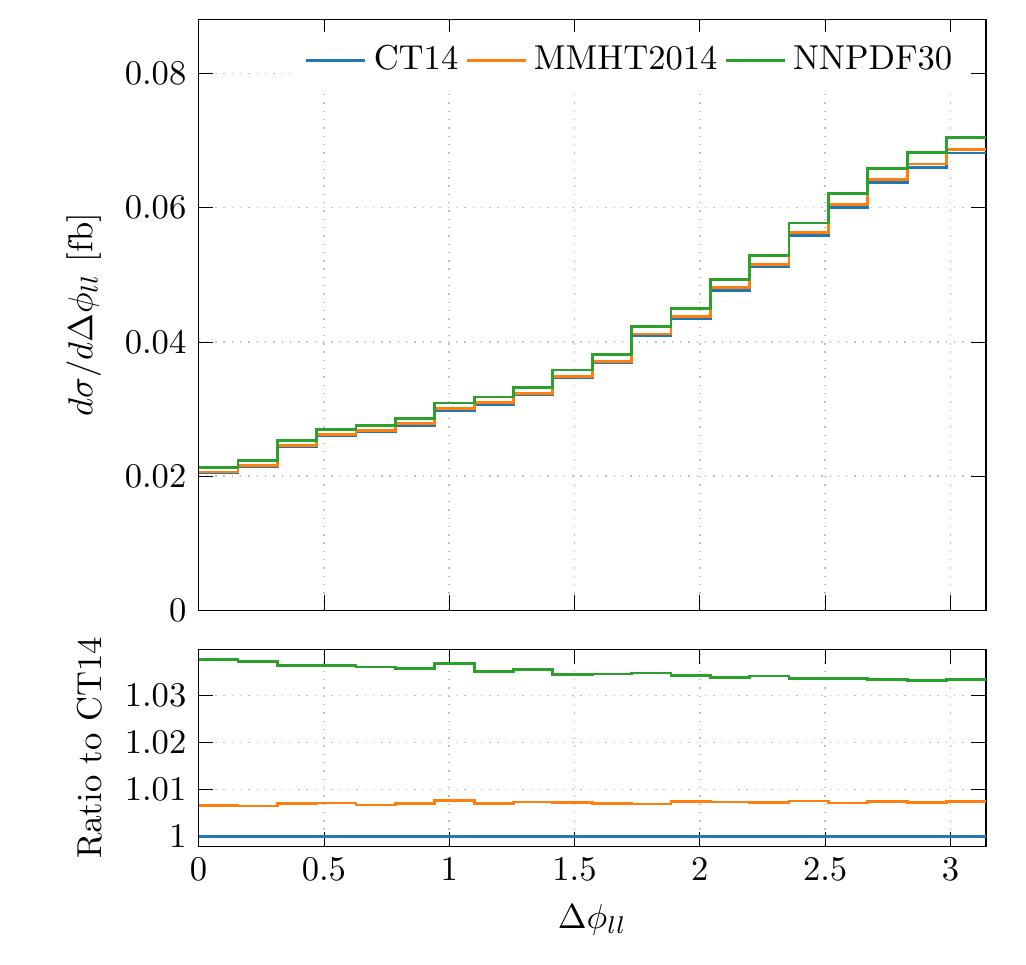}
  \includegraphics[width=0.49\textwidth]{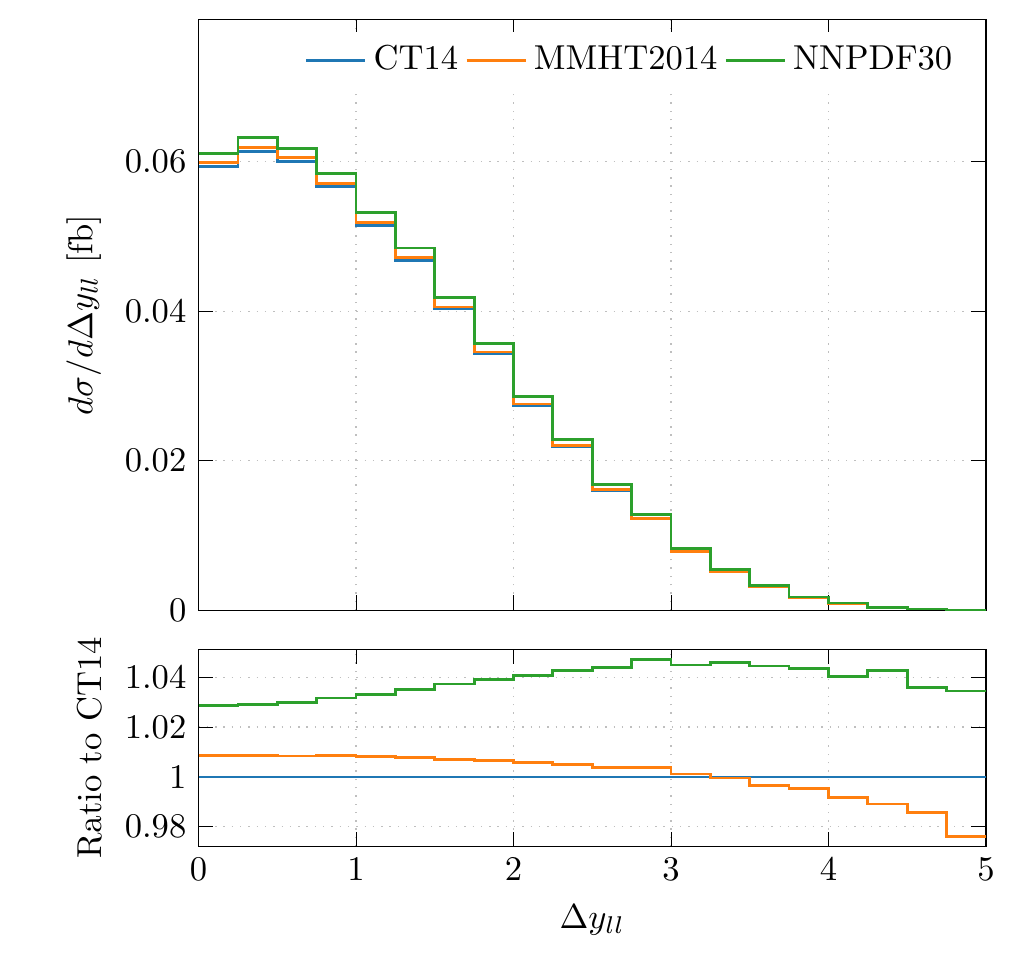}
 \end{center}
\vspace{-0.6cm}
\caption{\it
Differential cross section distributions for  $pp\to e^+ \nu_e
\mu^- \bar{\nu}_\mu b\bar{b} \, \nu_\tau \bar{\nu}_\tau +X$ at
the LHC run II with $\sqrt{s}=13$ TeV. The upper plots show absolute
NLO QCD predictions for three different PDF sets and for
$\mu_0=H_T/3$. Lower panels display the ratio to the default CT14
set. The following observables are presented: $p_T^{miss}$, $m_{\ell
\ell}$, (averaged) $p_{T, \, \ell}$, (averaged)  $y_\ell$ as well as $\Delta
\phi_{\ell\ell}$ and $\Delta y_{\ell\ell}$.  }
\label{fig:pds}
\end{figure}
\begin{table}[t!]
\begin{center}
\begin{tabular}{||c|c|c|c||}
  \hline \hline
 \textsc{Scale Choice}&
  $\sigma^{\rm NLO}_{\rm CT14}$ [fb]
  & $\sigma^{\rm NLO}_{\rm MMHT2014}$  [fb] &  $\sigma^{\rm NLO}_{\rm
                                              NNPDF3.0}$
                                              [fb]\\[0.2cm]
  \hline \hline
  $\mu_0=m_t+m_Z/2$              & $0.1266^{+0.0014 (1.1\%)}_{-0.0075 (5.9\%)}$
  & $0.1275^{+0.0014 (1.1\%)}_{-0.0076 (5.9\%)}$
                                            &
  $0.1309^{+0.0014 (1.1\%)}_{-0.0079 (6.0\%)}$                                            
  \\[0.2cm]
  \hline
  $\mu_0=H_T/3$              &
$0.1270^{+0.0009 (0.7\%)}_{- 0.0086 (6.8\%)}$
  &
  $0.1278^{+ 0.0009 (0.7\%)}_{-0.0089 (7.0\%)}$  
                                            &
$0.1312^{+0.0009 (0.7\%)}_{- 0.0090 (6.9\%)}$
  \\[0.2cm]
  \hline 
  $\mu_0=E_T/3$              &
$0.1272^{+ 0.0020 (1.6\%)}_{- 0.0086 (6.8\%)}$
  &
$0.1279^{+ 0.0020 (1.6\%)}_{- 0.0086 (6.8\%)}$
                                            &
$0.1313^{+ 0.0021 (1.6\%)}_{- 0.0090 (6.9\%)}$
  \\[0.2cm]
  \hline 
  $\mu_0=E^\prime_T/3$ &
$0.1268^{+ 0.0019 (1.5\%)}_{- 0.0081 (6.4\%)}$
  &
 $0.1280^{+ 0.0019 (1.5\%)}_{- 0.0082 (6.4\%)}$   
                                            &
$0.1315^{+ 0.0020 (1.5\%)}_{- 0.0086 (6.5\%)}$
  \\[0.2cm]
  \hline 
  $\mu_0=E^{\prime \prime}_T/3$ &
$0.1286^{+ 0.0013 (1.0\%)}_{- 0.0060 (4.7\%)}$
  &$0.1295^{+ 0.0013 (1.0\%)}_{- 0.0060 (4.7\%)}$
                                            &
$0.1330^{+ 0.0013 (1.0\%)}_{- 0.0063 (4.8\%)}$
  \\[0.2cm] 
\hline \hline 
 \end{tabular}
\end{center}
\caption{\label{tab:pdfs} \it
  NLO cross sections for the $pp\to e^+ \nu_e \mu^- \bar{\nu}_\mu
b\bar{b} \, \nu_\tau \bar{\nu}_\tau +X$ process at the LHC run II with
$\sqrt{s}=13$ TeV.  Results for three different PDF sets  are
presented. Also included are theoretical errors as obtained from the
scale variation.}
\end{table}

In this section, we complete our analysis of theoretical
uncertainties. The theoretical uncertainty as obtained from the scale
dependence of the cross section is not the only source of 
theoretical systematic uncertainties. Another source of uncertainties
comes from the parameterisation of PDFs. In the following we take the
PDF uncertainties to be the difference between our default PDF set
(CT14) and the other two PDF sets considered (MMHT14 and NNPDF3.0). In
this way we account for different theoretical assumptions that enter
into the parameterisation of the PDFs which are difficult to quantify
within the CT14 scheme. Moreover, the differences coming from NLO
results for various PDF sets are comparable (usually even higher) to
the individual estimates of PDF systematics. We have checked that this
is the case for similar processes, namely for $pp \to e^+ \nu_e \mu^-
\bar{\nu}_\mu b\bar{b} j+X$
\cite{Bevilacqua:2015qha,Bevilacqua:2016jfk} and $pp \to e^+ \nu_e
\mu^- \bar{\nu}_\mu b\bar{b}\gamma +X$ production
\cite{Bevilacqua:2018woc,Bevilacqua:2018dny}. Results with the
recomputed NLO QCD corrections to the $pp \to e^+ \nu_e \mu^-
\bar{\nu}_\tau b\bar{b} \nu_\tau \bar{\nu}_\tau +X$ production process
for MMHT14 and NNPDF3.0 PDF sets are summarised in Table
\ref{tab:pdfs}. Taken in a very conservative way, the PDF
uncertainties for the process under scrutiny and with
$\mu_R=\mu_F=\mu_0=m_t+m_Z/2$ are of the order of $0.0043$ fb $(3.4
\%)$. After symmetrisation they are reduced down to $ 0.0026$ fb
$(2.0\%)$. Our result for the integrated cross section at NLO in QCD
with the CT14 PDF set and for $\mu_0 = m_t+m_Z/2$ is, thus, given by
\begin{equation}
\sigma^{\rm NLO}_{pp\to e^+\nu_e\mu^-\bar{\nu}_\mu b\bar{b} \nu_\tau
  \bar{\nu}_\tau \,} \left(\mu_0=m_t+m_Z/2\right)
=  0.1266^{+0.0014 (1.1\%)}_{-0.0075 (5.9\%)} \, {\rm  [scales]}\,
{}^{+ 0.0009\, (0.7\%)}_{+0.0043\, (3.4\%)} \,{\rm [PDF]} \,  {\rm fb}\,.
\end{equation}
The PDF uncertainties are almost a factor of $2$ smaller than the
theoretical uncertainties due to the scale dependence. The latter
remain the dominant source of the theoretical systematics. The same
pattern is obtained for the other scale choices. For example for
$\mu_0=H_T/3$ and $\mu_0= E_T^{\prime\prime}/3$ we have
\begin{equation}
  \begin{split}
\sigma^{\rm NLO}_{pp\to e^+\nu_e\mu^-\bar{\nu}_\mu b\bar{b} \nu_\tau
  \bar{\nu}_\tau \,} \left(\mu_0=H_T/3\right)
&=  {0.1270}^{+0.0009 (0.7\%)}_{- 0.0086 (6.8\%)} \, {\rm  [scales]}\,
{}^{+ 0.0008\, (0.6\%)}_{+0.0042\, (3.3\%)} \,{\rm [PDF]} \,  {\rm
  fb}\,, \\[0.2cm]
\sigma^{\rm NLO}_{pp\to e^+\nu_e\mu^-\bar{\nu}_\mu b\bar{b} \nu_\tau
  \bar{\nu}_\tau \,} \left(\mu_0=E_T^{\prime\prime}/3\right)
&=  0.1286^{+ 0.0013 (1.0\%)}_{- 0.0060 (4.7\%)} \, {\rm  [scales]}\,
{}^{+0.0009 \, (0.7\%)}_{+0.0044\, (3.4\%)} \,{\rm [PDF]} \,  {\rm
  fb}\,.
\end{split}
\end{equation}
Lastly, we have examined PDF uncertainties for various
differential cross sections. In Figure \ref{fig:pds} NLO differential
distributions as a function of $p_T^{miss}$, $m_{\ell \ell}$, the
averaged $p_{T, \, \ell}$, the averaged $y_\ell$ as well as $\Delta
\phi_{\ell\ell}$ and $\Delta y_{\ell\ell}$ are shown as examples. The
upper panels present the NLO predictions for three different PDF sets
at the central scale value $\mu_R=\mu_F=\mu_0=H_T/3$.  The lower
panels of Figure \ref{fig:pds} give the ratio of the MMHT14 (NNPDF3.0)
PDF set to CT14. As we can observe for all observables shown the PDF
uncertainties are at most at the level of $4\%$, thus, well below the
uncertainties from the scale dependence. This result remains unchanged
regardless of whether the observable was dimensionful or not and
independent of the scale we have utilised.

To summarise this part, for $pp\to e^+\nu_e \mu^- \bar{\nu}_\mu
b\bar{b} \nu_\tau \bar{\nu}_\tau$ production at the LHC Run II with
$\sqrt{s}= 13$ TeV with the rather inclusive selection cuts that we
have employed and input parameters used, the size of PDF
uncertainties, both at the level of total and differential cross
sections, is substantially smaller than the size of theoretical errors
from the scale dependence. The latter remain the dominant component of
the final theoretical error for our predictions at NLO in QCD.

%
\section{Comparison to top anti-top pair production}
\label{sec:4}
%

From the experimental point of view both the $pp\to e^+ \nu_e \mu^-
\bar{\nu}_\mu b\bar{b}$ and $pp\to e^+ \nu_e \mu^- \bar{\nu}_\mu
b\bar{b} \, \nu_\tau \bar{\nu}_\tau$ production processes comprise the
same final states, two charged leptons, two bottom flavoured jets and
missing transverse momentum from undetected neutrinos. In the
following we would like to compare these two processes in order to see
the impact of the enlarged missing transverse momentum on the
kinematics of these final state. We start, however, by presenting the
results for the integrated cross sections for the $pp\to e^+ \nu_e
\mu^- \bar{\nu}_\mu b\bar{b}$ production process.  With our inclusive
cut selection LO predictions for two different scale choices
$\mu_0=m_t/2$ and $\mu_0=H_T/4$ as well as for the CT14 PDF set are
\begin{equation}
  \begin{split}
\sigma^{\rm LO}_{pp\to e^+\nu_e\mu^-\bar{\nu}_\mu b\bar{b}\, } ({\rm
  CT14}, \mu_0=m_t/2)
&= 1126^{+ 379\, (34\%)}_{-265\, (23\%)} \, {\rm fb}\,, \\[0.2cm]
\sigma^{\rm LO}_{pp\to e^+\nu_e\mu^-\bar{\nu}_\mu b\bar{b} \,} ({\rm
  CT14}, \mu_0=H_T/4)
&=  1067^{+348 \, (33\%)}_{-247\, (23\%)} \, {\rm fb}\,.
\end{split}
\end{equation}
At the NLO level in QCD they are given by 
\begin{equation}
  \begin{split}
\sigma^{\rm NLO}_{pp\to e^+\nu_e\mu^-\bar{\nu}_\mu b\bar{b}\, } ({\rm
  CT14}, \mu_0=m_t/2)
&= 1107^{+ 16\, (1.4\%)}_{-88\, (7.9\%)} \, {\rm fb}\,, \\[0.2cm]
\sigma^{\rm NLO}_{pp\to e^+\nu_e\mu^-\bar{\nu}_\mu b\bar{b} \,} ({\rm
  CT14}, \mu_0=H_T/4)
&=  1103^{+19 \, (1.7\%)}_{-58\, (5.2\%)} \, {\rm fb}\,.
\end{split}
\end{equation}  
Since also in this case we generate decays of the weak bosons to
different lepton generations only the complete $\ell^{\pm} \ell^{\mp}$
cross section (with $\ell_{1,2}=e,\mu$) can be obtained by multiplying
the above result with a lepton-flavour factor of $4$.
%
\begin{figure}[t!]
  \begin{center}
    \includegraphics[width=0.49\textwidth]{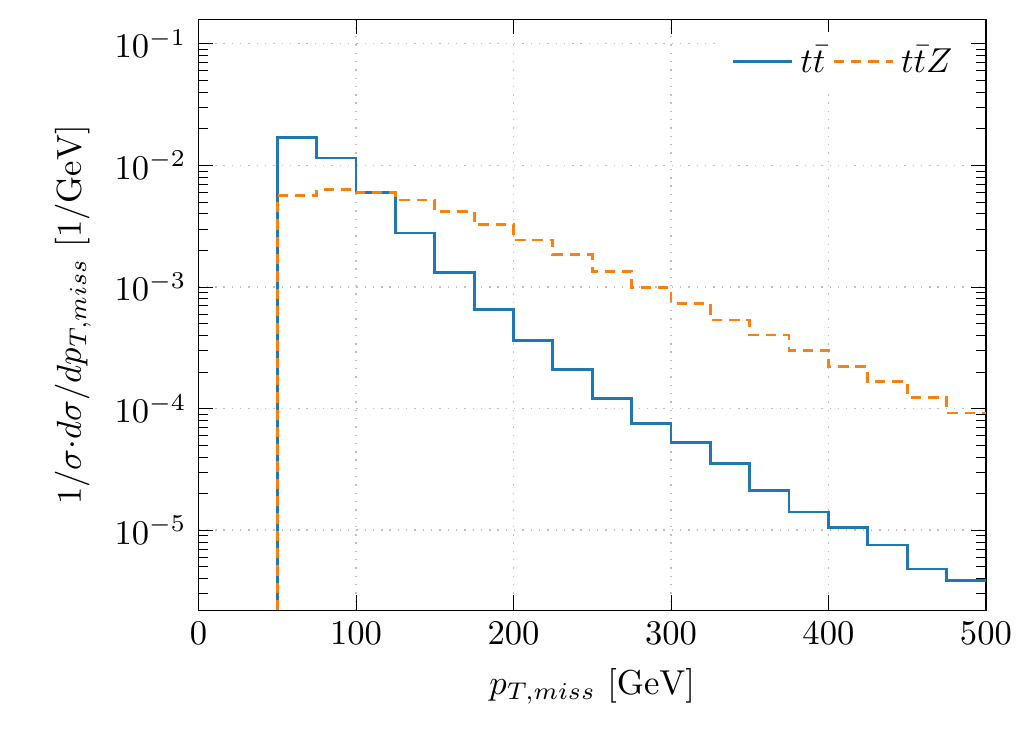}
    \includegraphics[width=0.49\textwidth]{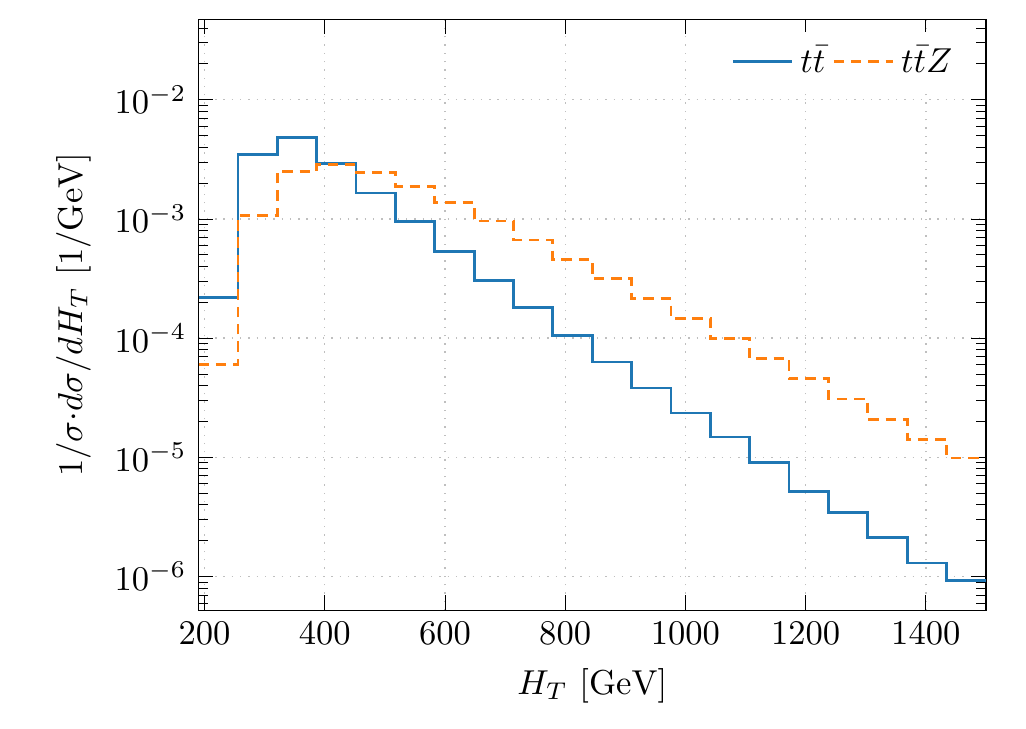}
      \includegraphics[width=0.49\textwidth]{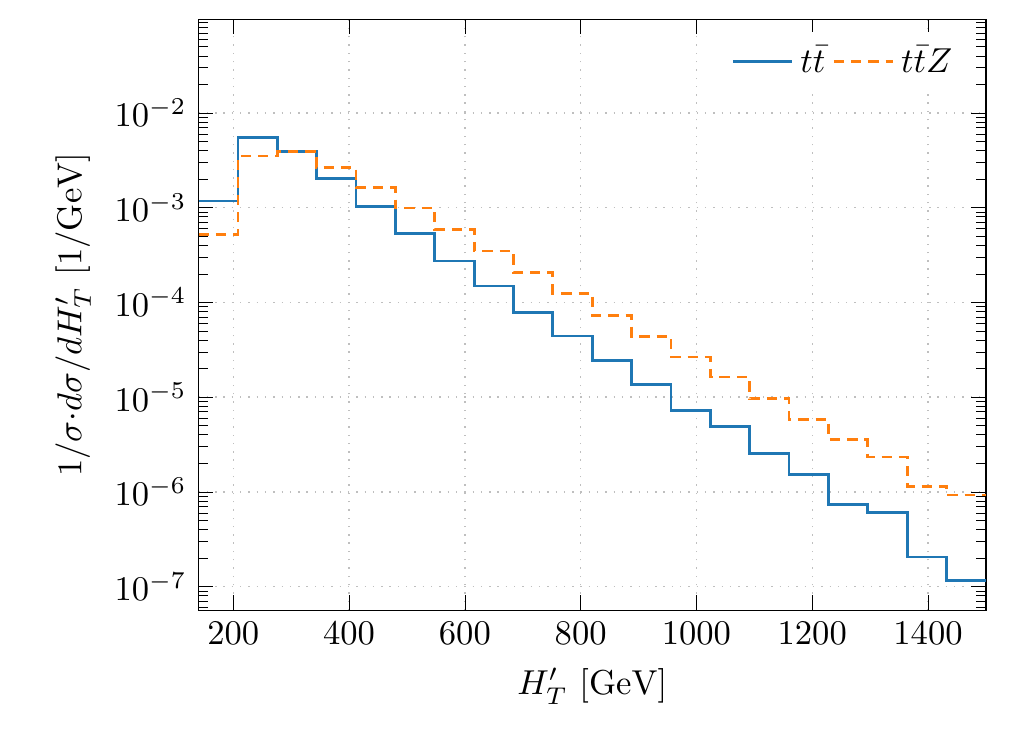}
      \includegraphics[width=0.49\textwidth]{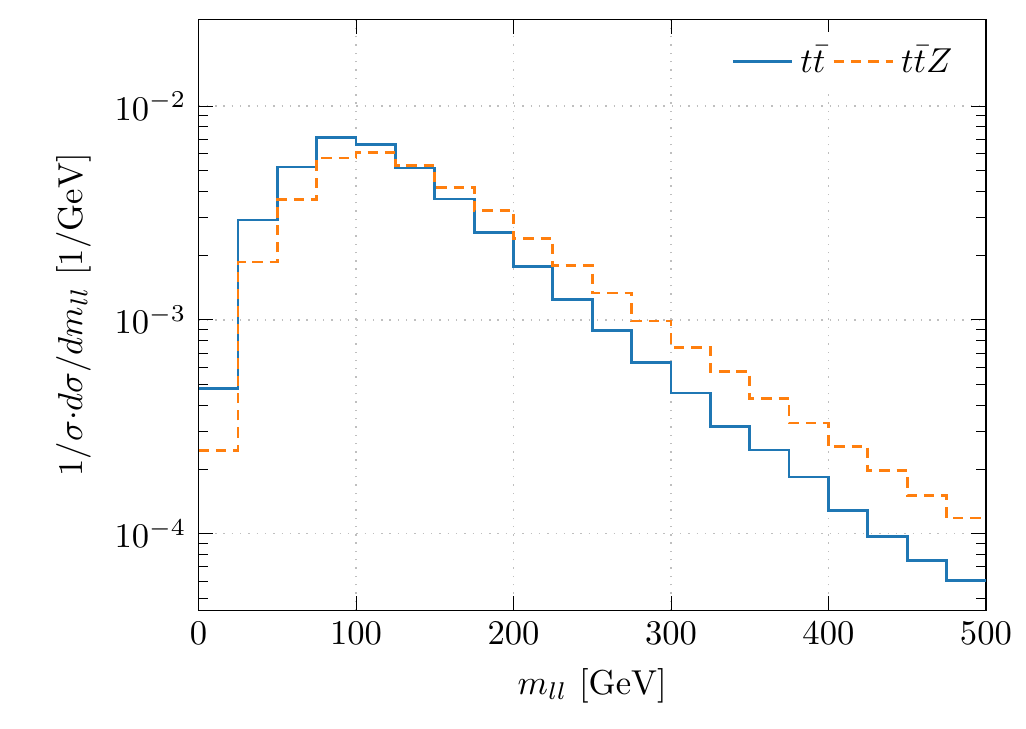}
  \includegraphics[width=0.49\textwidth]{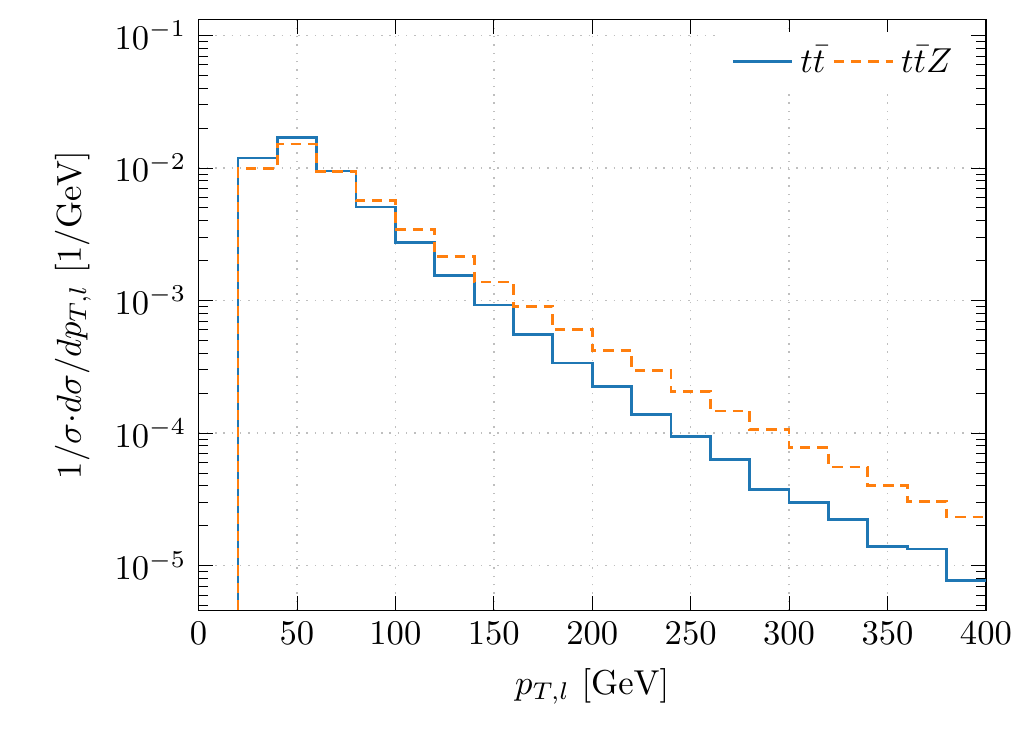}
        \includegraphics[width=0.49\textwidth]{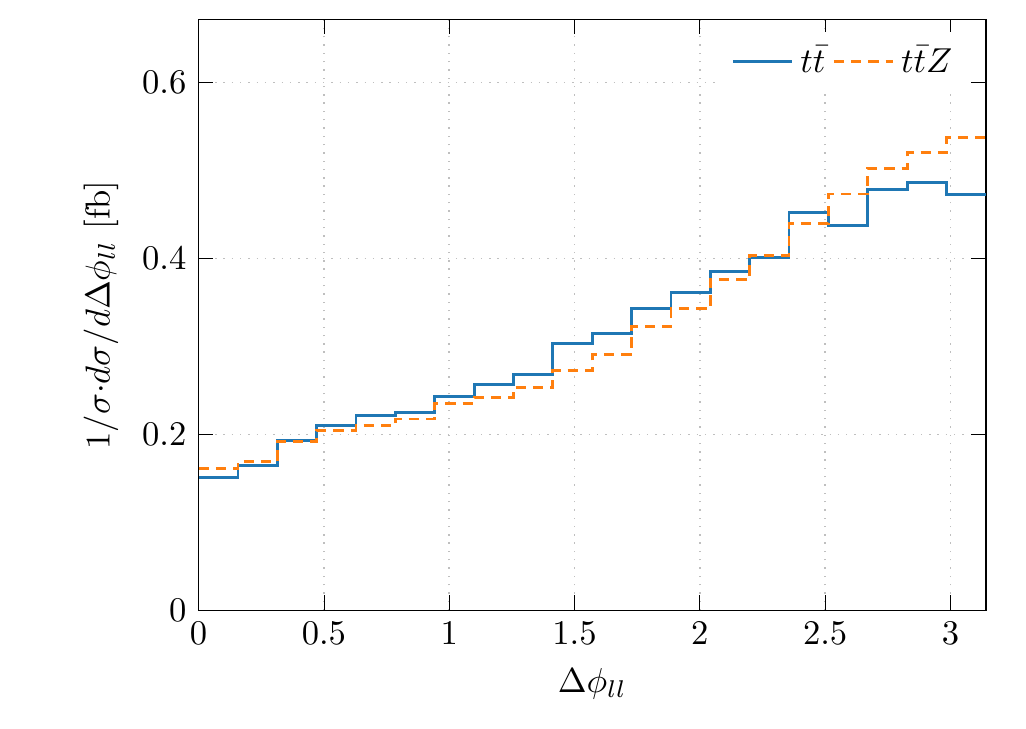}
  \includegraphics[width=0.49\textwidth]{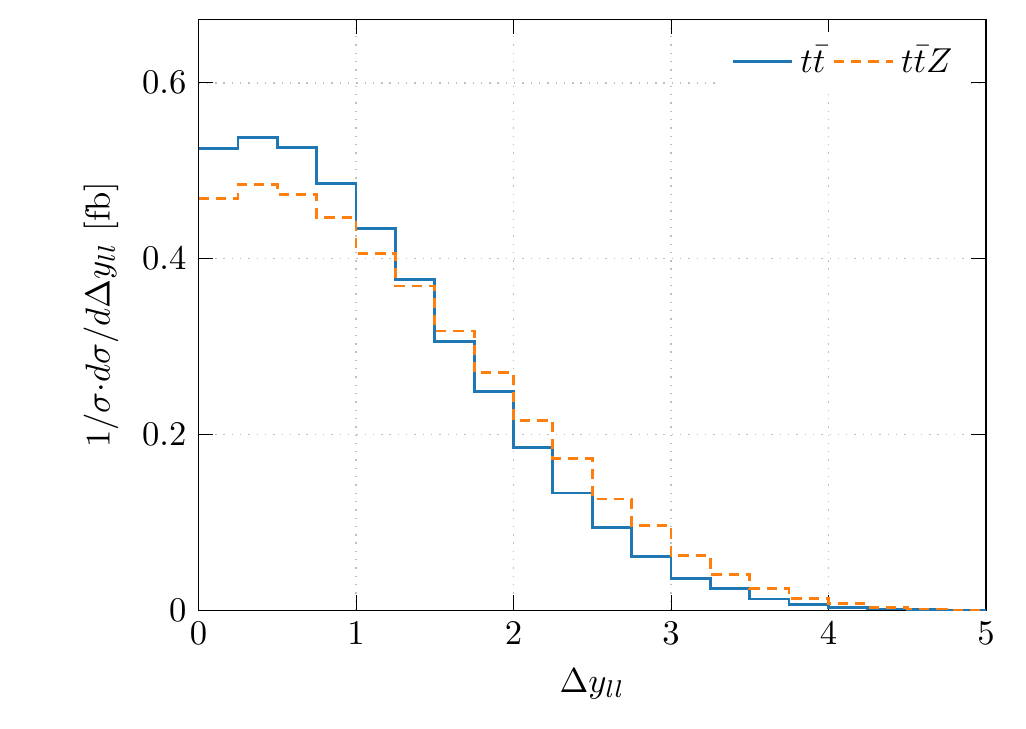}
  \includegraphics[width=0.49\textwidth]{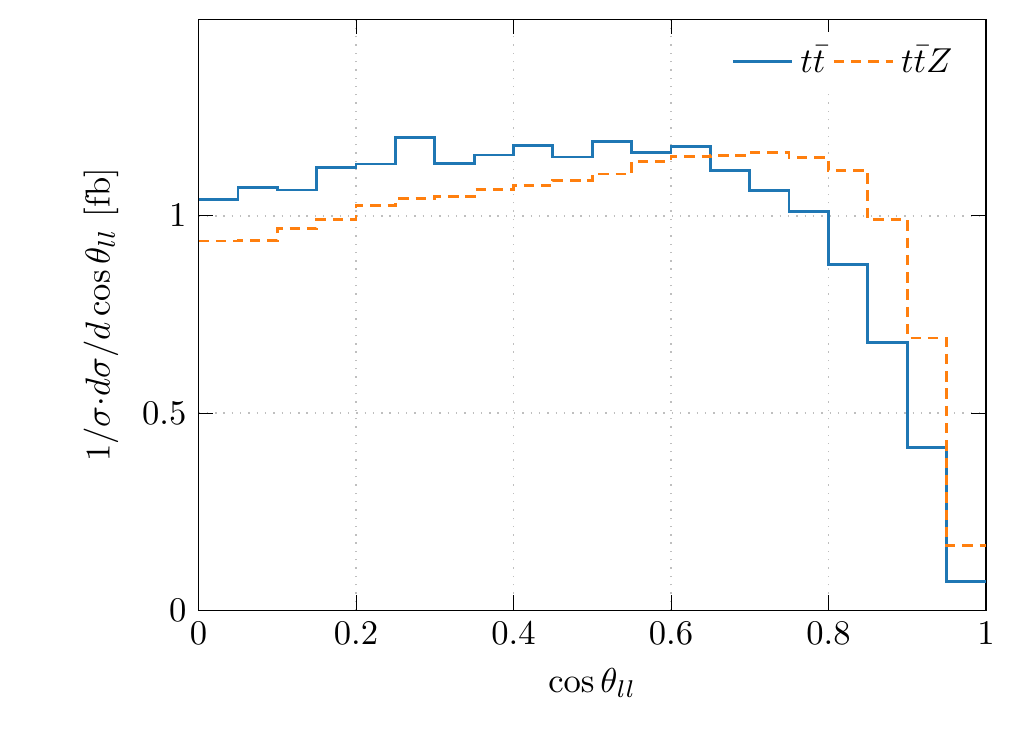}
\end{center}
\vspace{-0.6cm}
\caption{\it Comparison of the normalised NLO differential cross
sections for $pp\to e^+ \nu_e \mu^- \bar{\nu}_\mu b\bar{b} +X$ and
$pp\to e^+ \nu_e \mu^- \bar{\nu}_\mu b\bar{b} \, \nu_\tau
\bar{\nu}_\tau +X$ at the LHC with $\sqrt{s} = 13$ TeV. The following
distributions are shown: $p_T^{miss}$, $H_T$, $H_T^{\prime}$,
$m_{\ell\ell}$, (averaged) $p_{T,\, \ell}$, $\Delta \phi_{\ell\ell}$,
$\Delta y_{\ell\ell}$ and $\cos \theta_{\ell\ell}$. The NLO CT14 PDF
set is employed and the scale choices $\mu_0=H_T/3$ $(t\bar{t}Z)$ and
$\mu_0=H_T/4$ $(t\bar{t} \,)$ are utilised.}
\label{fig:normalised}
\end{figure}
%

We observe that the integrated cross section for top quark pair
production in the di-lepton top quark decay channel is 4 orders of
magnitude larger than the one for $pp\to e^+ \nu_e \mu^- \bar{\nu}_\mu
b\bar{b} \, \nu_\tau \bar{\nu}_\tau$.  As already mentioned typical
new physics scenarios predict cross sections for $ \ell^\pm \ell^\mp
b\bar{b} +p_T^{miss}$ that are of the order of femtobarns. Thus, very
exclusive and sophisticated cuts have to be employed to reduce the
size of the reducible top quark background process keeping a sizeable
amount of signal events at the same time. Such cut selection, that are
designed to diminish the double and single top quark resonance
contributions, would enhance the $W^+W^- b\bar{b}$ part in both
$t\bar{t}$ and $t\bar{t}Z$ background processes. Moreover, the signal
is expected to be a subtle excess over the SM backgrounds in the tails
of kinematic variables, e.g. in the invariant mass of two charged
leptons $m_{\ell\ell}$. Furthermore, shape differences in differential
cross sections between signal and background processes can potentially
be exploited to increase the signal to background ratio. Therefore, a
proper modelling of top quark decays, including QCD effects in the top
quark decay chain and incorporating the complete off-shell effects of
the top quark, is simply mandatory.

In the following we concentrate on shape differences between the two
main background processes $pp\to e^+ \nu_e \mu^- \bar{\nu}_\mu
b\bar{b}$ and $pp\to e^+ \nu_e \mu^- \bar{\nu}_\mu b\bar{b} \,
\nu_\tau \bar{\nu}_\tau$. To this end in Figure \ref{fig:normalised}
normalised differential distributions constructed from final states
for both $pp\to e^+ \nu_e \mu^- \bar{\nu}_\mu b\bar{b}$ and $pp\to e^+
\nu_e \mu^- \bar{\nu}_\mu b\bar{b} \, \nu_\tau \bar{\nu}_\tau$ are
depicted. They are given at NLO in QCD for the CT14 PDF set and for
the dynamical scale choice. Specifically, for the $pp\to e^+ \nu_e
\mu^- \bar{\nu}_\mu b\bar{b} \, \nu_\tau \bar{\nu}_\tau$ process
$\mu_R=\mu_F=\mu_0 = H_T /3$ is used and for $pp\to e^+ \nu_e \mu^-
\bar{\nu}_\mu b\bar{b}$ the scale choice $\mu_0= H_T /4$ is
utilised instead. For the total missing transverse momentum we notice
large differences between the two background processes. Assuming for
example, that it would be sufficient to consider the $t\bar{t}$
background only in new physics analyses in the $t\bar{t}+p_T^{miss}$
channel  is not  satisfactory or acceptable. We can observe that the
$p_T^{miss}$ observable, which is always employed to suppress the
top-like backgrounds, has a harder missing $p_T$ spectrum for $e^+
\nu_e \mu^- \bar{\nu}_\mu b\bar{b} \, \nu_\tau \bar{\nu}_\tau$ than in
the case of the $e^+ \nu_e \mu^- \bar{\nu}_\mu b\bar{b}$
background. In the latter case $p_T^{miss}$ is highly peaked towards
low values. In the former case the primary source of the long
$p_T^{miss}$ tail is the neutrinos from the $Z$ gauge boson decay.
Because the $pp\to e^+ \nu_e \mu^- \bar{\nu}_\mu b\bar{b}$ production
process does not exhibit long enough tails in the $p_T^{miss}$
distribution any final $S/B$ ratios as calculated with the help of
$pp\to e^+ \nu_e \mu^- \bar{\nu}_\mu b\bar{b}$ only can be grossly
overestimated. Consequently, limits on the signal strengths, which are
usually translated into constraints on the parameter space of new
physics models, might not be realistic. 

Large shape differences are also visible in the case of $H_T$, which
is not surprising since $p_T^{miss}$ is incorporated in the definition
of the observable.  We can further notice, however, that the shape of
various observables built out of the charged leptons and $b$-jets
only, i.e. out of visible final states, have been changed by the
enlarged $p_T^{miss}$. Shape differences can be noticed both for
dimensionful and dimensionless observables. In the case of $H_T^{
\prime }$, $m_{\ell\ell}$ and the (averaged) transverse momentum of
the charged lepton, $p_{T, \, \ell}$ the spectra are harder when the
additional contribution to $p_T^{miss}$ is included. In the case of
dimensionless observables we depict $\Delta \phi_{\ell\ell}$, $\Delta
y_{\ell\ell}$ and $\cos\theta_{\ell\ell}$ where shape differences are
the most pronounced and visible over the whole plotted range. Let us
say again at this point that both $H_T$ and $H_T^{ \prime }$ are very
often used to further suppress reducible top quark backgrounds in new
physics analyses, whereas $\Delta \phi_{\ell\ell}$, $\Delta
y_{\ell\ell}$ and $\cos\theta_{\ell\ell}$ are regularly employed
either to enhance sensitivity of the new physics signal or to verify
the hypothesis of scalar/vector nature of the new heavy resonances
that are associated with various BSM hypothesis.  Consequently, the
$pp\to \nu_e \mu^- \bar{\nu}_\mu b\bar{b} \, \nu_\tau \bar{\nu}_\tau$
irreducible background process has to be always taken into account and
carefully studied for the proper description of relevant observables
in the $t\bar{t}+p_T^{miss}$ channel.

%
\section{Summary and Conclusions}
\label{sec:5}
%
%

In this paper, we have presented the first complete NLO QCD prediction
for the $pp \to t\bar{t}Z(\to \nu_\tau\bar{\nu}_\tau)$ process in the
di-lepton top quark decay channel for the LHC run II energy of
$\sqrt{s}=13$ TeV. With an inclusive cut selection and for
$\mu_R=\mu_F=\mu_0=m_t+m_Z/2$ NLO QCD corrections reduce the
unphysical scale dependence by a factor of $6$ ($8$ after
symmetrisation of errors) and increase the total rate by about $12\%$
compared to the LO prediction.  The theoretical uncertainty of the NLO
cross section as estimated from scale dependence is $5.9\%$ ($3.5\%$
after symmetrisation). By comparison, the PDF uncertainties are
negligible. Taken in a very conservative way, they are of the order of
$3.4\%$. After symmetrisation, they are reduced down to
$2.0\%$. Consequently, the theoretical uncertainty resulting from the
scale variation remains the dominant source of theoretical
systematics.

Similar conclusions can be drawn from the results with $\mu_0=
H_T/3$ and $\mu_0=E^{\prime\prime}_T/3$. Specifically, NLO QCD
corrections of the order of $1\%$ and $5\%$ have been obtained
respectively for $\mu_0= H_T/3$ and $\mu_0=E^{\prime\prime}_T/3$.  Our
best NLO QCD predictions for the $pp\to e^+\nu_e\mu^-\bar{\nu}_\mu
b\bar{b} \nu_\tau \bar{\nu}_\tau$ process can be summarised as
\begin{equation}
  \begin{split}
\sigma^{\rm NLO}_{pp\to e^+\nu_e\mu^-\bar{\nu}_\mu b\bar{b} \nu_\tau
  \bar{\nu}_\tau \,} \left(\mu_0=H_T/3\right)
&=  {0.1270}^{+0.0009 (0.7\%)}_{- 0.0086 (6.8\%)} \, {\rm  [scales]}\,
{}^{+ 0.0008\, (0.6\%)}_{+0.0042\, (3.3\%)} \,{\rm [PDF]} \,  {\rm
  fb}\,, \\[0.2cm]
\sigma^{\rm NLO}_{pp\to e^+\nu_e\mu^-\bar{\nu}_\mu b\bar{b} \nu_\tau
  \bar{\nu}_\tau \,} \left(\mu_0=E_T^{\prime\prime}/3\right)
&=  0.1286^{+ 0.0013 (1.0\%)}_{- 0.0060 (4.7\%)} \, {\rm  [scales]}\,
{}^{+0.0009 \, (0.7\%)}_{+0.0044\, (3.4\%)} \,{\rm [PDF]} \,  {\rm
  fb}\,.
\end{split}
\end{equation}
The complete cross section for $pp\to \ell^+ \nu_\ell \ell^-
\bar{\nu}_\ell \, b\bar{b} \, \nu_\ell \bar{\nu}_\ell$, where
$\ell=e,\mu$ and $\nu_\ell=\nu_e,\nu_\mu,\nu_\tau$ can be obtained by
multiplying the above results by $12$. Despite the relatively small
cross section, good theoretical control over the $pp\to e^+ \nu_e
\mu^- \bar{\nu}_\mu b\bar{b} \, \nu_\tau \bar{\nu}_\tau$ production
process is phenomenologically relevant. This irreducible SM background
is of the order of $1.5$ fb at NLO in QCD. For comparison, typical
predictions for DM scenarios are also at a similar level. 

In a next step, we examined the size of NLO QCD corrections to various
differential distributions with the different scale choices that we
have proposed. We started with standard observables that describe
charged lepton and $b$-jet kinematics. We have thoroughly investigated
the following set of observables: (averaged) $p_{T,\, \ell}$,
$m_{\ell\ell}$, (averaged) $y_\ell$ and $\Delta R_{\ell\ell}$ as well
as (averaged) $p_{T,\, b}$, $m_{b\bar{b}}$, (averaged) $y_b$ and
$\Delta R_{b\bar{b}}$. Differential cross sections have shown large
differences in shape with respect to LO within our fixed-scale
setting, i.e. for $\mu_0=m_t+m_Z/2$. In particular, large negative
corrections have been clearly seen in the tails of several
distributions for dimensionful observables. Thus, an accurate
description of the shapes of observables can only be given via a full
NLO QCD computation in this case. Adopting $\mu_0=H_T/3$ and
$\mu_0=E^{\prime\prime}_T/3$ dynamical scale choices, resulted in
moderate higher order QCD corrections up to $10\%-15\%$. The NLO
theoretical uncertainties for the leptonic and $b$-jet observables as
estimated from scale variation were of the order of $\pm10\%$ ($\pm
5\%$ after symmetrisation).  Combining information about the size of
NLO QCD corrections and the NLO QCD theoretical uncertainties, we
concluded that either scale $\mu_0=H_T/3$ or $\mu_0=E^{\prime
\prime}/3$ may be employed at the differential level for an adequate
description of the standard observables in the $e^+\nu_e\, \mu^-
\bar{\nu}_\mu \, b\bar{b} \, \nu_\tau \bar{\nu}_\tau$ production
process at the LHC with a centre of mass system energy of
$\sqrt{s}=13$ TeV in the presence of rather inclusive cuts on the
measured final states.

Moving forward, we employed our recommended dynamical scale choices to
discuss the size of NLO QCD corrections to a few observables that are
particularly relevant in the context of dark matter searches.  Among
others, we have identified six observables, three dimensionful $E_T,
H_T$ and $H_T^{\prime}$ as well as three dimensionless $\Delta
y_{\ell\ell}, \Delta \phi_{\ell\ell}$ and
$\cos\theta_{\ell\ell}$. Substantial NLO QCD corrections up to $35\%$
$(\pm20\%)$ have been obtained for dimensionful (dimensionless)
observables. Overall, the differential ${\cal K}$-factors show major
changes in the shape of the observables. On the other hand, NLO QCD
theoretical uncertainties up to $\pm 20\%$ $(\pm 10\%)$ have been
estimated from scale variation. Well behaved as they are at NLO in
QCD, these leptonic observables can now be safely utilised to probe
new physics at the LHC.

Among all infrared-safe observables in $e^+\nu_e\,
\mu^-\bar{\nu}_\mu\, b\bar{b}\nu_\tau\, \bar{\nu}_\tau$ production,
the total missing transverse momentum plays a special role. The
observation of an access in $p_T^{miss}$ represents the most important
signature in various BSM and DM models. Thus, we investigated this
observable separately. We observed substantial NLO QCD corrections up
to $57\%$ and $48\%$ when our recommended scale choice, based either
on $\mu_0=H_T/3$ or $\mu_0=E_T^{\prime\prime}/3$, was
employed. Predictions based on the fixed scale choice
$\mu_0=m_t+m_Z/2$, however, received NLO QCD corrections up to $27\%$
only and showed shape distortions up to $19\%$. This behaviour is to
be contrasted with the behaviour for other observables, where the
dynamical scale choice guaranteed reduced shape distortions. In order
to understand why the fixed scale choice performed better for the
$p_T^{miss}$ observable, we analysed the double differential NLO cross
section distribution expressed as a function of $p_T^{miss}$ and
$m_{t\bar{t}}$. Furthermore, we investigated two additional
observables: the transverse momentum of the $Z$ boson reconstructed
from its invisible decay products $(p_{T, \,Z})$ as well as the
missing transverse momentum restricted to the invisible particles
coming from the decays of the top quarks only $(p_T^{\prime \,
miss})$. Our differential analysis revealed that in the case of
$p_T^{miss}$ and $p_T^{\prime \, miss}$ the proposed dynamic scale
choices resulted in too large scales and the fixed scale choice was
simply more adequate.

In a next step, we studied the theoretical uncertainty related to the
parameterisation of PDFs. For all observables that we have scrutinised
the PDF uncertainties have been substantially smaller than the
theoretical uncertainties from the scale dependence. The latter
remains the dominant source of the final theoretical error for our
predictions at NLO in QCD.

Finally, because $pp\to e^+ \nu_e \mu^- \bar{\nu}_\mu b\bar{b}$ and
$pp\to e^+ \nu_e \mu^- \bar{\nu}_\mu b\bar{b} \, \nu_\tau
\bar{\nu}_\tau$ comprise the same final states (two charged leptons,
two bottom flavoured jets and missing transverse momentum from
undetected neutrinos) we compared the two production processes to
quantify the impact of the enlarged missing transverse momentum on the
kinematics of the final state. Substantial shape differences have been
observed both for dimensionful and dimensionless observables. Both
kinds of observables have often been employed to enhance sensitivity
of the new physics signal or to verify the hypothesis of scalar/vector
nature of the new heavy resonances that are associated with various
BSM hypotheses.  For example, since the $pp\to e^+ \nu_e \mu^-
\bar{\nu}_\mu b\bar{b}$ production process does not exhibit long
enough tails in the $p_T^{miss}$ distribution, any final $S/B$ ratios
as calculated with the help of $pp\to e^+ \nu_e \mu^- \bar{\nu}_\mu
b\bar{b}$ only, can be grossly overestimated. As a result, limits on
the signal strengths, which are usually translated into constraints on
the parameter space of new physics models, might not be very
realistic. Consequently, the $pp\to \nu_e \mu^- \bar{\nu}_\mu b\bar{b}
\, \nu_\tau \bar{\nu}_\tau$ irreducible background process must be
additionally taken into account in searches of new physics in the
$t\bar{t}+p_T^{miss}$ channel. Good theoretical control over the
irreducible SM background is, therefore, a fundamental prerequisite
for a correct interpretation of possible signals of new physics that
may arise in this channel.

On the technical side let us mention that all results have been
generated with the help of the \textsc{Helac-NLO} MC framework. The
results  are available as event files in the form of either LHEFs or ROOT
Ntuples. These might be directly used for experimental analyses at the
LHC as well as for obtaining accurate SM predictions in BSM studies.
The Ntuple files are available upon request.

\acknowledgments

The work of M.W. and T.W.  was supported in part by the German
Research Foundation (DFG) Individual Research Grant: {\it "Top-Quarks
under the LHCs Magnifying Glass: From Process Modelling to Parameter
Extraction"} and in part by the DFG Collaborative Research
Centre/Transregio project CRC/TRR 257: {\it "P3H - Particle Physics
  Phenomenology after the Higgs Discovery''}.

The work of H.B.H. has
received funding from the European Research Council (ERC) under the
European Union's Horizon 2020 research and innovation programme (grant
agreement No 772099) and partial support from Rutherford Grant
ST/M004104/1.

The research of G.B. was supported by grant K 125105 of the National
Research, Development  and Innovation Office in Hungary.

Support by a grant of the Bundesministerium f\"ur Bildung und
Forschung (BMBF) is additionally acknowledged.

Simulations were performed with computing resources granted by RWTH
Aachen University under project \texttt{rwth0414}.

\end{document}